\documentclass[showpacs,preprintnumbers,amsmath,amssymb]{revtex4}

\usepackage{graphicx}
\usepackage{dcolumn}
\usepackage{bm}
\usepackage{url}
\usepackage{epsfig}
\usepackage{multirow}
\usepackage{tensor}
\usepackage{empheq}
\usepackage{amsmath}
\usepackage{color}
\definecolor{myblue}{rgb}{.8, .8, 1}

\def\be{\begin{equation}}
\def\ee{\end{equation}}
\def\ba{\begin{eqnarray}}
\def\ea{\end{eqnarray}}

\newcommand{\fr}[2]{\frac{#1}{#2}}

\def\ga{\mathrel{\raise.3ex\hbox{$>$\kern-.75em\lower1ex\hbox{$\sim$}}}}
\def\la{\mathrel{\raise.3ex\hbox{$<$\kern-.75em\lower1ex\hbox{$\sim$}}}}

\begin{document}

\title{Lagrangian perturbations theory : Third-order solution for general dark energy models}


\author{Seokcheon Lee}
\affiliation{School of Physics, Korea Institute for Advanced Study, Heogiro 85, Seoul 130-722, Korea}


\begin{abstract}
We investigate the Lagrangian perturbation theory of a general dark energy models with a constant equation of state, $\omega$, and derive the fitting form of the fastest growing mode solutions up to the third order. These fitting forms are less than few \% errors compared to the numerical calculation. We also correct the solutions of Einstein-de Sitter and the open universe with the proper initial conditions and the correct consideration for the growing mode solutions in order to compare with the general dark energy models. Even though these fitting forms are limited to the constant equation of state models, one can apply these to the time-varying $\omega$ by interpolating between models with the constant $\omega$. These fitting forms can be used for the correct prediction for the two point correlation function and can give the proper predictions for the different dark energy models.

\end{abstract}


\maketitle

\section{Introduction}
\setcounter{equation}{0}
The formation and the evolution of the large scale structures (LSS) in the Universe through the gravitational instability is one of the important probes of the cosmological parameters \cite{LSS}. One might be able to constrain the nature of dark energy through future galaxy surveys aim at a percent precision on a broad range of scales. One can rely on the linear perturbation theory on large scales where the amplitude of the density fluctuations is small, whereas numerical simulations or phenomenological models, such as the halo model are necessitate to investigate the small scale LSS which incorporate non-linear effects \cite{0206508}. The analytical investigations beyond the linear theory are suitable for the accurate theoretical predictions to the intermediate scales observations, such as the baryonic acoustic oscillations (BAO) \cite{9709112}.

There have been several analytic techniques for the quasi-linear perturbative expansion in order to solve the equations for an irrotational and pressureless fluid of cold dark matter (CDM) \cite{13112724}. Two major methods are the standard (or Eulerian) perturbation theory (SPT) and the Lagrangian perturbation theory (LPT). In SPT, the density and the velocity field of the fluid are the perturbed quantities which need to be small in order to satisfy the perturbative expansion \cite{0112551}. The perturbed quantity is the deviation of the particle trajectory field from the homogeneous background expansion in LPT \cite{9609036, 0412025}. Thus, LPT does not rely on the smallness of the density and velocity fields. However, most of these approaches are based on the single stream approximation and neglect the shell-crossing effects.

The solutions for higher orders of the LPT on Friedmann Lema\^{i}tre Robertson Walker (FLRW) background have been investigated \cite{9609036}. For irrotational flows, the general second order solution is found \cite{BE}, the third order one with slaved initial conditions is obtained \cite{9309055}, and the fourth order one is also solved \cite{12034260, 12108306}. These solutions are obtained for the Einstein-de Sitter universe and/or the open universe. However the Universe is dominated by the dark energy at present epoch and thus one needs to find the higher order solutions based on the dark energy with CDM ($\omega$CDM) universe. The third order solutions in Eulerian perturbation theory for the general dark energy model was investigated \cite{08061437}

Lagrangian resummation theory (LRT) has been developed which investigate both non-linear biasing and redshift space distortions \cite{07112521, 08071733, Err, 11024619}. From these method, one can predict BAO peaks in both real and redshift spaces \cite{08122905, 09091802, 12090780, 13066660, 13073285}. Even though LRT improves the theoretical prediction for the quasi-linear observable, it is not so accurate compared to the realistic value because one adopts the LPT higher order solutions based on EdS universe instead of $\omega$CDM.

In the present paper, we reexamine the fastest growing mode solutions for the LPT up to the third order for the EdS and open universe. Even though there have been the known solutions for these models, they are given by incorrect initial conditions and thus provides the improper coefficients. From solutions for the EdS model with the correct initial condition, one can obtain the exact numerical LPT solutions for flat $\omega$CDM universe. We provide the approximate analytic solutions for $\omega$CDM models up to the third order with less than 5 \% errors. We are preparing the dark energy dependent the power spectrum and other statistical observables incorporating these solutions.

In section 2 we briefly review LPT. The differential equations order by order for the general $\omega$CDM models are given. Then, in section 3 we investigate the solution for the each order for the EdS and the open universe with the correct initial conditions. We also obtain the analytic approximation solution of the displacement vector up to the third order for $\omega$CDM models. We make conclusions in section 4.

\section{Lagrangian Perturbation Theory}
\setcounter{equation}{0}
It is convenient to define the physical distance as $\vec{r} = a(t) \vec{x}$ where $a(t)$ is the expansion scale factor and $\vec{x}$ is the comoving Eulerian coordinates.
The gravitational evolution of structure for a collisionless (cold) dust in subhorizon scales corresponding to those of the large scale structure at present is described by the Newtonian equations (mass conservation, Euler, and Poisson equation, respectively)
\ba \fr{\partial \delta}{\partial t} + \fr{1}{a} \vec{\nabla} \cdot \Bigl[ (1 + \delta) \vec{v} \Bigr] &=& 0 \, , \label{mass} \\
\fr{\partial \vec{v}}{\partial t} + \fr{1}{a} \Bigl( \vec{v} \cdot \vec{\nabla} \Bigr) \vec{v} + \fr{\dot{a}}{a} \vec{v} &=& - \fr{1}{a} \vec{\nabla} \phi - \fr{1}{a \rho} \nabla_{j} (\rho \sigma_{ij} ) \, , \label{Euler} \\
\nabla^2 \phi - 4 \pi G a^2 \bar{\rho}_m \, \delta &=& 0 \, , \label{Poisson} \ea
where $\delta(\vec{x},t)$ is the density contrast, $\vec{v}(\vec{x},t) = a \dot{\vec{x}}$ is the peculiar velocity, $\phi(\vec{x},t)$ is the peculiar
gravitational potential, $\bar{\rho}_m$ is the mean matter energy density, $\sigma_{ij}$ is the stress tensor \cite{LSS, 0112551, 13112724}. Note that the mass conservation equation
couples the zeroth moment ($\rho$) to the first one ($\vec{v}$) of the distribution function, the Euler equation couples the first moment ($\vec{v}$)
to the second one ($\sigma_{ij}$), and so on. 
One can close the hierarchy by postulating an ansatz for the stress tensor $\sigma_{ij}$, i.e. the equation of state of the cosmological fluid \cite{0112551}. For this purpose, one can assume the irrotationality condition $\nabla \times \vec{v} = 0$ in addition to the above equations (\ref{mass}) - (\ref{Poisson}). In Eulerian perturbation theory, one expands the density contrast as
\be \delta(\vec{x},t) \equiv \sum_{n=1}^{\infty} \epsilon_{n}(\vec{x}) \delta^{(n)}(t) \label{deltaEul} \, ,\ee
where $\epsilon$ is just a book-keeping device with $\epsilon \ll 1$. 
From the second order, the behavior of $\delta$ is no longer local. The density perturbation at one spot depends on the initial perturbation at other
places through the peculiar velocity. 

Instead of working on Eulerian non-linear perturbation theory which studies the dynamics of density and velocity fields, it is possible to develop non-linear perturbation theory in the Lagrangian scheme by following the trajectories of particles or fluid elements \cite{Buchert, Moutarde, BJCP}. This is possible because the choice of fields representing the cosmic quantities is not unique and Lagrangian approach is a change of the coordinate system itself. According to the Lagrangian point of view, the path of each fluid is followed during its evolution. Each fluid particle is labeled by its initial coordinate, $\vec{q} \equiv \vec{x}(t_0)$ and the position of the $\vec{q}$-particle at later time $t$ may be described by
\be \vec{x}(\vec{q},t) = \vec{q} + \vec{S}(\vec{q},t) \, , \label{vecx} \ee
where $\vec{S}$ is the displacement vector. Thus, the motion of the fluid element may be completely described by the displacement $\vec{S}$.
A slight perturbation of the Lagrangian particle paths carries a large amount of non-linear information about the corresponding Eulerian
evolved observable, since the Lagrangian picture is intrinsically non-linear in the density field. This overcomes the difficulty of the
Eulerian approach by allowing the large density contrast $\delta$. 
During the highly non-linear evolution, many particles coming from the different original positions will tend to arrive at the same
Eulerian place. 
Since the initial mass density field is sufficiently uniform, one can assume that the Eulerian mass density $\rho_{m}(\vec{x},t)$
at any given time $t$ satisfies the continuity relation
\be \rho_{m}(\vec{x},t) d^3 x = \bar{\rho}_{m}(\vec{x}_0) \Bigl( 1 + \delta (\vec{x},t) \Bigr) d^3 x = \bar{\rho}_{m}(\vec{q}) d^3 q \label{deltam}
\, . \ee 
If one uses $\vec{x}(t_0) = \vec{q}$, then one obtains \be 1 + \delta(\vec{x},t) = \textrm{J}(\vec{q},t)^{-1} \label{deltam2} \, , \ee where $\textrm{J}(\vec{q},t)$ is the determinant of the Jacobian matrix ${\cal J}$ of the transformation from Lagrangian to Eulerian space, $\textrm{J}(\vec{q},t) \equiv {\rm det} \Bigl(\fr{\partial \vec{x}}{\partial \vec{q}} \Bigr) \equiv \textrm{det} {\cal J}$. If one wants to express the Euler equation and the Poisson equation given by Eqs. (\ref{Euler}) and (\ref{Poisson}) in terms of $\vec{S}$, then one needs to change the spatial derivatives operator $\nabla_{\vec{x}}$ into the differentiation with respect to the Lagrangian position $\vec{q}$
\be \nabla_{\vec{x}} \equiv \fr{\partial}{\partial x_{\alpha}} = \fr{\partial q_{\beta}}{\partial x_{\alpha}} \fr{\partial}{\partial q_{\beta}} \equiv \Bigl( J_{\alpha \beta} \Bigr)^{-1} \nabla_{\beta} \, , \label{ct} \ee  where $\nabla = \nabla_{\vec{q}}$ and the Jacobian matrix $J_{\alpha \beta}$ is given by
\be J_{\alpha \beta} \equiv {\cal J} = \fr{\partial x_{\alpha}}{\partial q_{\beta}} = \delta_{\alpha \beta} +
\fr{\partial S_{\alpha}}{\partial q_{\beta}} \equiv \delta_{\alpha \beta} + S_{\alpha \beta} = {\cal I} + {\cal S} \label{J} \, , \ee
where ${\cal I}$ is the identity matrix and ${\cal S}$ is a $3 \times 3$ matrix whose elements are
$S_{\alpha\beta}$ called the deformation tensor. In general, the deformation tensor is not
symmetric, i.e. $S_{\alpha \beta} \neq S_{\beta \alpha}$. $S_{\alpha \beta}$ is symmetric if and only if the displacement vector $\vec{S}$ is an
irrotational field in the Lagrangian space \cite{9406016}.
Thus, one obtains the inverse Jacobian matrix as
\be {\cal J}^{-1} = \fr{1}{\textrm{J}} \Bigl( (1+ \nabla \cdot S) \delta_{\alpha\beta} -S_{\alpha\beta} +S^{c}_{\alpha\beta} \Bigr)
\label{Jm1} \, , \ee where $S^{c}_{\alpha\beta}$ is an element of the cofactor matrix ${\cal S}^{c}$. If one takes the divergence on the Euler
equation (\ref{Euler}) using the above relations between the Eulerian and the Lagrangian space, then one obtains \cite{9406016, 9604078}
\be \Bigl( (1+ \nabla \cdot S) \delta_{\alpha\beta} -S_{\alpha\beta} +S^{c}_{\alpha\beta} \Bigr) \Bigl(\fr{d^2}{d t^2}  S_{\beta\alpha} + 2 H \fr{d }{dt} S_{\beta\alpha} \Bigr) = 4 \pi G \rho_{m} (\textrm{J}-1) \, . \label{Gii3} \ee
In addition to the above Eq. (\ref{Gii3}), one can impose the irrotationality of the peculiar velocity \cite{12034260}
\be \epsilon_{\alpha\beta\gamma} \dot{S}_{\gamma\beta} = \epsilon_{\alpha\beta\gamma} S_{\mu\beta} \dot{S}_{\mu\gamma} + S_{\alpha\nu} \epsilon_{\nu\beta\gamma} \Bigl( S_{\mu\beta} \dot{S}_{\mu\gamma} - \dot{S}_{\gamma\beta} \Bigr) \, . \label{irro} \ee
Now one expands the displacement vector $\vec{S}$ according to the Lagrangian perturbative prescription \be \vec{S}(t, \vec{q}) = D[t] \vec{S}^{(1)}(\vec{q}) + E[t] \vec{S}^{(2)}(\vec{q}) + F[t] \vec{S}^{(3)}(\vec{q}) + F_{T}[t] \vec{T}^{(3)} + \cdots \, . \label{Sexp} \ee This explicit separation with respect to the spatial and temporal coordinates for each order is not an assumption but a property of the perturbative Lagrangian description for an Einstein-de Sitter universe \cite{9406016}. The solutions at each order can be non-separable functions of $t$ and $\vec{q}$ for the general model, but we will adopt the above ansatz for our models. Thus, if one adopts the above expression for $\vec{S}$ in Eq. (\ref{Sexp}), then one can expand the right hand side term of Eq. (\ref{Gii3}) as
\ba \textrm{J} - 1 &=& S_{11} + S_{22} + S_{33} + S_{11}S_{22} + S_{11}S_{33} + S_{22}S_{33} - S_{12}S_{21} - S_{13}S_{31} - S_{23}S_{32} \nonumber \\ &+& S_{11}S_{22}S_{33} - S_{11}S_{23}S_{32} + S_{12}S_{23}S_{31} - S_{12}S_{21}S_{33} + S_{13}S_{21}S_{32} - S_{13}S_{22}S_{31} \nonumber \\ &\equiv& \mu_{1}(\vec{S}) + \mu_{2}(\vec{S}) + \mu_{3}(\vec{S}) \nonumber \\ &=& D \mu_{1}(S^{(1)}) + E \mu_{1}(S^{(2)}) + D^2 \mu_{2}(S^{(1)},S^{(1)})  + F \mu_{1}(S^{(3)}) \label{detJ} \\ &+& DE \mu_{2}(S^{(1)},S^{(2)}) +  DE \mu_{2}(S^{(2)},S^{(1)}) + D^3 \mu_{3}(S^{(1)}) \, , \nonumber \ea
where $\mu_{a}^{(n)}$ are defined by
\ba \mu_{1}(S^{(n)}) &\equiv& S_{ii}^{(n)} \label{mu1} \, , \\
\mu_{2}(S^{(n)},S^{(m)}) &\equiv& \fr{1}{2} \Bigl( S_{ii}^{(n)} S_{jj}^{(m)} - S_{ij}^{(n)} S_{ji}^{(m)} \Bigr) \label{mu2} \, , \\
\mu_{3}(S^{(n)}) &\equiv& {\rm det} S_{ij}^{(n)} \label{mu3} \, . \ea
Thus, one can obtain Lagrangian Poisson equation order by order (from the linear to the third orders)
\ba \ddot{D} + 2H \dot{D} - 4 \pi G \rho_{m} D &=& 0 \, , \label{Deq} \\
\ddot{E} + 2H \dot{E} - 4 \pi G \rho_{m} E &=& - 4 \pi G \rho_{m}  D^2 \, ,  {\rm if} \, \mu_{1}( S^{(2)} ) = \mu_{2}(S^{(1)},S^{(1)}) \, , \label{Eeq} \\
\ddot{F}_{a} + 2H \dot{F}_{a} - 4 \pi G \rho_{m} F_{a} &=& -8 \pi G \rho_{m} D^3 \,  , {\rm if} \, \mu_{1}( S^{(3)} ) = \mu_{3}(S^{(1)}) \, , \label{Faeq} \\
\ddot{F}_{b} + 2H \dot{F}_{b} - 4 \pi G \rho_{m} F_{b} &=& - 8 \pi G \rho_{m}  D (E - D^2)  \, , {\rm if} \, \mu_{1}(S^{(3)}) = \mu_{2}(S^{(1)},S^{(2)}) = \mu_{2}(S^{(2)},S^{(1)}) \, , \label{Fbeq} \ea
where dots represent the derivatives with respect to the cosmic time $t$ and $\mu_{2}(S^{(1)},S^{(2)}) = \mu_{2}(S^{(2)},S^{(1)})$ is satisfied for any tensor \cite{12034260}. Also, one can obtain the one more equation for the irrotationality
\be \ddot{F}_{T} + 2 H \dot{F}_{T} = - 4 \pi G \rho_{m} D^3 \, , {\rm if} \, \epsilon_{\alpha\beta\gamma} T_{\gamma\beta}^{(3)} = \epsilon_{\alpha\beta\gamma} S_{\mu\beta}^{(1)} S_{\mu\gamma}^{(2)} \, . \label{FTeq} \ee

\section{LPT solutions}
\setcounter{equation}{0}

We now solve the dynamical equations for the temporal part of the displacement vector $\vec{S}$ order by order given by Eqs. (\ref{Deq}) - (\ref{Fbeq}). We investigate the solutions for three different models, Einstein-de Sitter (EdS), open universe, and the universe of the general dark energy with the constant equation of state ($\omega$CDM). Even though the solutions for both EdS and open universe are well known, we reinvestigate those models on purpose. We can obtain the initial conditions of the fastest growing mode solutions for the general dark energy models from those of EdS because both models are close to each other at early epoch. We also check the so-called BJCP solutions for the open universe with the proper initial conditions to correct them \cite{BJCP}. We obtain the fitting forms of the fastest growing solutions for $\omega$CDM. We show the errors of these fitting forms in the appendix.

\subsection{First order time component}
Even though the exact first order solutions for the considered models are well known, we reinvestigate these solutions by solving Eq. (\ref{Deq})
\be \ddot{D} + 2H \dot{D} - 4 \pi G \rho_{m} D = 0 \, . \label{1DE} \ee
Now we show the solutions of the above equation for the different cosmological models.  \\

i) case I : EdS \\
\\
\noindent By replacing the cosmic time $t$ with $\tau$ as $d \tau = \fr{\alpha}{a^2} d t$,  the above Eq. (\ref{1DE}) is rewritten by
\be \fr{d^2 D}{d \tau^2} - \fr{4 \pi G \rho_{m}}{\alpha^2} a^4 D = 0  \,\, \rightarrow \,\, \fr{d^2 D}{d x^2} - \fr{6}{x^2} D = 0 \, , \label{1DEdS} \ee where $x = \fr{\tau}{\tau_0} = \fr{1}{\sqrt{a}}$. The solution of the above equation can be written as
\be D = c_{aD} g_{1a} + c_{bD} g_{1b} \, , \label{DyEdS} \ee where $c_{aD}$ and $c_{bD}$ are the integral constants and
\ba g_{1a} &=& x^{-2} = a \, , \label{g1aEdS} \\
g_{1b} &=& x^{3} = a^{-\fr{3}{2}} \, . \label{g1bEdS} \ea
$g_{1a}$ is a growing mode solution and $g_{1b}$ is the decaying one. One can set $c_{aD} = 1$ from the initial condtion $g_{1a}(a_i) = a_i$. \\
\begin{center}
\begin{figure}
\vspace{1.5cm}
\centerline{\epsfig{file=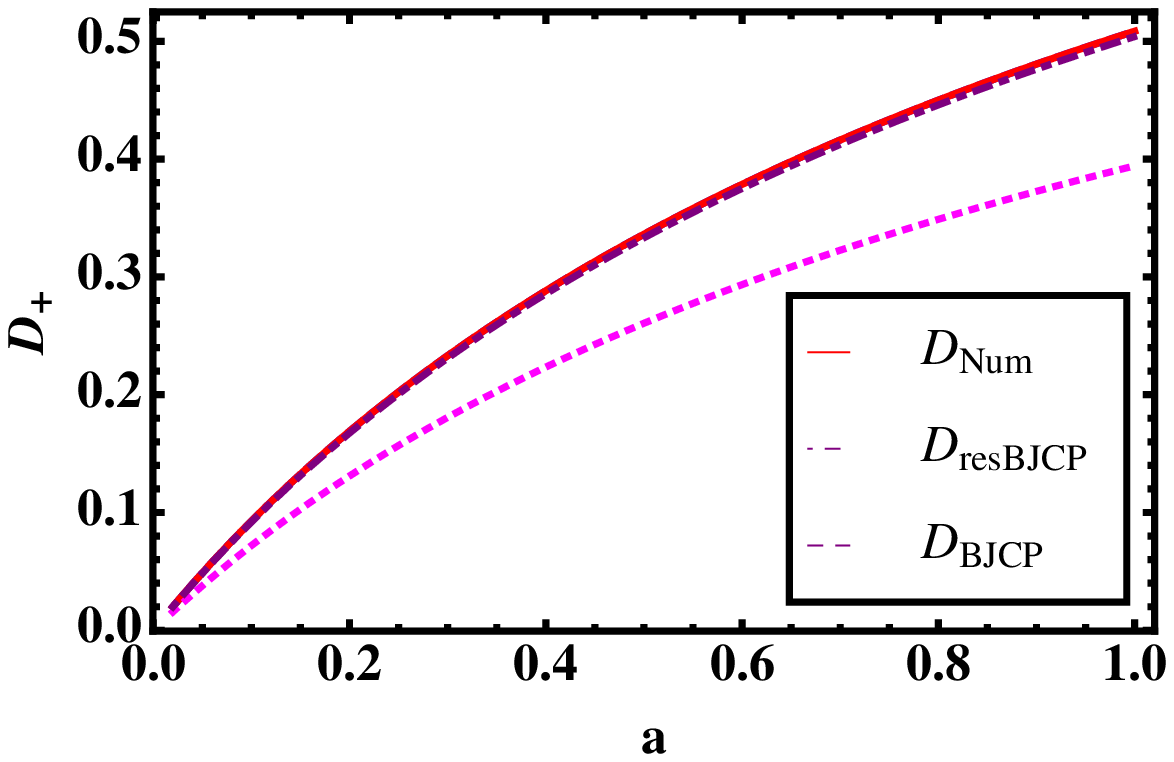, width=7.0cm} \epsfig{file=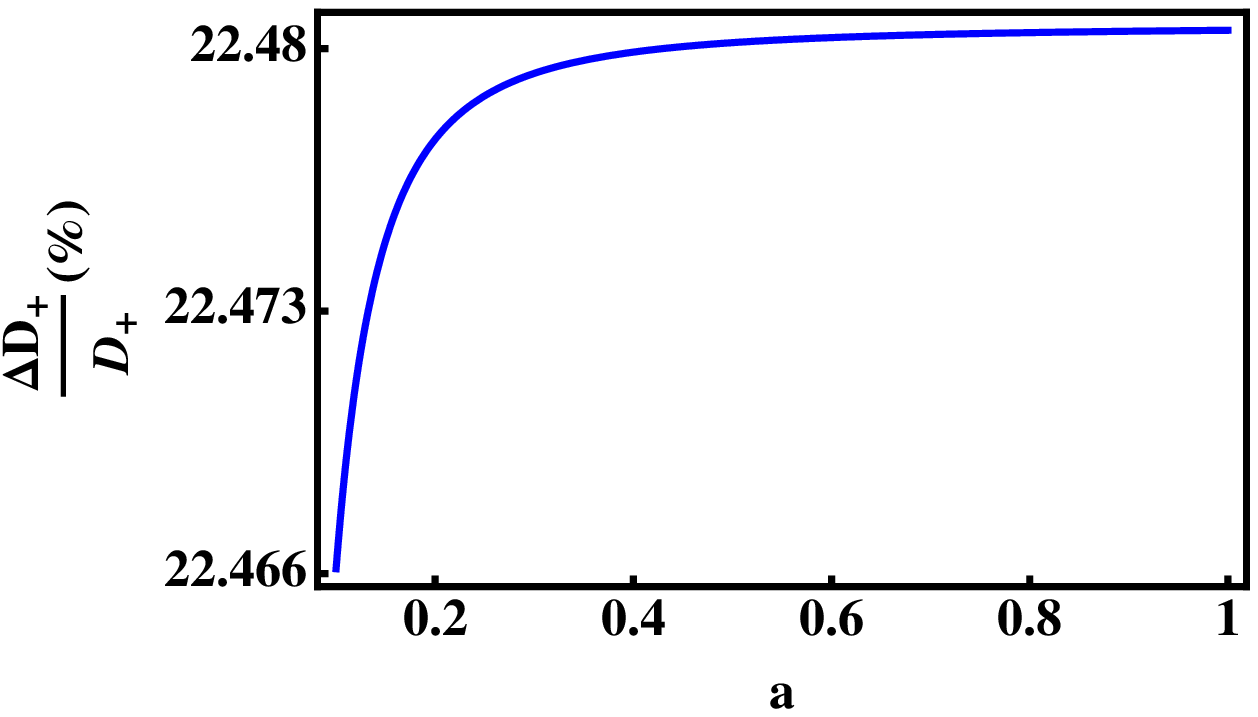, width=7.6cm}}
\vspace{-0.5cm}
\caption{The first order solution of the fastest growing mode for the open universe. a) Solid line represents the numerical solution. The analytic solution with correct initial conditions is shown as dotdashed line which is almost overlapped with solid line. The dotted line shows the evolution of $D_{\textrm{BJCP}}$. There exits discrepancy between the correct growing solution and the incorrect one. b) Error between the correct solution and the incorrect one.} \label{fig1}
\end{figure}
\end{center}

ii) case II : Open Universe \\
\\
One can repeat the above process for the open universe. Again by defining $d \tau = \fr{\tau_i}{a^2} dt$, one obtains
\be \sqrt{1 + \fr{A}{a}} = \fr{\tau}{\tau_i} \equiv y \, , \label{tau3} \ee where $A \equiv \fr{\Omega_{m0}}{1 - \Omega_{m0}}$. Thus, one gets
\be a = \fr{A}{y^2 - 1} \,\, , \,\, a H = -\fr{2}{A} y \, , \label{a} \ee
and
\be 1 - \Omega_{m} = y^{-2} \, . \label{1mOmega} \ee
If one replaces the variable $y$ with $t$, then the above equation (\ref{1DE}) becomes
\be \fr{d^2 D}{d \tau^2} - \fr{ 4 \pi G \rho_m a^4}{c^2 \tau_i^2} D = \fr{1}{\tau_i^2} \Biggl[ \fr{d^2 D}{d y^2} - \fr{6}{y^2 - 1} D \Biggr] = 0 \label{growthOpen} \, . \ee The analytic solution of the above Eq (\ref{growthOpen}) is well known as BJCP and given by
\ba g_{1a} &=& 1 + 3(y^2 -1) \Bigl(1 + y L(y) \Bigr)  = 1 + 3 \fr{A}{a} \Bigl(1 -L(a) \sqrt{\fr{A}{a} + 1} \,\Bigr)  \label{g1aDopen} \, , \\
g_{1b} &=& y (y^2 -1)  = - \fr{A}{a} \sqrt{\fr{A}{a} + 1} \, ,\label{g1bDopen} \ea where $L(y) = \fr{1}{2} \ln \Bigl[\fr{y-1}{y+1} \Bigr]$ and
$L(a) = \fr{1}{2} \ln \Bigl[\fr{\sqrt{\fr{A}{a} + 1}+1}{\sqrt{\fr{A}{a} + 1}-1} \Bigr]$.
In the references, one separates $g_{1a}$ and $g_{1b}$ as a growing and a decaying mode solution, respectively. However, this is not true. In general case, we are not able to separate the growing mode from the decaying one. One can check this from the above solutions Eqs. (\ref{g1aDopen}) and (\ref{g1bDopen}). Both $g_{1a}$ and $g_{1b}$ are growing as $a$ increases even though $g_{1b}$ is negative. The growing mode solution of Eq. (\ref{growthOpen}) is obtained by using two initial conditions.
\ba D(a_i) &=& a_{i} = c_{aD} g_{1a}(a_i) + c_{bD} g_{1b}(a_i) \, ,
\label{Dyi} \\
\fr{d D}{d a} \Bigl|_{a_i} &=& 1 = c_{aD} \fr{d g_{1a}}{d a} \Bigl|_{a_i} + c_{bD} \fr{d g_{1b}}{d a} \Bigl|_{a_i} \, ,
\label{dDdas} \ea where we use $a_i$ is the initial epoch. We adopt $a_i = \fr{1}{50}$ which gives the proper sub-horizon behavior for the growth factor. Thus, we can obtain the correct $c_{aD}$ and $c_{bD}$ from these initial conditions. In BJCP, they ignore this fact and just choose the $g_{1a}$ as the growing mode solution with $c_{aD} = 1$. We show the evolutions of $D_{+}$ and $D_{\textrm{BJCP}}$ and the difference between them in Fig. \ref{fig1}. We choose $\Omega_{m0} = \fr{1}{3}$ in this figure. The error is about 23 \% at the present epoch. We also show the rescaled BJCP solution by normalized the coefficient of $g_{1a}$ to satisfy
\be r_{aD} g_{1a} (a_i) = a_i \, . \label{raD} \ee In this case, the error between the exact solution and the rescaled BJCP is less than 1 \% for
$ \fr{1}{50} \leq a \leq 1$. Later, we will consider the fastest second order solution and in this case we need to use the $r_{aD}$ instead of
$c_{aD}$ in order to get the correct evolution of the second order solution. \\
\begin{figure}
\centering
\vspace{1.5cm}
\begin{tabular}{cc}
\epsfig{file=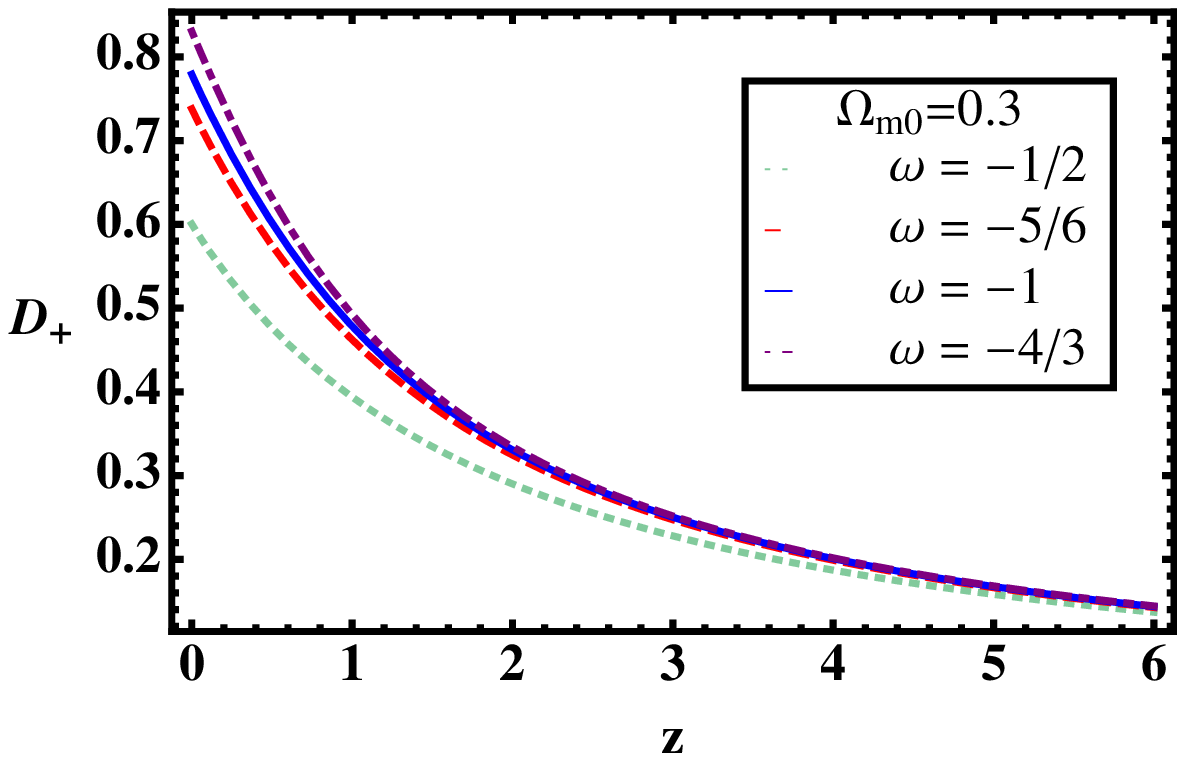,width=0.5\linewidth,clip=} &
\epsfig{file=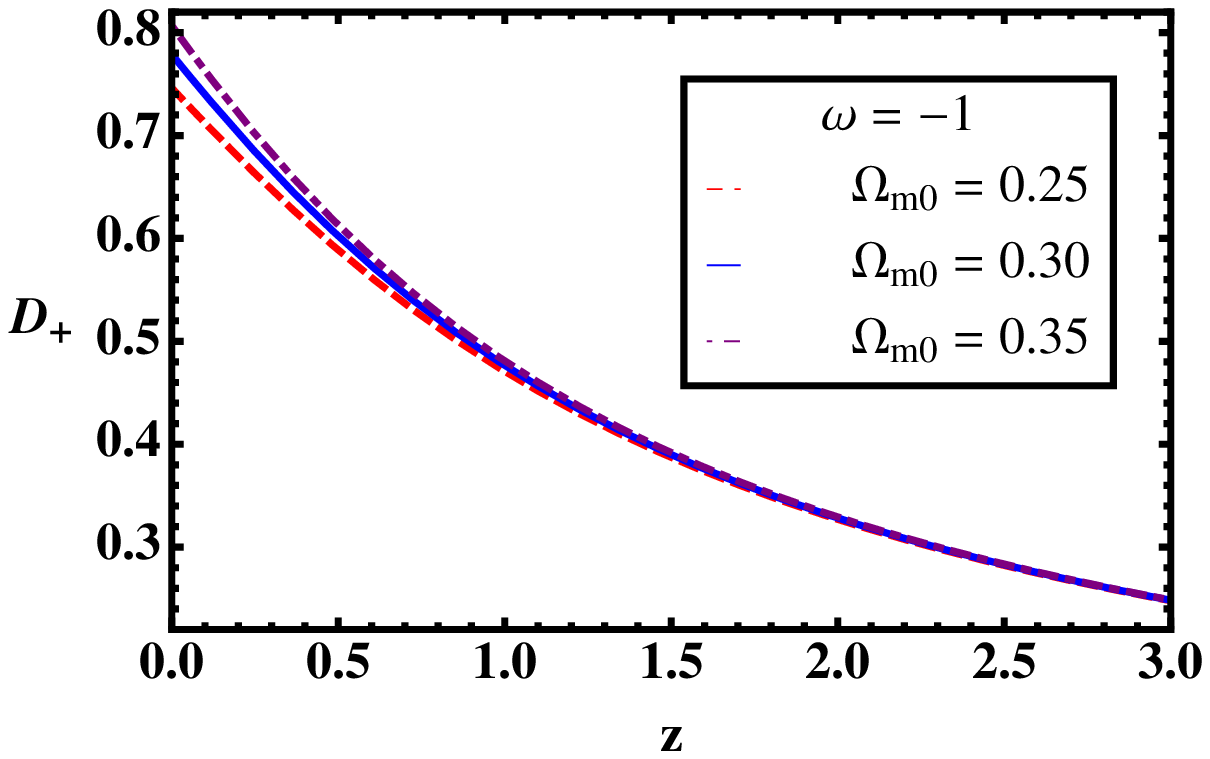,width=0.5\linewidth,clip=} \\
\epsfig{file=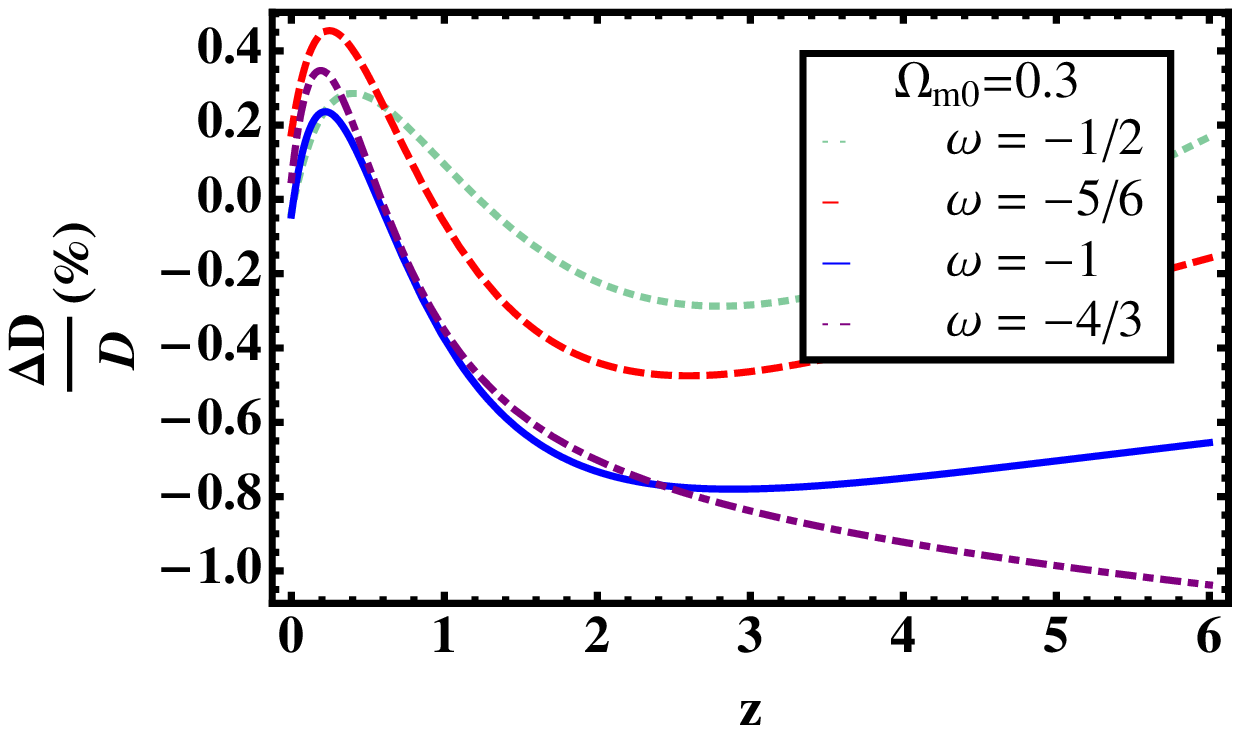,width=0.5\linewidth,clip=} &
\epsfig{file=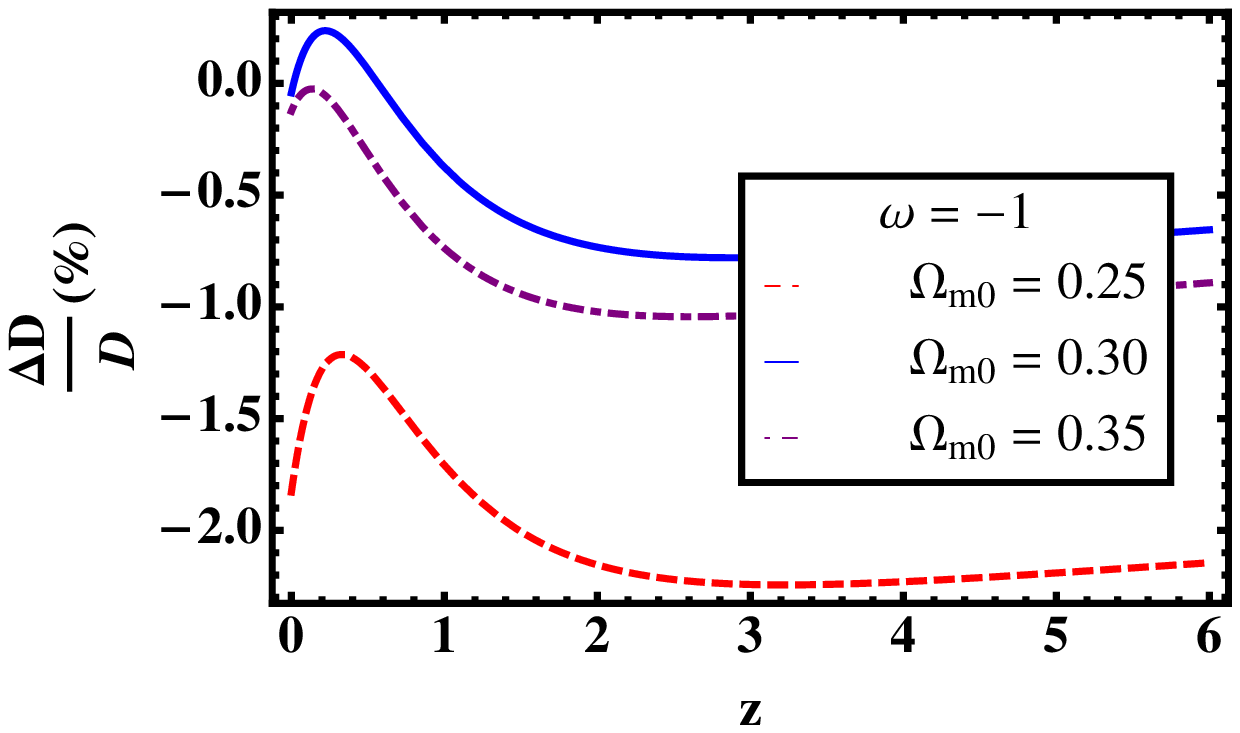,width=0.5\linewidth,clip=} \\
\end{tabular}
\vspace{-0.5cm}
\caption{{\bf First row} : The evolution of $D_{+}$ for $\omega$CDM models. a) The evolution of $D_{+}$ as a function of $z$ with the different values of $\omega$ for the fixed $\Omega_{m0} = 0.3$. The dotdashed, solid, dashed, and dotted lines correspond to $\omega = -\fr{4}{3}, -1, -\fr{5}{6}$, and $-\fr{1}{2}$, respectively. b) $D_{+}(z)$ for the different values of $\Omega_{m0}$ with $\omega = -1$. The dotdashed, solid, and dashed lines represent $\Omega_{m0} = 0.35, 0.3$, and $0.25$, respectively. {\bf Second row} : The errors of the first order fitting form as a function of $z$. c) $\fr{\Delta D}{D}$ for the same models as in a). d) $\fr{\Delta D}{D}$ dependence on $\Omega_{m0}$ for the different values of $\omega$ as in b). } \label{fig2}
\end{figure}

iii) case III : $\omega$CDM \\
\\
In this case, one can rewrite the above Eq. (\ref{1DE}) as
\be Y \fr{d^2 D}{d Y^2} + \Biggl[ 1 + \fr{1}{6 \omega} - \fr{1}{2(Y+1)} \Biggr] \fr{d D}{d Y} - \fr{1}{6 \omega^2} \fr{1}{Y+1} D = 0 \label{1DEw} \, , \ee where $Y \equiv A a^{3 \omega}$. The solution of the above equation is found as \cite{09061643,09072108}
\ba g_{1a} &=&  F\Bigl[ -\fr{1}{3 \omega}, \fr{1}{2 \omega}, \fr{1}{2} + \fr{1}{6\omega}, - A a^{3\omega} \Bigr] \label{g1aGw} \, , \\
g_{1b} &=& a^{\fr{3\omega - 1}{2}} F \Bigl[\fr{1}{2} - \fr{1}{2 \omega}, \fr{1}{2} + \fr{1}{3\omega}, \fr{3}{2}- \fr{1}{6\omega}, -A a^{3\omega} \Bigr] \label{g1bGw} \, , \ea where $F$ is the hypergeometric function. If one uses the transformation formulas for the hypergeometric function \cite{Abramowitz}, then one obtains
\ba g_{1a} &=& a F \Bigl[- \fr{1}{3 \omega}, \fr{1}{2} - \fr{1}{2\omega}, 1 - \fr{5}{6\omega}, - \fr{a^{-3\omega}}{A} \Bigr] \label{g1aGw2} \, , \\
g_{2a} &=& a^{-\fr{3}{2}} F \Bigl[\fr{1}{2 \omega}, \fr{1}{2} + \fr{1}{3\omega}, 1 + \fr{5}{6\omega}, -\fr{a^{-3\omega}}{A} \Bigr] \label{g1bGw2} \, . \ea The above solution can be interpreted as the solution of Eq. (\ref{1DEw}) after replace the variable $Y = -Z^{-1}$. With this replacement the above Eq. (\ref{1DEw}) becomes
\be Z \fr{d^2 D}{d Z^2} + \Biggl[ 1 - \fr{1}{6 \omega} + \fr{Z}{2(Z-1)} \Biggr] \fr{d D}{d Z} + \fr{1}{6 \omega^2} \fr{1}{Z(Z-1)}  D = 0 \label{1DEwZ} \, . \ee When $\omega = -\fr{1}{3}$, this is the same as the open universe in the case II. With using the initial conditions of the growth factor given by Eqs. (\ref{Dyi}) and (\ref{dDdas}), one can obtain the integral constants. Thus, we can regard the open universe as the specific case of $\omega$CDM.

Even though we already obtain the exact analytic solution for the first order in $\omega$CDM, we also obtain the fitting form of $g_{1a}$
\be g_{1a} = B_{D} Y^{P_{D}} (1+Y)^{Q_{D}} \label{g1afit} \, , \ee
where
\ba B_{D} &=& (-6 \omega)^{0.347 - 0.260A + 0.035 \omega} + 1.220A -1.191 \, , \label{BD} \\
P_{D} &=& - (-14.88 \omega)^{-0.80} \, , \label{PD} \\
Q_{D} &=& - (-3.74 \omega)^{-1.15} \, . \label{QD} \ea
\begin{figure}
\centering
\vspace{1.5cm}
\begin{tabular}{cc}
\epsfig{file=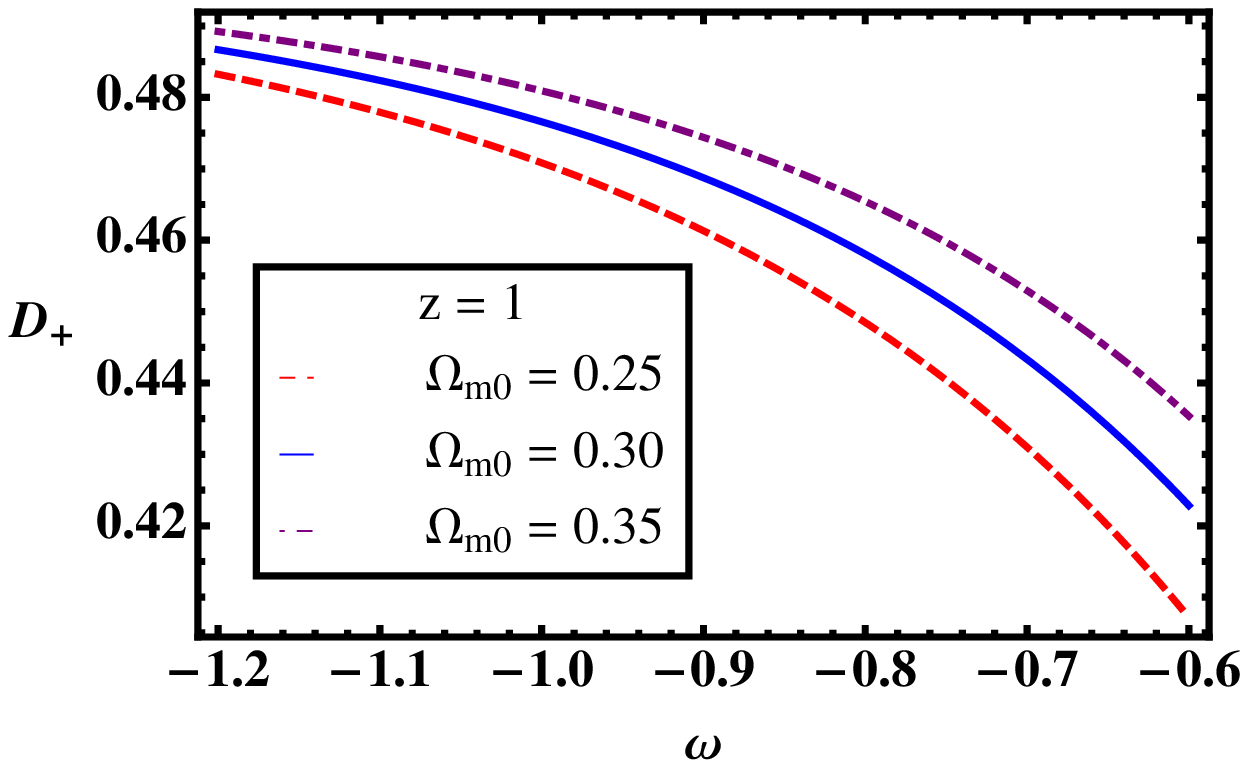,width=0.5\linewidth,clip=} &
\epsfig{file=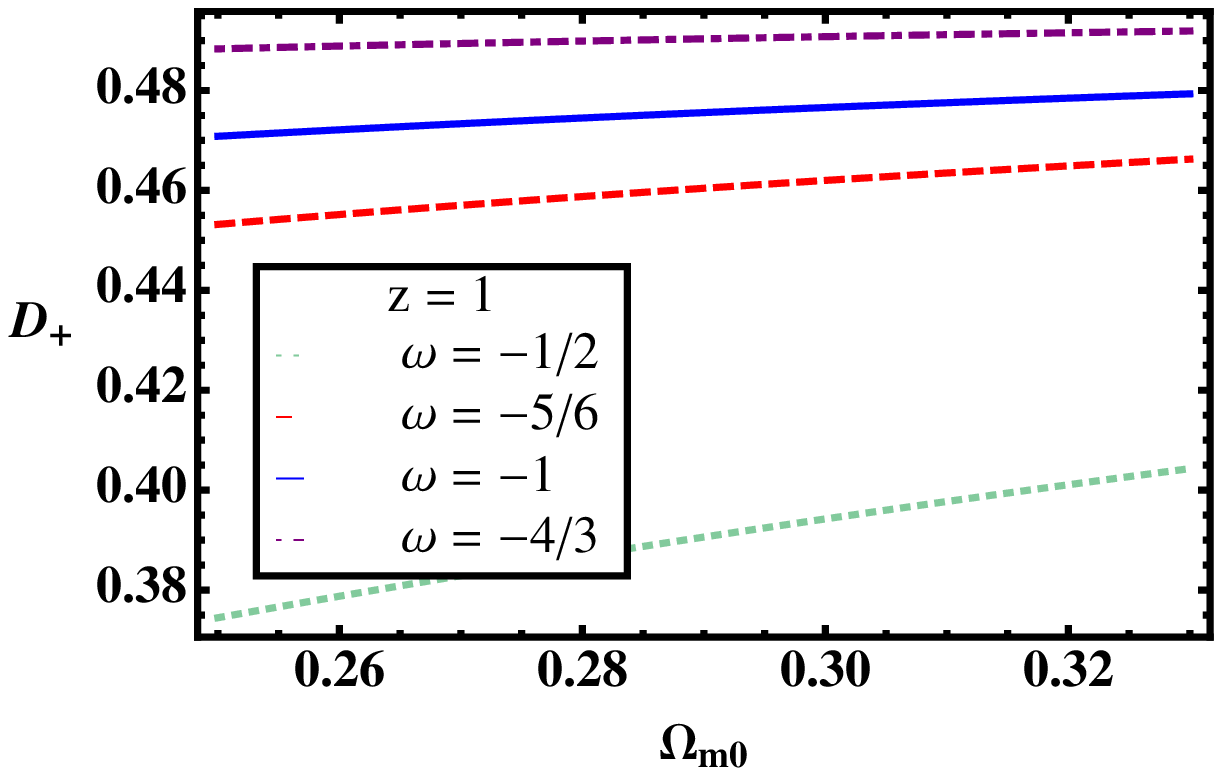,width=0.5\linewidth,clip=} \\
\epsfig{file=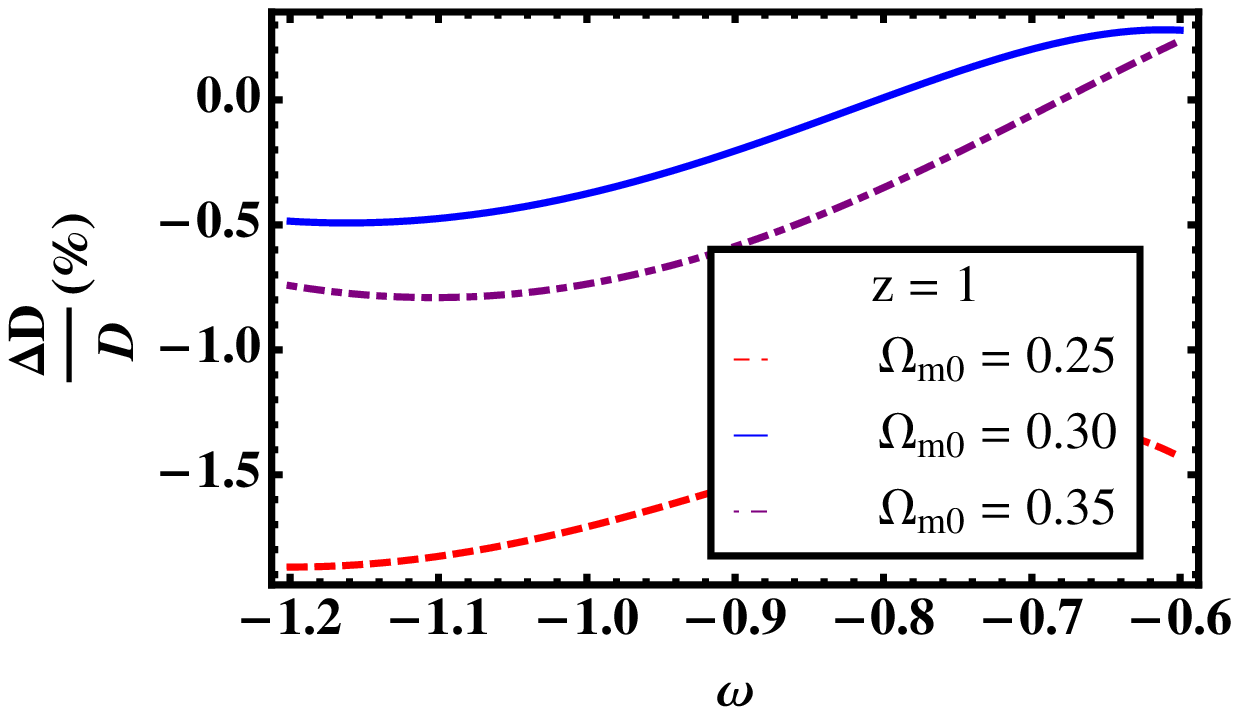,width=0.5\linewidth,clip=} &
\epsfig{file=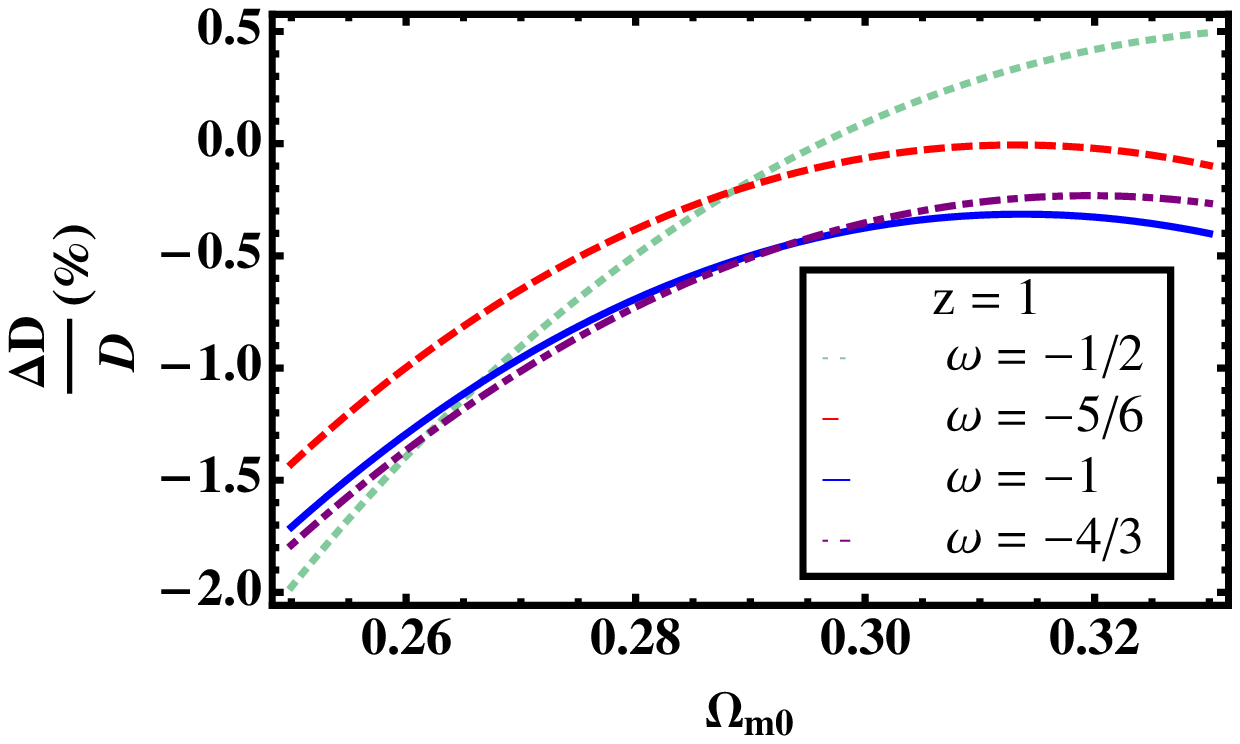,width=0.5\linewidth,clip=} \\
\end{tabular}
\vspace{-0.5cm}
\caption{{\bf First row} : The values of $D_{+}$ for the different cosmological parameters at $z=1$. a) $D_{+}$ dependence on $\omega$ for the different values of $\Omega_{m0}$. Dotdashed, solid, and dashed lines correspond to $\Omega_{m0} = 0.35, 0.30$, and $0.25$, respectively. b) $D_{+}$ dependence on $\Omega_{m0}$ for the different values of $\omega$. Dotted, dashed, solid, and dotdashed lines correspond to $\omega = -\fr{1}{2}, -\fr{5}{6}, -1$, and $-\fr{4}{3}$, respectively. {\bf Second row} : The errors of the first order fitting form as a function of $\omega$ and $\Omega_{m0}$ at $z=1$. c) $\fr{\Delta D}{D}$ dependence on $\omega$ for the different values of $\Omega_{m0}$. d) $\fr{\Delta D}{D}$ dependence on $\Omega_{m0}$ for the different values of $\omega$. } \label{fig3}
\end{figure}

We demonstrate the evolution of $D_{+}$ for the different cosmological models in the first row of Fig. \ref{fig2}. In the left panel of the first row, we show the evolutions of $D_{+}$ for the different values of $\omega$ when $\Omega_{m0} = 0.3$. The dotted, dashed, solid, and dotdashed lines correspond to $\omega = -\fr{1}{2}, -\fr{5}{6}, -1$, and $-\fr{4}{3}$, respectively. One obtains the larger $D_{+}$ for the smaller value of $\omega$. This is due to the fact that there has been the more matter density in the past for the smaller values of $\omega$ to give the larger values of $D_{+}$ at present. At present, the value of $D_{+}$ varies from 0.83 to 0.6 for $-\fr{4}{3} \leq \omega \leq -\fr{1}{2}$. In the right panel, we also show the evolutions of $D_{+}$ for the different values of $\Omega_{m0}$ for $\Lambda$CDM model. The dotted, solid, and dotdashed lines correspond $\Omega_{m0} = 0.25, 0.30$, and $0.35$, respectively. As one expects, the bigger the $\Omega_{m0}$, the larger the $D_{+}$. $D_{+}$ varies from $0.75$ to $0.81$ for $0.25 \leq \Omega_{m0} \leq 0.35$. It changes about 8 \% when $\Omega_{m0}$ varies from $0.25$ to $0.35$. In the second row of Fig.\ref{fig2}, we show the errors of the fitting form as a function of the redshift $z$ for the different models. In the first column, we show the errors of the fitting form for the different values of $\omega$ when $\Omega_{m0} = 0.3$. The dotted, dashed, solid, and dotdashed lines correspond to $\omega = -\fr{1}{2}, -\fr{5}{6}, -1$, and $-\fr{4}{3}$, respectively. The errors are less than $1$ \% for all models up to $z \leq 6$. In the second column, we check the errors with the different $\Omega_{m0}$ values for the $\Lambda$CDM model using the same notation as that of the first row. The errors are less than $1$ \% for $\Omega_{m0} = 0.30$ and $0.35$. The error can be about $2$ \% for $\Omega_{m0} = 0.25$.

We also investigate the dependence of $D_{+}$ on the different cosmological parameters at the specific $z$. In the first row of Fig. \ref{fig3}, we show the values of $D_{+}$ as a function of $\omega$ and $\Omega_{m0}$. In the left panel, we fix the redshift at $z = 1$ and check the dependence of $D_{+}$ on $\omega$ for the different values of $\Omega_{m0}$. The dashed, solid, and the dotdashed lines correspond to $\Omega_{m0} = 0.25, 0.30$, and $0.35$, respectively. Again, the bigger the $\omega$, the smaller the $D_{+}$ for each $\Omega_{m0}$. Also we notice that the change rate is steeper for the smaller value of $\Omega_{m0}$. In the right panel, we show the property of $D_{+}$ as a function of $\Omega_{m0}$ for the different $\omega$ models at $z = 1$. Again, the dotted, dashed, solid, and dotdashed lines correspond to $\omega = -\fr{1}{2}, -\fr{5}{6}, -1$, and $-\fr{4}{3}$, respectively. One interesting point is that $D_{+}$ dependence on $\Omega_{m0}$ becomes weaker as $\omega$ decreases. For example, $D_{+}$ for $\omega = -1$ model changes only 2 \% when $\Omega_{m0}$ varies from $0.25$ to $0.35$. However, $D_{+}$ changes about 10 \% for the same variation of $\Omega_{m0}$ when $\omega = -\fr{1}{2}$. The errors of the fitting form at the specific $z$ for the different cosmological models are also studied. In the second row of Fig. \ref{fig3}, we investigate the errors of the fitting form for the different values of $\omega$ and $\Omega_{m0}$. In the first column, we check the dependence of errors on $\omega$ for the different values of $\Omega_{m0}$. Except $\Omega_{m0} = 0.25$, the errors are less than $1$ \% for $-1.2 \leq \omega \leq - 0.6$. In the second column, we show the errors as a function of $\Omega_{m0}$ for the different $\omega$ models at $z = 1$. The error can be as large as 2 \% when $\Omega_{m0} = 0.25$.

\subsection{Second order time component}

In general, the second order equation satisfies,
\be \ddot{E} + 2H \dot{E} - 4 \pi  G \rho_m E = - 4 \pi G \rho_m D^2 \, , \label{GEEq} \ee where $D$ is the solution of the first order equation. Unlike the first order solution, the second order solution constrains both the spatial and temporal dependence as shown in Eq. (\ref{Eeq}). In the appendix, we prove that the solution $E$ can be obtained from the specific relation with the first order solution for EdS case. Again, we consider the solutions of this equation for the different cosmological models. We need these solutions to obtain the higher order solutions. \\

i) case I : EdS \\
\\
Again, one can rewrite the above equation by using $x$
\be \fr{d^2 E}{d x^2} - \fr{6}{x^2} E = -\fr{6}{x^2} D^2 \label{EdSEEq} \, . \ee After replacing $D$ given by Eq. (\ref{DyEdS}) into the above equation, one obtains
\be E = -\fr{3}{7} c_{aD}^2 g_{1a}^2 + 2 c_{aD} c_{bD} g_{1a} g_{1b} -\fr{1}{4} c_{bD}^2 g_{1b}^2 + e_{a} g_{1a} + e_{b} g_{1b} \label{EyEdS} \, ,\ee
where $e_{a}$ and $e_{b}$ are the integral constants for the homogeneous solution of the above Eq. (\ref{EdSEEq}). For the fastest growing solution,
one ignores the term including $g_{1b}$. Thus, the fastest growing mode solution of the second order becomes
\be E_{+}(a) = -\fr{3}{7} c_{aD}^2 a^{2} + e_{a} a \label{EapEdS} \, . \ee Now one can consider the initial condition of the above solution. At the early epoch, only the first order term contributes to the perturbation and thus one can put $E{+}(a_i) = 0$. From this initial condition, one can link the $e_{a}$ with the initial epoch as
\be e_{a} = \fr{3}{7} a_i \label{caE} \, , \ee where we use $c_{aD} = 1$. With the above relation, the second order solution is rewritten as
\be E_{+}(a) = -\fr{3}{7} a^{2} + \fr{3}{7} a_i a \label{EapEdS2} \, . \ee  Thus, one can obtain the initial condition of $\fr{dE_{+}}{da}$ as
\be \fr{dE_{+}}{da} \Bigl|_{a_i} = -\fr{3}{7} a_{i} \label{dEapEdSda} \, . \ee Both $E{+}(a_i) = 0$ and Eq. (\ref{dEapEdSda}) can be used as the
initial conditions of the general dark energy models for the numerical calculation. \\

\begin{center}
\begin{figure}
\vspace{1.5cm}
\centerline{\epsfig{file=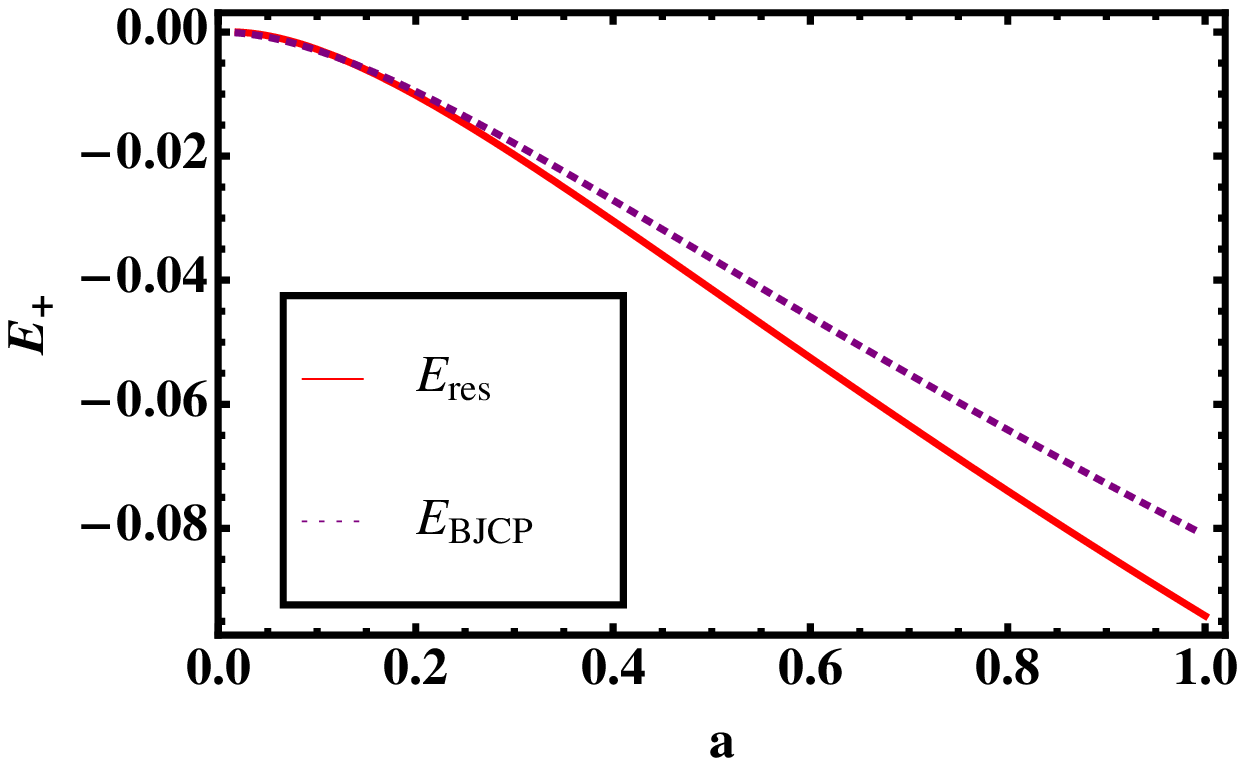, width=7.0cm} \epsfig{file=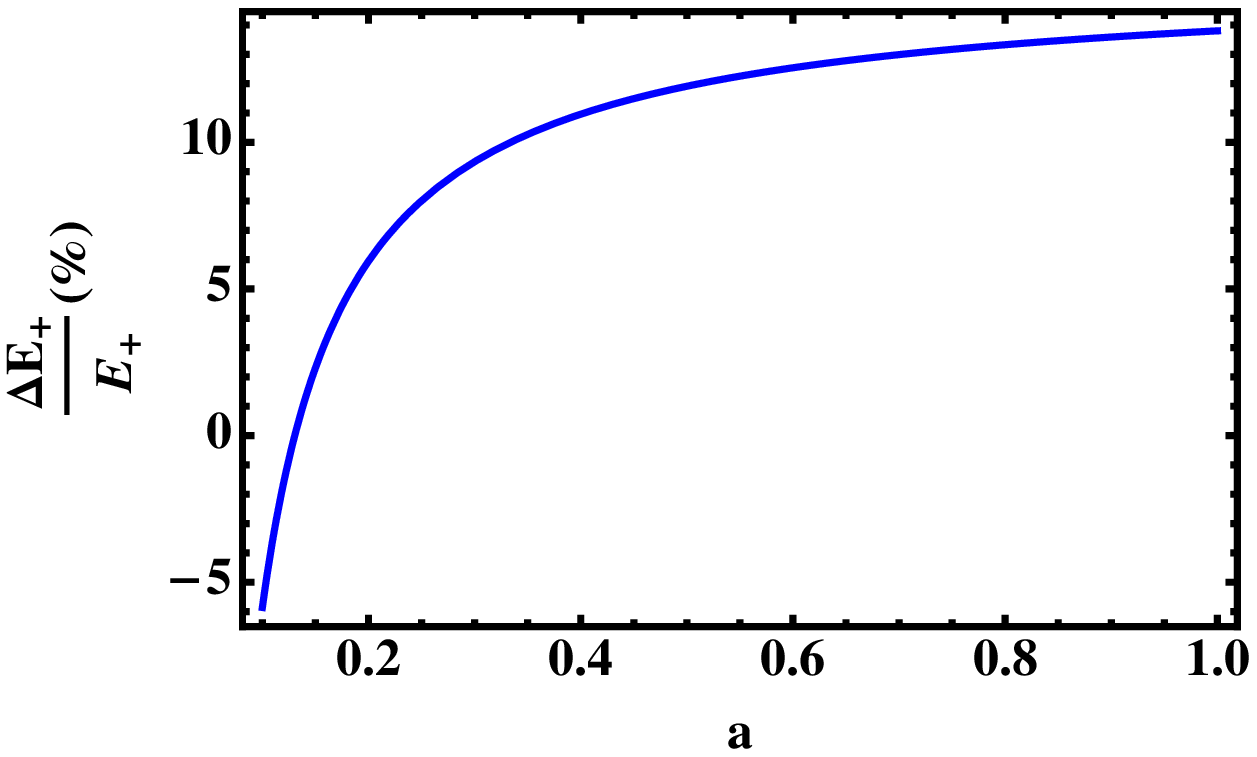, width=7.2cm}}
\vspace{-0.5cm}
\caption{ The second order fastest growing solution for the open universe. a) The solid line represents the evolution of $E_{+}$ with the correct rescaled coefficient. The dotted line depicts $E_{\textrm{BJCP}}$.  b)Error between the correct solution and BJCP one.} \label{fig4}
\end{figure}
\end{center}

ii) case II : Open Universe \\
\\
In this case, the second order perturbation equation (\ref{GEEq}) becomes
\be \fr{d^2 E}{dy^2} - \fr{6}{y^2-1} E = -\fr{6}{y^2-1} D^2 \, , \label{OpenEeq} \ee where $D$ is the solution of the first order perturbation equation given by Eqs. (\ref{g1aDopen}) and (\ref{g1bDopen}). One can obtain the general solution of the above equation
\be E = c_{aE} g_{2a} + c_{bE} g_{2b} + c_{cE} g_{2c} + e_{a} g_{1a} + e_{b} g_{1b} \, , \label{OpenE} \ee where $c_{iE}$s with $i = a, b, c$ are the coefficients of the particular solution which can be determined by the integral constants of the first order solution  and $e_{i}$s are the constants of integration for the homogeneous solutions. The particular solutions are obtained as
\ba g_{2a} &=& 1 - \fr{9}{4} (y^2 -1 ) \Bigl[ y + (y^2-1) L(y) \Bigr]^2 \, , \label{Openg2a} \\
g_{2b} &=& -\fr{1}{4} (y^2-1)^3 \, , \label{Openg2b} \\
g_{2c} &=& -\fr{3}{4} (y^2-1) \Bigl[y^3 + y + (y^2-1)^2 L(y) \Bigr] \, . \label{Openg2c} \ea
From the above solutions, one can find that
\be c_{aE} = c_{aD}^2 \,\,\, , c_{bE} = c_{bD}^2 \,\,\, , c_{cE} = 2 c_{aD}c_{bD} \, . \label{Opencs} \ee
The coefficients of the particular solution are determined from the integral constants of the first order solutions. Thus, one needs to determine the integral constants of the homogeneous solutions. If one just considers the contribution from the first part of the first order solution, then one can ignore the three terms including $g_{1b}$ and $g_{2b}$ in the above Eq. (\ref{OpenE}) and the solution becomes
\be E_{+} = c_{+aE} g_{2a} + e_{a} g_{1a} \, , \label{OpenEP} \ee where $c_{+aE} = r_{aD}^2$ and $r_{aD}$ is given by Eq. (\ref{raD}). We replace the coefficient $c_{aE}$ into $c_{+aE}$ as we mentioned before. If we only consider the first part of the first order solution, then we need to rescaled the coefficient by using the initial condition. Thus, we can obtain $e_{a}$ by using the fact that
\be E_{+} (a_i) = c_{+aE} g_{2a} (a_i) + e_{a} g_{1a} (a_i) = 0 \,\,\, \rightarrow \,\,\, e_{a} = -\fr{g_{2a}(a_i)}{g_{1a}(a_i)} r_{aD}^2 \, .  \label{laE} \ee
Now we can compare the correct solution given by Eq. (\ref{OpenEP}) with BJCP solution given as
\be E_{\textrm{BJCP}} = -\fr{1}{2} - \fr{9}{2} (y^2 - 1) \Biggl( 1 + y L + \fr{1}{2} \Bigl[ y + (y^2 - 1) L \Bigr]^2 \Biggr) \, . \label{Ebjcp} \ee
As we already see in the first order solution $g_{1a}$, the correct solution should have the proper constant of integration to satisfy
$D_{+}(a_i) = a_i$. In BJCP, they ignore this fact and this causes the improper behavior for the second order solution. We show the evolutions of $E_{+}$ with the correct rescaled coefficients and $E_{\textrm{BJCP}}$ and the differences between them in Fig. \ref{fig4}. We show the evolutions of both the correctly rescaled second order solution $E_{\textrm{res}}$ and the $E_{\textrm{BJCP}}$ in the left panel of Fig. \ref{fig4}. The solid and dashed lines correspond to $E_{\textrm{res}}$ and $E_{\textrm{BJCP}}$, respectively. As we can see $E_{\textrm{BJCP}}$ is overestimated compared to the correct second order solution. We again adopt $\Omega_{m0} = \fr{1}{3}$ in this figure. We also show the difference between $E_{\textrm{res}}$ and $E_{\textrm{BJCP}}$ in the right panel of Fig. \ref{fig4}. The error between them is about 15 \% at the present epoch. Even though we show the exact solution, we also provide the approximate solution for this model.
\be E_{open}(a) = B_{Eo} Y^{P_{Eo}} (1 + Y)^{Q_{Eo}} \, , \label{Eopen} \ee
where
\ba B_{Eo} &=& -(-6 \omega_{13})^{-2.96 + 2.51 A} + 0.158 \, , \label{BEo} \\
P_{Eo} &=& -(-6.64 \omega_{13})^{-1.24 + 0.81 A} \, , \label{PEo} \\
Q_{Eo} &=& -(-1.85 \omega_{13})^{-1.68 + 1.01 A} \, , \label{QEo} \ea
where $\omega_{13} \equiv -\fr{1}{3}$. This fitting form has less than 1 \% error for $0.25 \leq \Omega_{m0} \leq 0.35$ up to $z \leq 3$. \\
\begin{figure}
\centering
\vspace{1.5cm}
\begin{tabular}{cc}
\epsfig{file=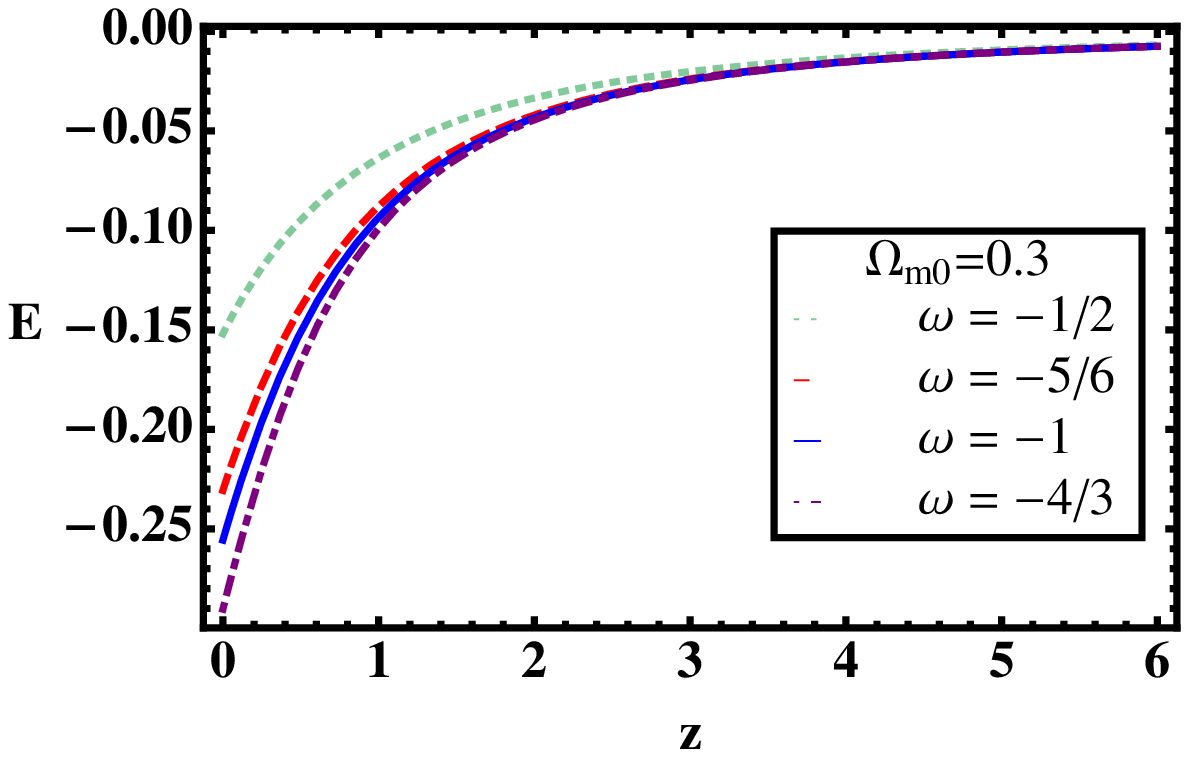,width=0.5\linewidth,clip=} &
\epsfig{file=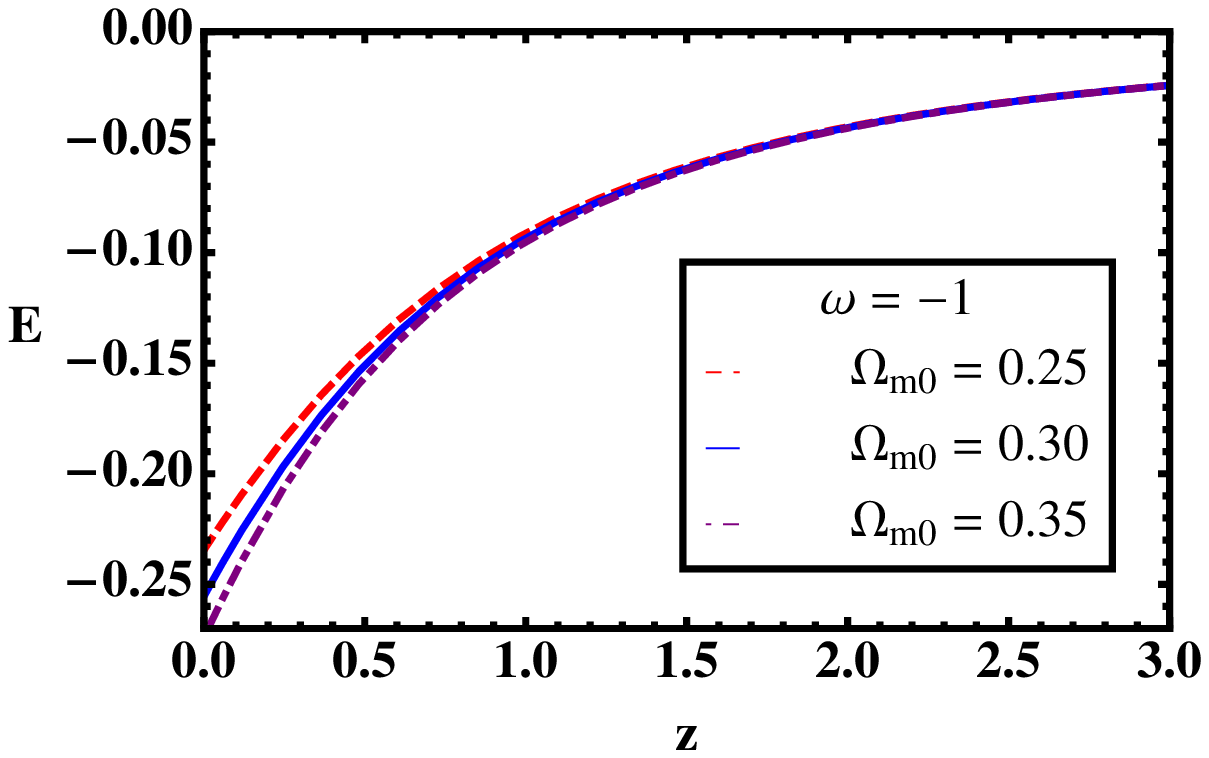,width=0.5\linewidth,clip=} \\
\epsfig{file=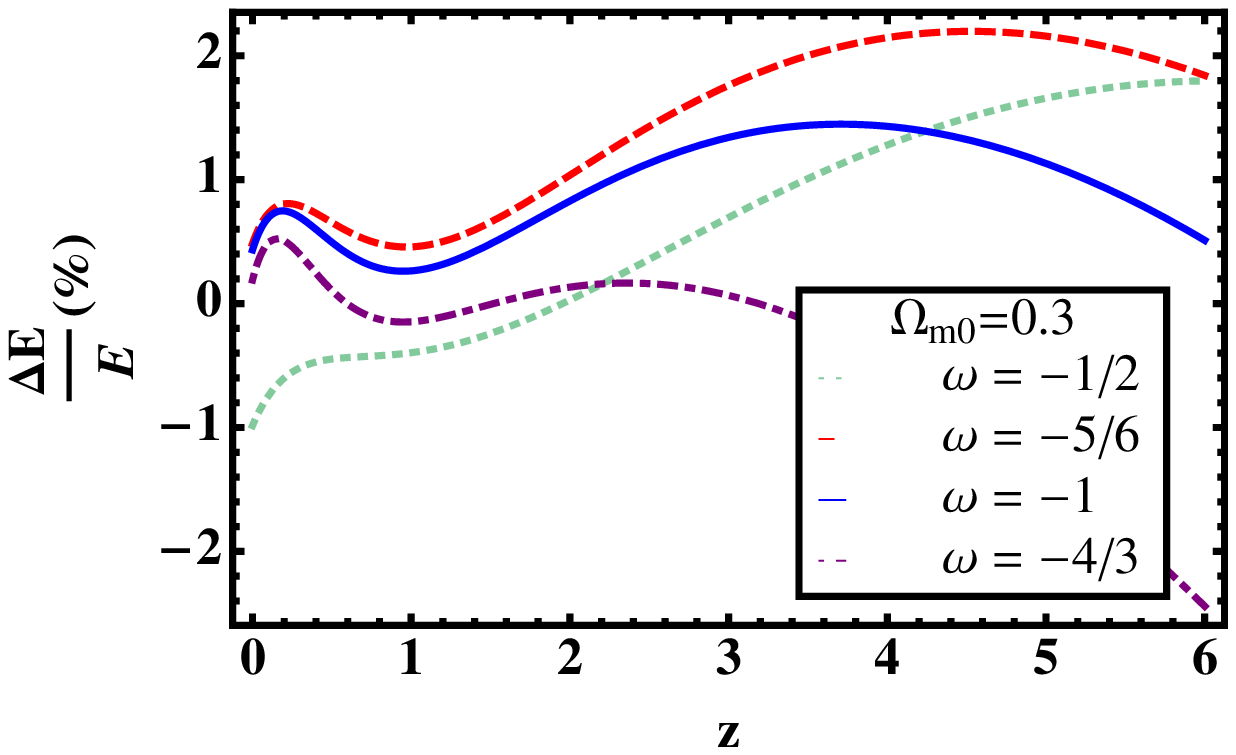,width=0.5\linewidth,clip=} &
\epsfig{file=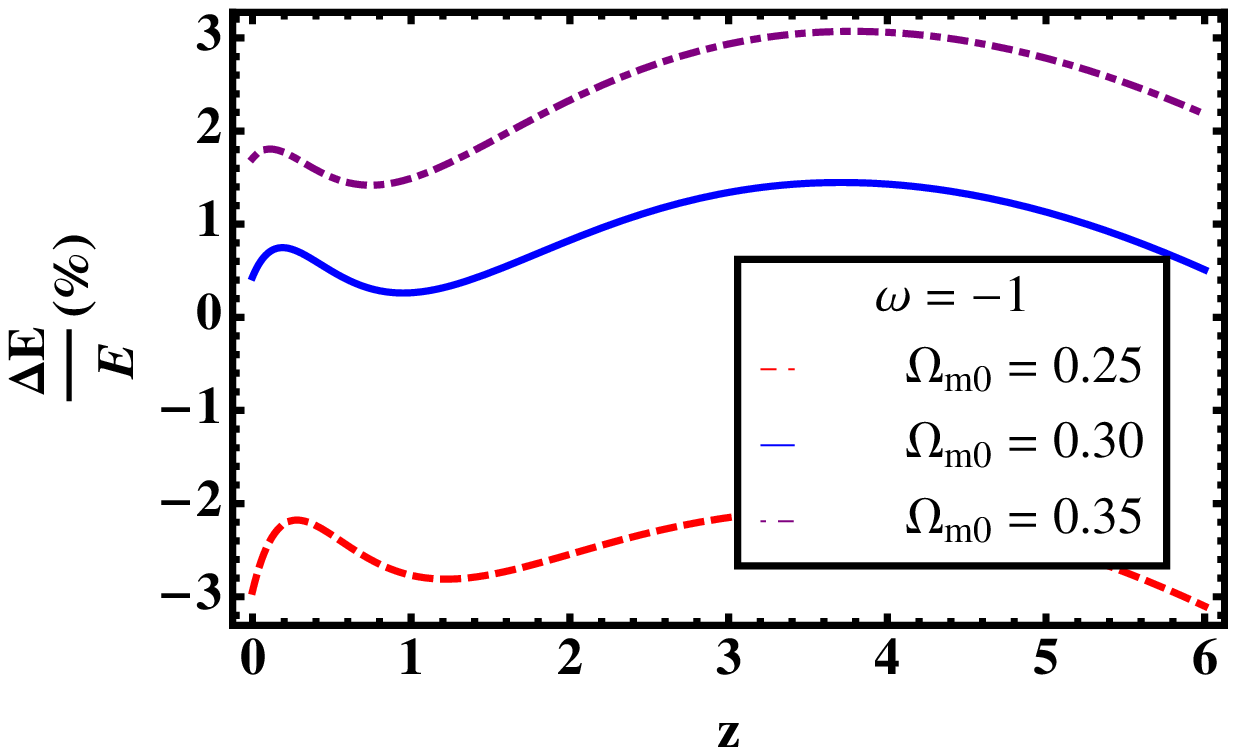,width=0.5\linewidth,clip=} \\
\end{tabular}
\vspace{-0.5cm}
\caption{ {\bf First row} : The evolution and accuracy of $E$ for the different $\omega$CDM models. a) $E(z)$ with the different values of $\omega$ for $\Omega_{m0} = 0.3$. The dotdashed, solid, dashed, and dotted lines correspond to $\omega = -\fr{4}{3}, -1, -\fr{5}{6}$, and $-\fr{1}{2}$, respectively. b) $E(z)$ with the different values of $\Omega_{m0}$ when $\omega = -1$. The dotdashed, solid, and dashed lines represent $\Omega_{m0} = 0.35, 0.3$, and $0.25$, respectively. {\bf Second row} : The errors of the second order fitting form as a function of $z$. c) $\fr{\Delta E}{E}$ for the different values of $\omega$ when $\Omega_{m0} = 0.3$. d) $\fr{\Delta E}{E}$ for the different values of $\Omega_{m0}$ when $\omega = -1$.} \label{fig5}
\end{figure}

\begin{figure}
\centering
\vspace{1.5cm}
\begin{tabular}{cc}
\epsfig{file=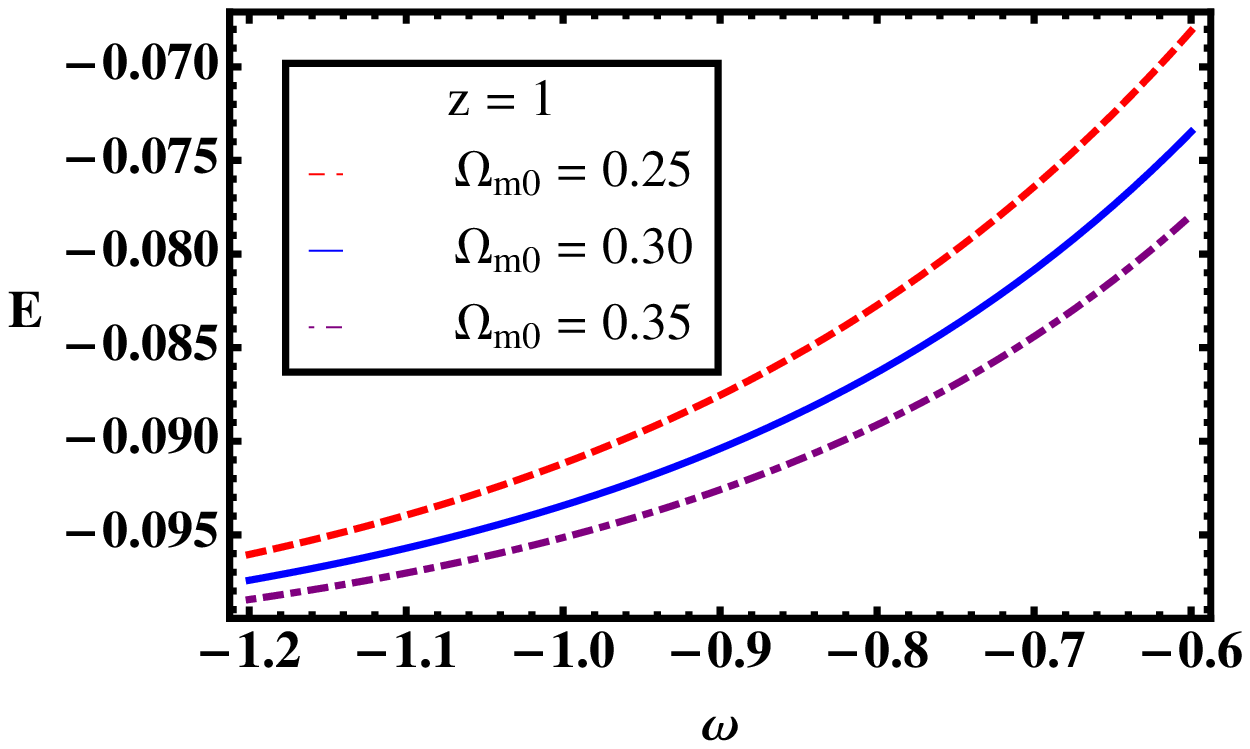,width=0.5\linewidth,clip=} &
\epsfig{file=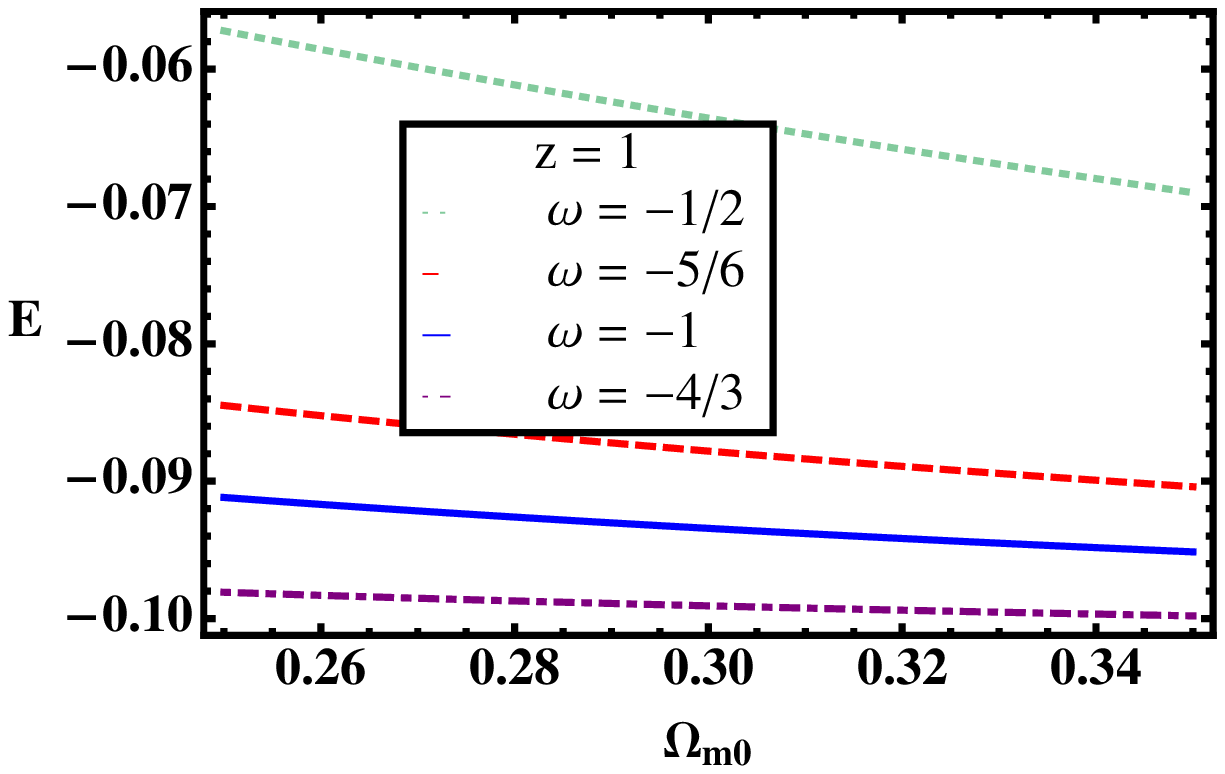,width=0.5\linewidth,clip=} \\
\epsfig{file=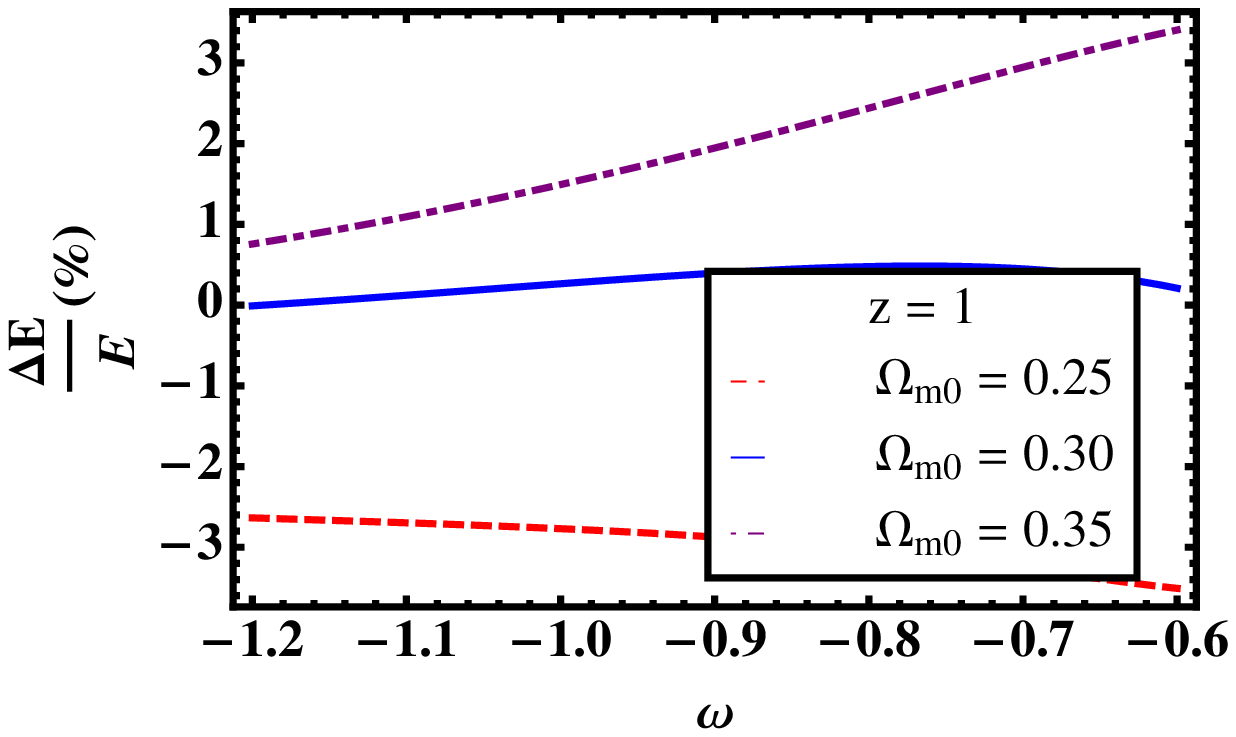,width=0.5\linewidth,clip=} &
\epsfig{file=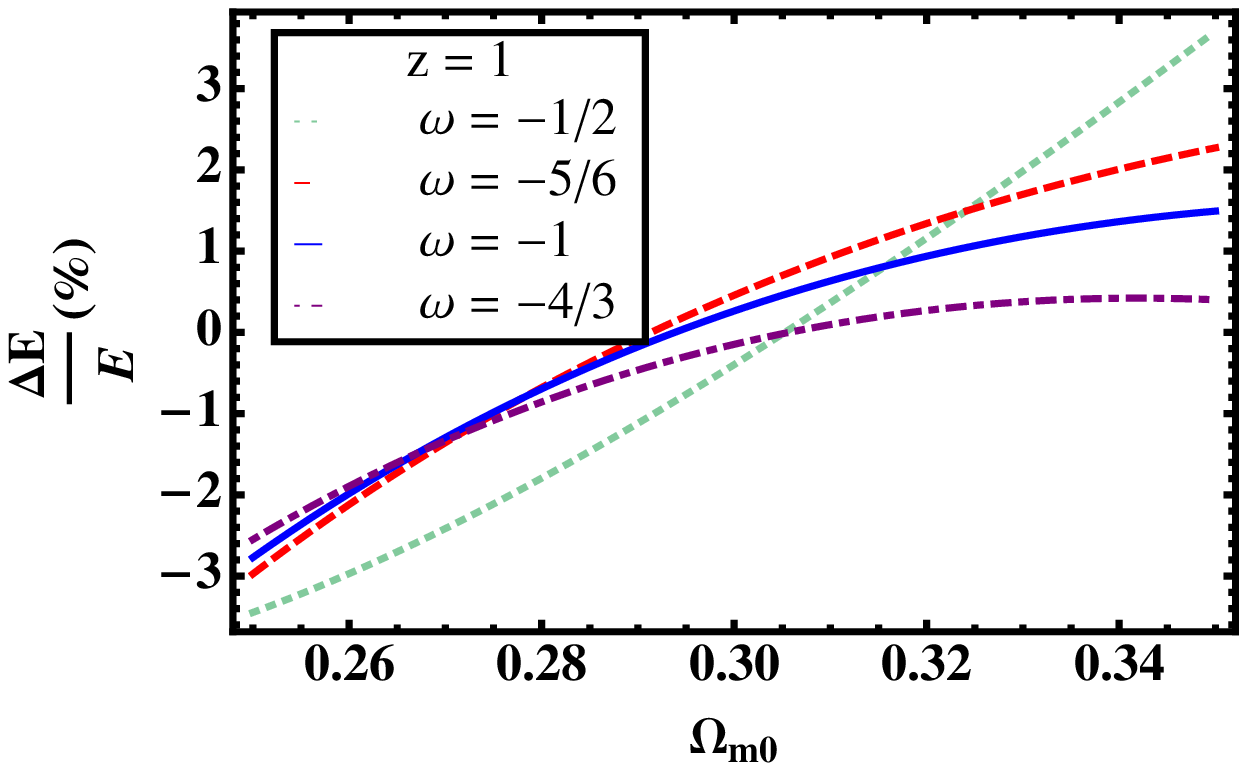,width=0.5\linewidth,clip=} \\
\end{tabular}
\vspace{-0.5cm}
\caption{{\bf First row} : The values of $E$ for the different cosmological parameters at $z=1$. a) $E$ dependence on $\omega$ for the different values of $\Omega_{m0}$. Dotdashed, solid, and dashed lines correspond to $\Omega_{m0} = 0.35, 0.30$, and $0.25$, respectively. b) $E$ dependence on $\Omega_{m0}$ for the different values of $\omega$. Dotted, dashed, solid, and dotdashed lines correspond to $\omega = -\fr{1}{2}, -\fr{5}{6}, -1$, and $-\fr{4}{3}$, respectively. {\bf Second row} : The errors on the fitting form at $z = 1$ for the different cosmological parameters. c) The dependence of errors on $\omega$ for the different values of $\Omega_{m0}$. d) The errors as a function of $\Omega_{m0}$ for the different values of $\omega$.} \label{fig6}
\end{figure}
ii) case III : $\omega$CDM \\
\\
In this case, one can rewrite the above Eq. (\ref{1DE}) as
\be Y \fr{d^2 E}{d Y^2} + \Biggl[ 1 + \fr{1}{6 \omega} - \fr{1}{2(Y+1)} \Biggr] \fr{d E}{d Y} - \fr{1}{6 \omega^2} \fr{1}{Y+1} E = -\fr{1}{6 \omega^2} \fr{1}{Y+1} D^2 \label{2DEw} \, , \ee where $D$ is the first order solution. The homogeneous solution is same as the first order one with the different integral constants and the nonhomogeneous solution should be obtained from the first order solutions. There is no analytic solution for the above equation but we can obtain the approximate fitting form of the fastest growing mode solution. If we adopt the initial conditions Eqs. (\ref{EapEdS2}) and (\ref{dEapEdSda}), we can obtain the numerical solution of the above Eq. (\ref{2DEw}). Similar to the approximate first order solution, we obtain the fitting form of $E_{+}$ for the fastest growing solution as
\be E(a) = B_{E} Y^{P_E} (1+Y)^{Q_E} \, , \label{Ewa} \ee where we abbreviate subscript $+$ and
\ba B_{E} &=& -(-6 \omega)^{0.178 - 0.122A + 0.011 \omega} + 1.242 -0.618 A \label{BE} \, , \\
P_{E} &=& -(- 6.64 \omega)^{-0.80} \label{PE} \, , \\
Q_{E} &=& - (-1.85 \omega)^{-1.18} \label{QE} \, . \ea From now on, we will drop the subscript $+$ for the fastest growing solution.

From the above fitting form, we obtain several properties of the fastest growing solution $E$. First, the signature of $E$ is opposite to that of $D$. Thus, it decreases as a function of time. Second, as $\omega$ decreases, so does $E$. One can understand this because the dark energy with the negative $\omega$ acts like the negative pressure. Third, $E$ decreases as $\Omega_{m0}$ increases. These properties allow us to study the behaviors of $E$ for the various cosmological parameters compared to those of EdS. We show the time evolution of $E$ for the different cosmological models in the first row of Fig. \ref{fig5}. In the left panel, we show $E(z)$ for the different values of $\omega$ when $\Omega_{m0} = 0.3$. The dotted, dashed, solid, and dotdashed lines correspond to $\omega = -\fr{1}{2}, -\fr{5}{6}, -1$, and $-\fr{4}{3}$, respectively. The present values of $E$ vary from $-0.15$ to $-0.29$ for $\omega = -\fr{1}{2}$ and $-\fr{4}{3}$, respectively. In the right panel, we show the evolutions of $E$ for the different $\Omega_{m0}$ values when we consider the $\Lambda$CDM model. The dotted, solid, and dotdashed lines correspond $\Omega_{m0} = 0.25, 0.30$, and $0.35$, respectively. We obtain $-0.27 \leq E \leq -0.23$ for $0.25 \leq \Omega_{m0} \leq 0.35$. The fitting form for the second order fastest growing solution is given by Eqs. (\ref{Ewa}) - (\ref{QE}). We investigate the accuracy of this fitting form for the different cosmological models. In the second row of Fig. \ref{fig5}, we show the errors of the fitting form as a function of the redshift $z$ for the different models. In the first column, we show the $\fr{\Delta E}{E}$ for the different values of $\omega$ when we fix $\Omega_{m0} = 0.3$. The errors are less than $2$ \% for all considered models when $z \leq 4$. In the second column, we check the errors of the fitting form for the different $\Omega_{m0}$ values for the $\Lambda$CDM model. The errors are less than $3$ \% for all the considered $\Omega_{m0}$ values.

We also investigate the dependence of $E$ on the cosmological parameters at the specific $z$. In the first row of Fig. \ref{fig6}, we show the values of $E$ as a function of $\omega$ and $\Omega_{m0}$ at the specific redshift $z$. In the left panel, we fix the redshift $z = 1$ and check the dependence of $E$ on $\omega$ for the different values of $\Omega_{m0}$. The dashed, solid, and the dotdashed lines correspond to $\Omega_{m0} = 0.25, 0.30$, and $0.35$, respectively. For $\Omega_{m0} = 0.35$, $E$ varies from -0.078 to -0.098 when $\omega$ changes from -0.6 to -1.2. We also obtain $-0.096 \leq E \leq -0.068$ for $-1.2 \leq \omega \leq -0.6$ when $\Omega_{m0} = 0.25$. Thus, we can conclude that $E$ dependence on $\omega$ becomes weaker as $\Omega_{m0}$ increases. In the right panel, we show the dependence of $E$ on $\Omega_{m0}$ for the different $\omega$ models at $z = 1$. Again, the dotted, dashed, solid, and dotdashed lines correspond to $\omega = -\fr{1}{2}, -\fr{5}{6}, -1$, and $-\fr{4}{3}$, respectively. For $\omega = -\fr{1}{2}$, $E$ varies from -0.057 to -0.069 when $\Omega_{m0}$ changes from 0.25 to 0.35. $E$ changes from -0.096 to -0.098 when $\Omega_{m0}$ changes from 0.25 to 0.35 for $\omega = -\fr{4}{3}$. This case $E$ is almost constant for the different values of $\Omega_{m0}$. Thus, we can conclude that $E$ dependence on $\Omega_{m0}$ becomes weaker as $\omega$ decreases. The errors on the fitting form at $z = 1$ for the different cosmological models are shown in the second row of Fig. \ref{fig6}. We show the errors on the fitting form as a function of $\omega$ and $\Omega_{m0}$. In the first column, we show the dependence of errors on $\omega$ for the different values of $\Omega_{m0}$. The errors are about less than $3$ \% for $-1.2 \leq \omega \leq - 0.7$. In the second column, we show the errors as a function of $\Omega_{m0}$ for the different $\omega$ models at $z = 1$. The errors are less than 3 \% for all $\Omega_{m0}$ except $\omega = -\fr{1}{2}$.

\subsection{Third order time component}

\begin{figure}
\centering
\vspace{1.5cm}
\begin{tabular}{cc}
\epsfig{file=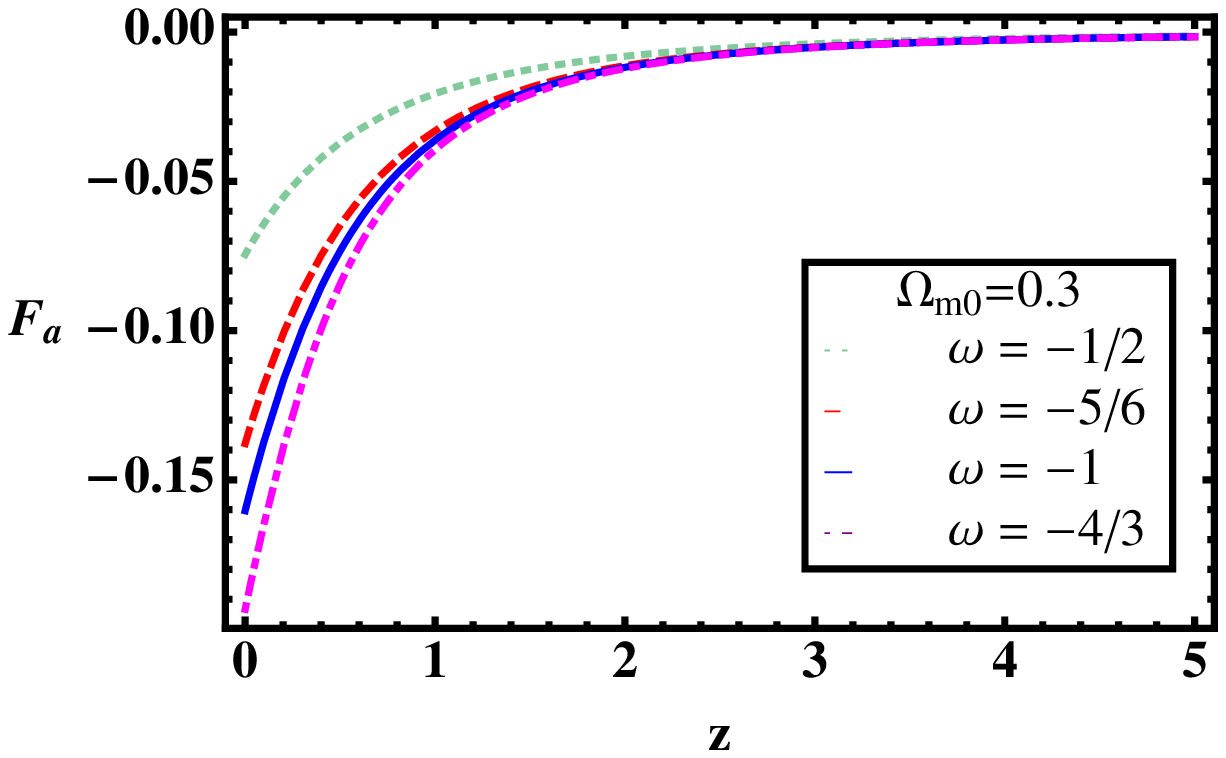,width=0.5\linewidth,clip=} &
\epsfig{file=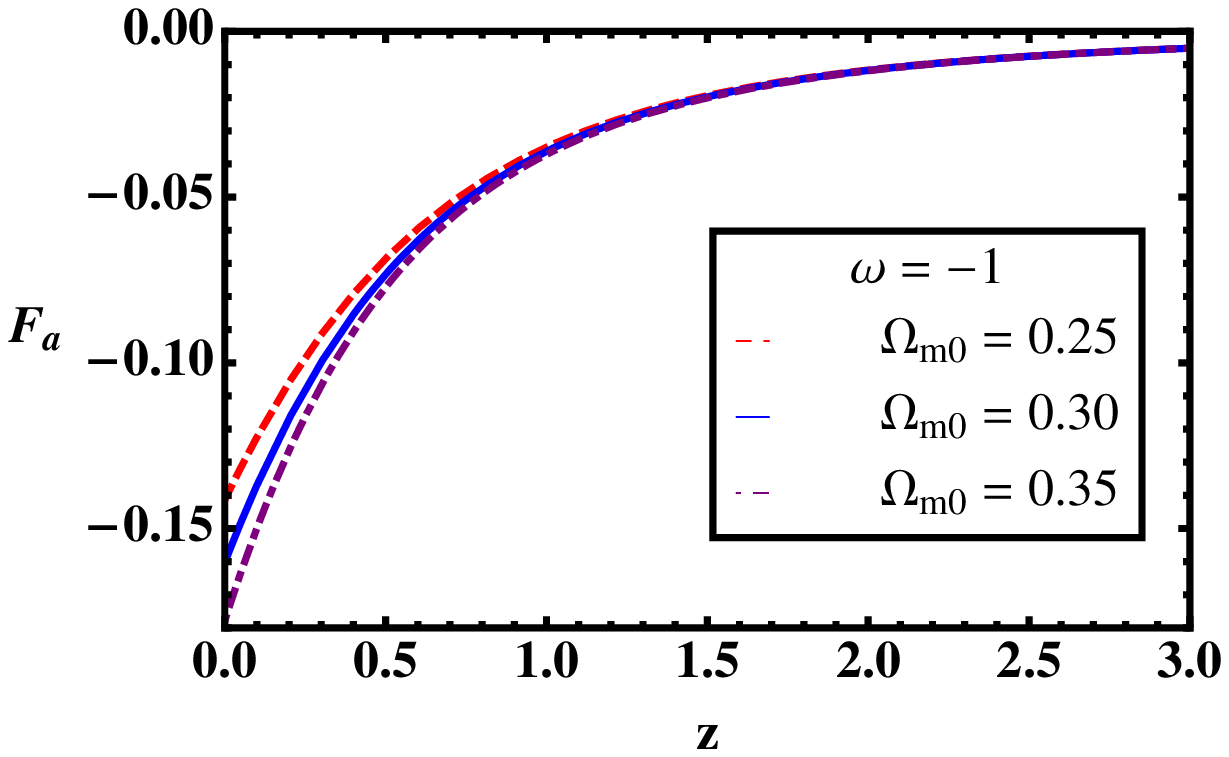,width=0.5\linewidth,clip=} \\
\epsfig{file=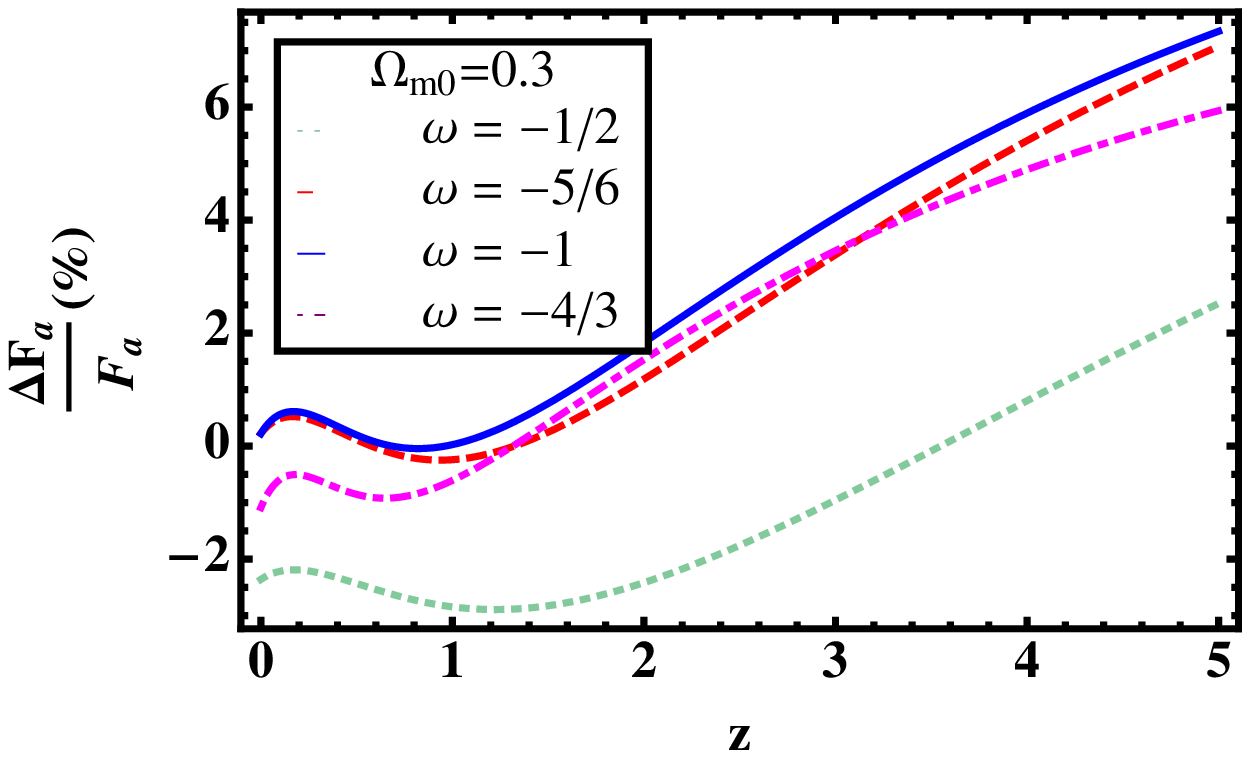,width=0.5\linewidth,clip=} &
\epsfig{file=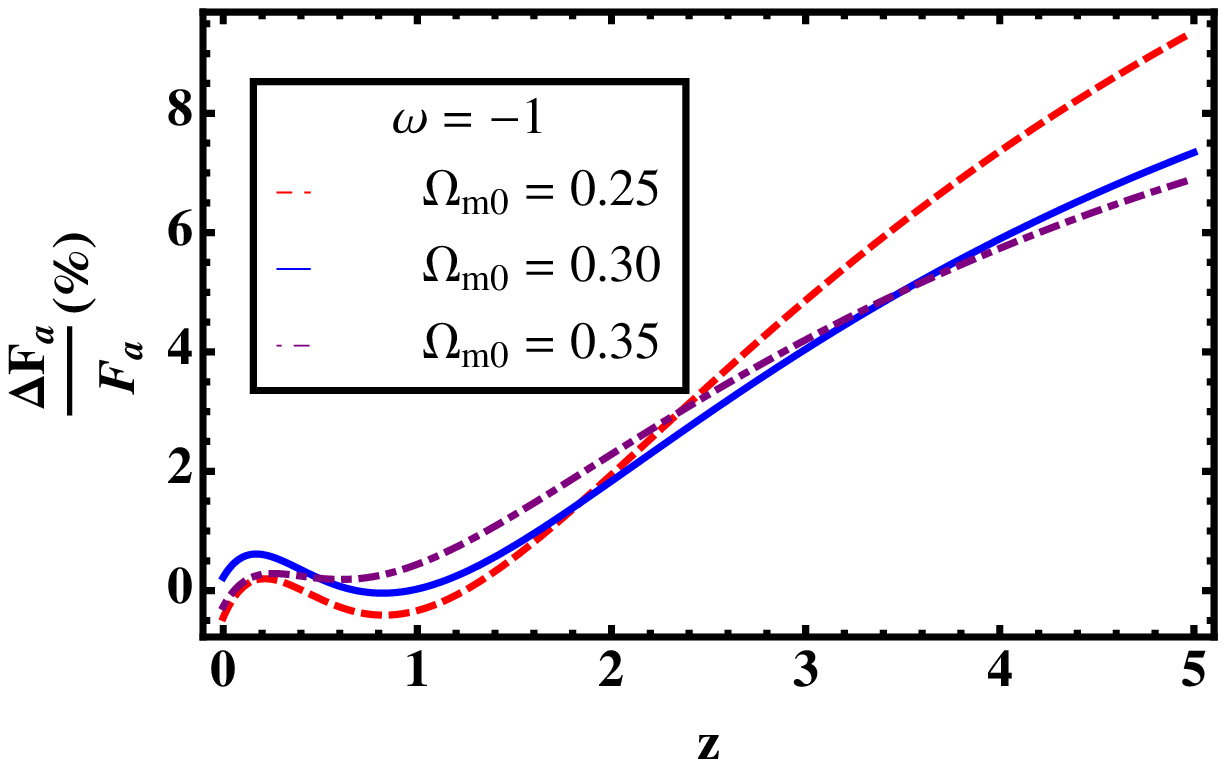,width=0.5\linewidth,clip=} \\
\end{tabular}
\vspace{-0.5cm}
\caption{{\bf First row} : The evolution of $F_{a}$ for the different $\omega$CDM models. a) $F_{a}(z)$ with the different values of $\omega$ when $\Omega_{m0} = 0.3$. The dotdashed, solid, dashed, and dotted lines correspond to $\omega = -\fr{4}{3}, -1, -\fr{5}{6}$, and $-\fr{1}{2}$, respectively. b) $F_{a}(z)$ with the different value of $\Omega_{m0}$ for $\omega = -1$. The dotdashed, solid, and dashed lines represent $\Omega_{m0} = 0.35, 0.3$, and $0.25$, respectively. {\bf Second row} : The errors of the third order fitting form $F_{a}$ as a function of $z$. c) $\fr{\Delta F_{a}}{F_{a}}$ for the different values of $\omega$ when $\Omega_{m0} = 0.3$. d) $\fr{\Delta F_{a}}{F_{a}}$ for the different values of $\Omega_{m0}$ when $\omega = -1$.} \label{fig7}
\end{figure}

In general, the temporal component of the third order equation splits into two systems,
\ba \ddot{F}_a + 2H \dot{F}_a - 4 \pi  G \rho_m F_a &=& - 8 \pi G \rho_m D^3 \, ,\label{GFaEq} \\
 \ddot{F}_b + 2H \dot{F}_b - 4 \pi  G \rho_m F_b &=& - 8 \pi G \rho_m D \Bigl( E - D^2 \Bigr) \, ,\label{GFbEq} \\
 \ddot{F}_{T} + 2 H \dot{F}_{T} &=& - 4 \pi G \rho_{m} D^3 \, , \label{GFTEq} \ea where $D$ and $E$ are the solutions of the first and the second order equations, respectively. Because we have the analytic forms of both $D$ and $E$, we can obtain the approximate analytic forms of $F_a$, $F_b$, and $F_{T}$ for the general dark energy models. We investigate the solutions for the different cosmological models. \\

i) case I : EdS \\
\\
First, we rewrite the equation for the $F_a$ by using x
\be \fr{d^2 F_{a}}{d x^2} - \fr{6}{x^2} F_{a} = -\fr{12}{x^2} D^3  \, . \label{DEFay} \ee
By using $D$, we obtain
\be F_{a} = -\fr{1}{3} c_{aD}^3 g_{1a}^3 + 9 c_{aD}^2 c_{bD} g_{1a}^2 g_{1b} - 6 c_{aD} c_{bD}^2 g_{1a} g_{1b}^2 -\fr{2}{11} c_{bD}^3 g_{1b}^3 + f_{a}^{(a)} g_{1a} + f_{b}^{(a)} g_{1b} \, , \label{Fa} \ee where $f_{a}^{(a)}$ and $f_{b}^{(a)}$ are the integral constants of the homogeneous solution of $F_{a}$. If we consider the contribution of the fastest growing solution only, then the $F_{a}$ becomes
\be F_{a+} (a) = -\fr{1}{3} c_{aD}^3 a^3 + f_{a}^{(a)} a \, . \label{Fap} \ee
We also adopt the previous consideration to obtain the initial conditions. From $F_{a+}(a_i) = 0$, one obtains
\be f_{a}^{(a)} = \fr{1}{3} a_i^2 \, , \label{faa} \ee where we use $c_{aD} = 1$ again. With this $f_{a}^{(a)}$, one gets the initial condition for $\fr{d F_{a+}}{da}$ as
\be \fr{d F_{a+}}{d a} \Bigl|_{a_i} = -\fr{2}{3} a_i^2 \, . \label{dFadai} \ee

We can repeat the same process for the $F_b$. The differential equation for $F_b$ becomes
\be \fr{d^2 F_{b}}{d x^2} - \fr{6}{x^2} F_{b} = -\fr{12}{x^2} D ( E - D^2)  \, . \label{DEFby} \ee
The solution for $F_b$ is given by
\ba F_{b} &=& \fr{10}{21} c_{aD}^3 g_{1a}^3 - \fr{6}{7} c_{aD} e_{a} g_{1a}^2 - \fr{30}{7} c_{aD}^2 c_{bD} g_{1a}^2 g_{1b} + 2 (c_{aD} e_{b} + c_{bD} e_{a}) g_{1a} g_{1b} \nonumber \\ &+& \fr{5}{2} c_{aD} c_{bD}^2 g_{1a} g_{1b}^2 -\fr{1}{2} c_{bD} e_{b} + \fr{5}{22} c_{bD}^3 g_{1b}^3 + f_{a}^{(b)} g_{1a} + f_{b}^{(b)} g_{1b} \, , \label{Fb} \ea where $f_{a}^{(b)}$ and $f_{b}^{(b)}$ are the integral constants of the homogeneous solution of $F_{b}$. Again the fastest growing solution is given by
\be F_{b+}(a) = \fr{10}{21} c_{aD}^3 g_{1a}^3 - \fr{6}{7} c_{aD} e_{a} g_{1a}^2 + f_{a}^{(b)} g_{1a} \, . \label{Fbp} \ee From $F_{b+}(a_i) = 0$, one obtains \be f_{a}^{(b)} = -\fr{16}{147} a_i^2 \, . \label{fab} \ee where we use $e_{a} = \fr{3}{7} a_i$. From this, initial condition for $\fr{d F_{b+}}{da}$ is given by
\be \fr{d F_{b+}}{d a} \Bigl|_{a_i} = \fr{86}{147} a_i^2 \, . \label{dFbdai} \ee

We write the differential equation for the $F_T$ by using the same parameter $x$
\be \fr{d^2 F_{T}}{d x^2} = -\fr{6}{x^2} D^3 \, . \label{DEFTy} \ee
The solution for $F_{T}$ is given by
\be F_{T} = -\fr{1}{7} c_{aD}^3 g_{1a}^3 - 9 c_{aD}^2 c_{bD} g_{1a}^2 g_{1b} - \fr{3}{2} c_{aD} c_{bD}^2 g_{1a} g_{1b}^2 - \fr{1}{12} c_{bD}^3 g_{1b}^3
+ f_{a}^{(T)} + f_{b}^{(T)} \fr{1}{\sqrt{a}} \, , \label{FT} \ee where $f_{a,b}^{(T)}$ are the integral constants of the homogeneous solution of
$F_{T}$. The fastest growing solution is
\be F_{T+}(a) = -\fr{1}{7} c_{aD}^3 g_{1a}^3 + f_{a}^{(T)} \, . \label{FTp} \ee From $F_{T+}(a_i) = 0$, one obtains
\be f_{a}^{(T)} = \fr{1}{7} a_{i}^3 \, . \label{f1T} \ee From this, the initial condition for $\fr{d F_{F+}}{da}$ is given by
\be \fr{d F_{T+}}{d a} \Bigl|_{a_i} = -\fr{3}{7} a_i^2 \, . \label{dFTdai} \ee
\begin{figure}
\centering
\vspace{1.5cm}
\begin{tabular}{cc}
\epsfig{file=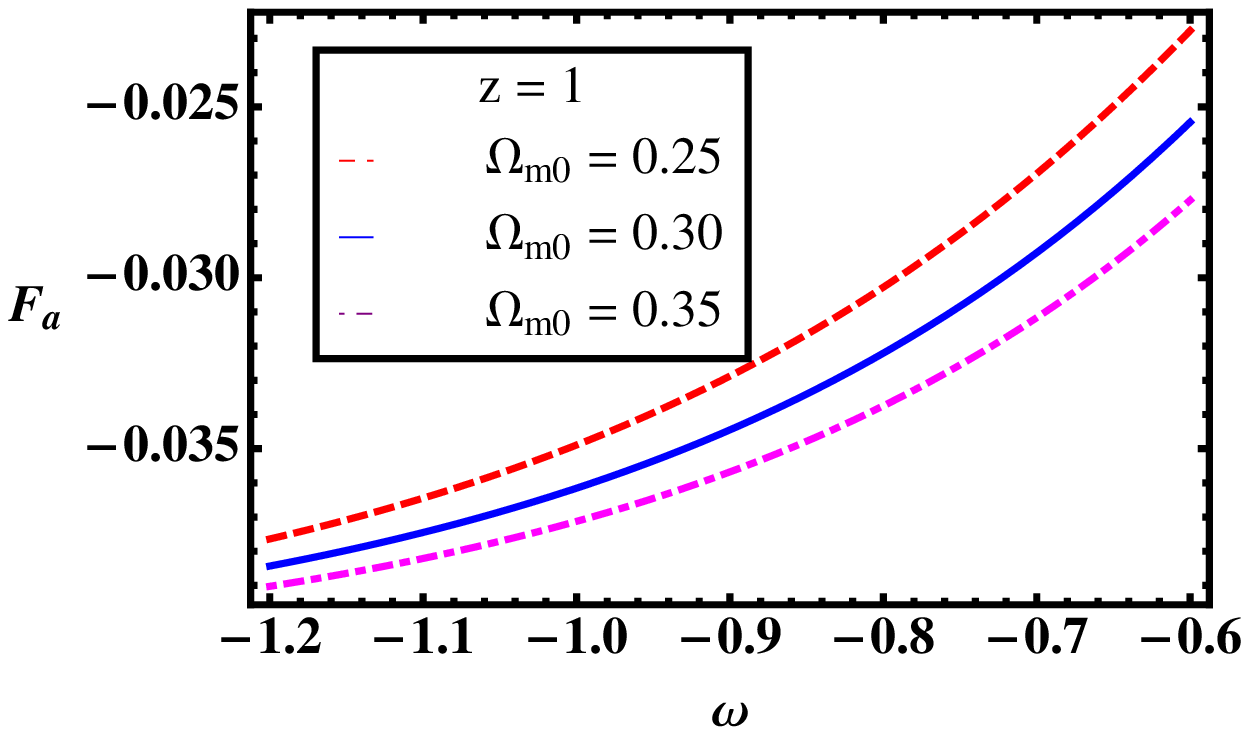,width=0.5\linewidth,clip=} &
\epsfig{file=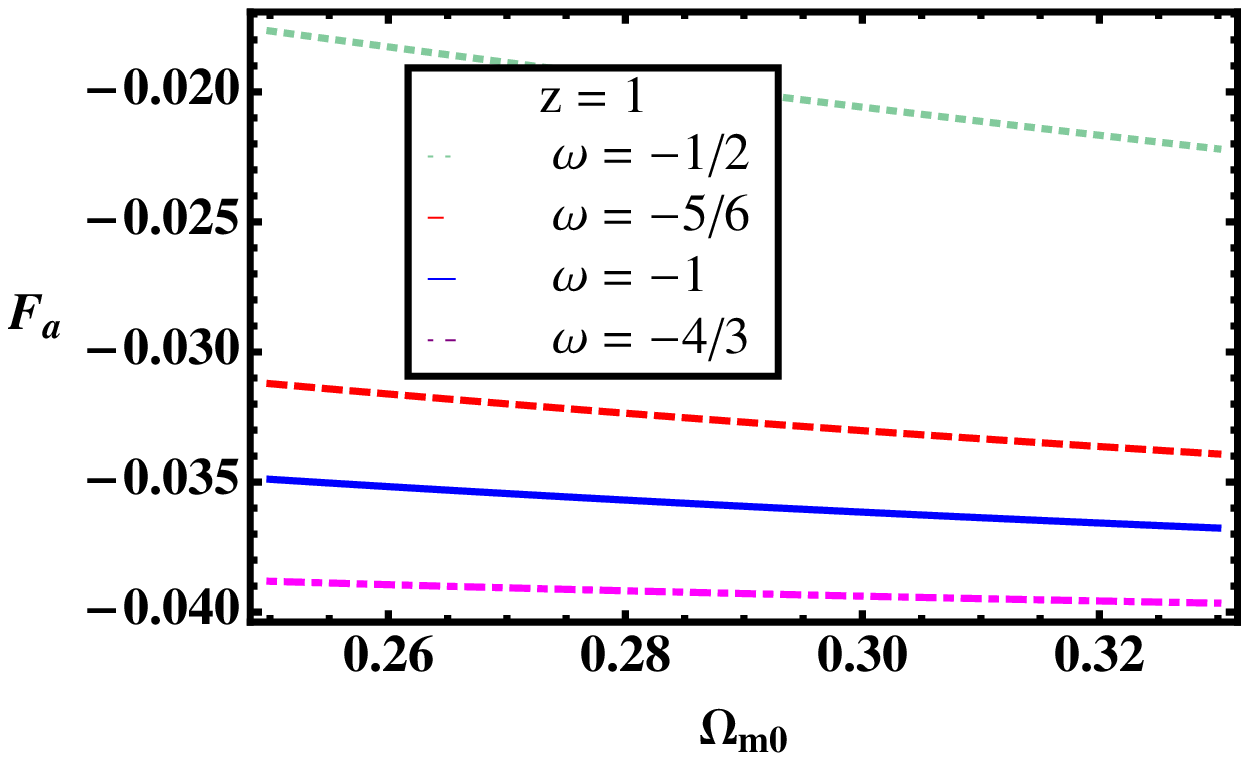,width=0.5\linewidth,clip=} \\
\epsfig{file=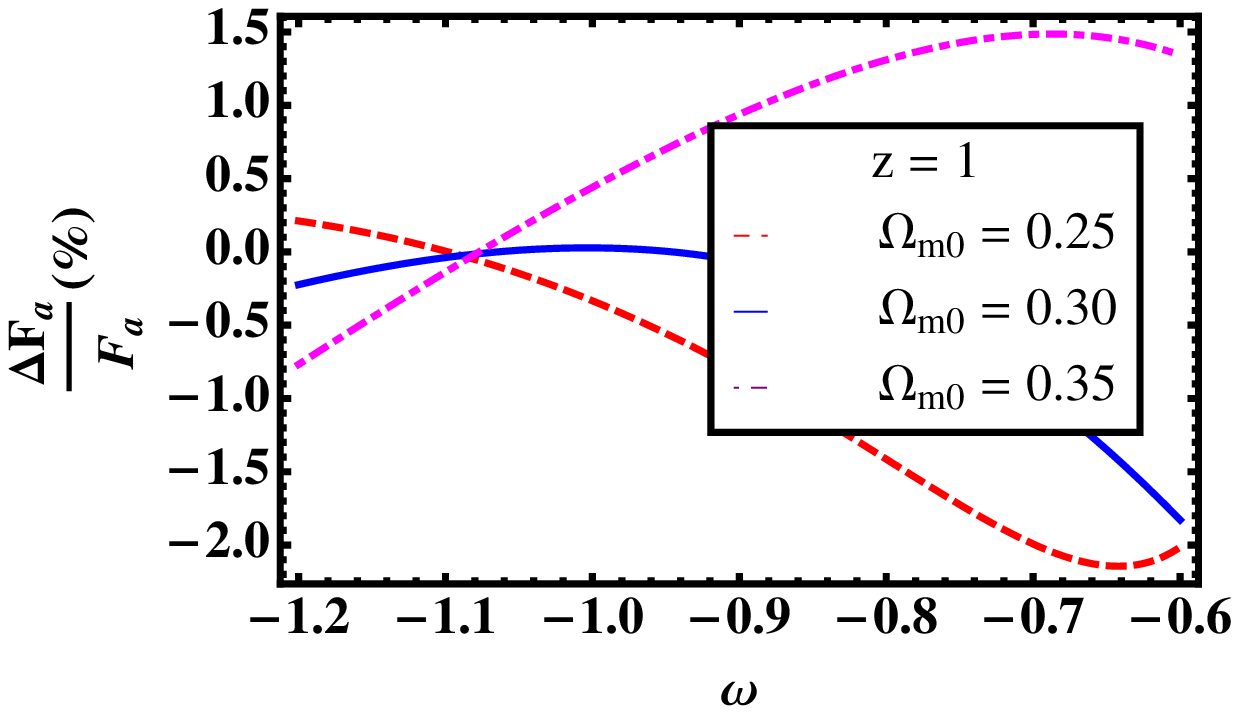,width=0.5\linewidth,clip=} &
\epsfig{file=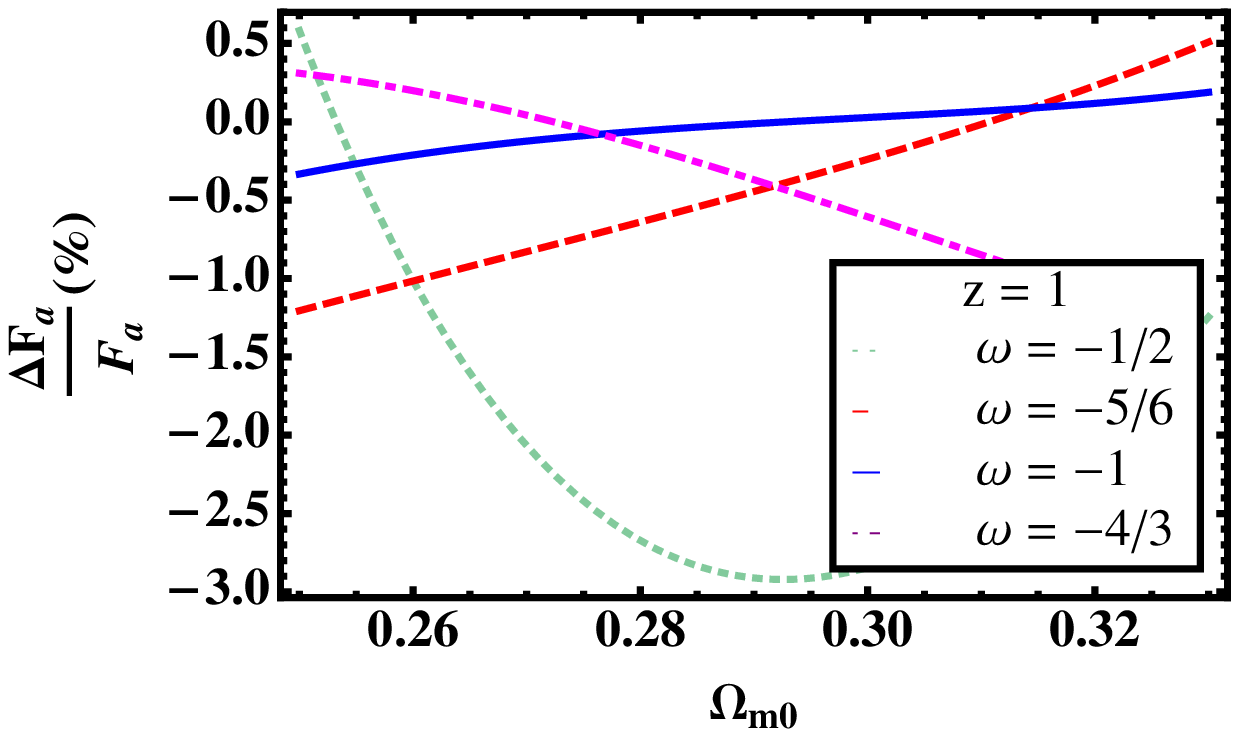,width=0.5\linewidth,clip=} \\
\end{tabular}
\vspace{-0.5cm}
\caption{{\bf First row} : The values of $F_a$ for the different cosmological parameters at $z=1$. a) $F_a$ dependence on $\omega$ for the different values of $\Omega_{m0}$. Dotdashed, solid, and dashed lines correspond to $\Omega_{m0} = 0.35, 0.30$, and $0.25$, respectively. b) $F_a$ dependence on $\Omega_{m0}$ for the different values of $\omega$. Dotted, dashed, solid, and dotdashed lines correspond to $\omega = -\fr{1}{2}, -\fr{5}{6}, -1$, and $-\fr{4}{3}$, respectively. {\bf Second row} : The errors on the fitting form at $z = 1$ for the different cosmological models. c) The dependence of errors on $\omega$ for the different values of $\Omega_{m0}$. d) The errors as a function of $\Omega_{m0}$ for the different values of $\omega$.} \label{fig8}
\end{figure}
ii) case II ; Open Universe \\
\\

The third order perturbation equations are rewritten as
\ba  \fr{d^2 F_{a}}{d y^2} - \fr{6}{y^2 - 1} F_{a} &=& -\fr{12}{y^2 - 1} D^3 \, , \label{DEFaoy} \\
 \fr{d^2 F_{b}}{d y^2} - \fr{6}{y^2 - 1} F_{b} &=& -\fr{12}{y^2 - 1 } D ( E - D^2) \, , \label{DEFboy} \\
 \fr{d^2 F_{T}}{d y^2} &=& -\fr{6}{y^2 - 1 } D^3 \, . \label{DEFToy} \ea As we show in the previous subsection, this is the special case of general dark energy model with equation of state $\omega = -\fr{1}{3}$. Thus, we will show the solutions of the above equations in the next case.

iii) case III : $\omega$CDM \\
\\

In this case, one can rewrite the Eq. (\ref{GFaEq}) as
\be Y \fr{d^2 F_a}{d Y^2} + \Biggl[ 1 + \fr{1}{6 \omega} - \fr{1}{2(Y+1)} \Biggr] \fr{d F_a}{d Y} - \fr{1}{6 \omega^2} \fr{1}{Y+1} F_{a} = - \fr{1}{3 \omega^2} \fr{1}{Y+1} D^3 \, . \label{DEFayGDE} \ee

We obtain the fitting form of the fastest growing solution $F_{a}$ as
\be F_a (a) = B_{Fa} Y^{P_{Fa}} (1+Y)^{Q_{Fa}} \, , \label{Fawa} \ee where
\ba B_{Fa} &=& -(-6 \omega)^{0.092} + 1.166 -0.333 A \label{BFa} \, , \\
P_{Fa} &=& -\Bigl( (- 5.86 + 4 A) \omega \Bigr)^{-0.71 - \fr{A}{6}} \label{PFa} \, , \\
Q_{Fa} &=& - \Bigl( ( -1.13 - \fr{A}{2}) \omega \Bigr)^{-1.23 + \fr{A}{6}} \label{QFa} \, . \ea

From the above fitting form Eq. (\ref{Fawa}), we obtain several properties of the fastest growing solution $F_{a}$.
First, the signature of $F_{a}$ is opposite to that of $D$. Thus, it decreases as a function of time. Second, as $\omega$ decreases,
so does $F_{a}$. Third, $F_{a}$ decreases as $\Omega_{m0}$ increases. These properties are similar to those of $E$ in the previous subsection.
It is natural because $E$ and $F_{a}$ have the source terms which is proportional to $D^2$ and $D^3$, respectively. We investigate the behaviors
of $F_{a}$ for the various cosmological parameters. We show the time evolution of $F_{a}$ for the different cosmological models in the first
row of Fig. \ref{fig7}. In the left panel, we show $F_{a}(z)$ for the different values of $\omega$ when $\Omega_{m0} = 0.3$.
The dotted, dashed, solid, and dotdashed lines correspond to $\omega = -\fr{1}{2}, -\fr{5}{6}, -1$, and $-\fr{4}{3}$, respectively. The present
values of $F_{a}$ vary from $-0.07$ to $-0.19$ for $\omega = -\fr{1}{2}$ and $-\fr{4}{3}$, respectively.  In the right panel, we show the
evolutions of $F_{a}$ for the different values of $\Omega_{m0}$ when we consider the $\Lambda$CDM model. The dotted, solid, and dotdashed lines
correspond $\Omega_{m0} = 0.25, 0.30$, and $0.35$, respectively. We obtain $-0.18 \leq F_a \leq -0.14$ for $0.25 \leq \Omega_{m0} \leq 0.35$.
In the second row of Fig. \ref{fig7}, we show the errors of the fitting form as a function of the redshift $z$ for the different models. In the first column, we show the $\fr{\Delta F_{a}}{F_{a}}$ for the different values of $\omega$ when we fix $\Omega_{m0} = 0.3$. The errors are less than $4$ \% for all models when we consider $z \leq 3$. In the second column, we check the errors of the fitting form for the different $\Omega_{m0}$ values when we consider the $\Lambda$CDM model. Again, the errors are less than $4$ \% for all the considered $\Omega_{m0}$ values up to $z \leq 3$.

\begin{figure}
\centering
\vspace{1.5cm}
\begin{tabular}{cc}
\epsfig{file=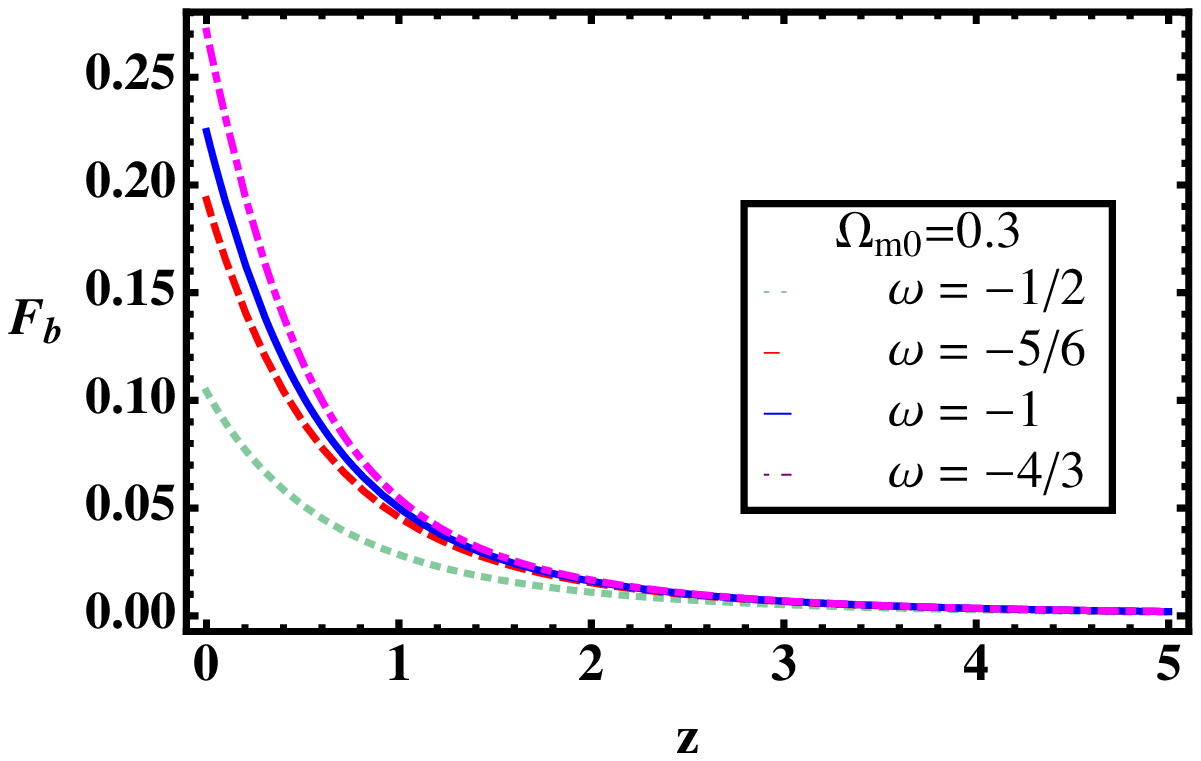,width=0.5\linewidth,clip=} &
\epsfig{file=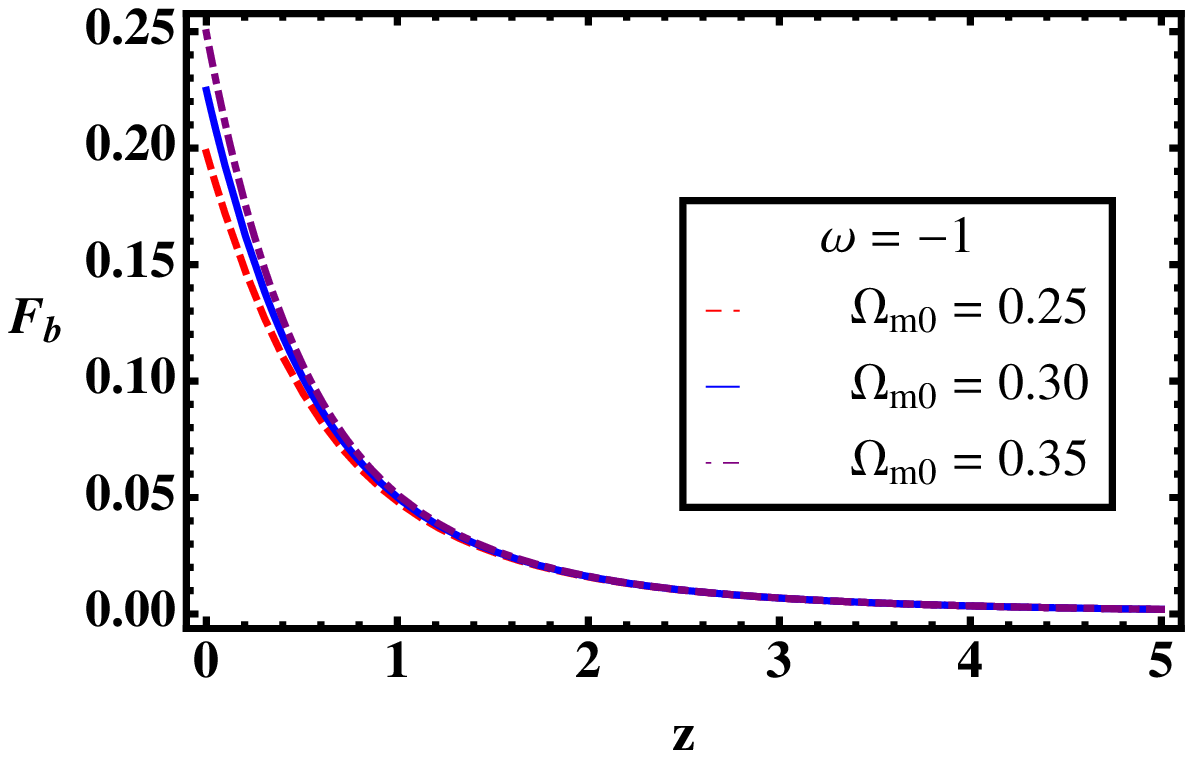,width=0.5\linewidth,clip=} \\
\epsfig{file=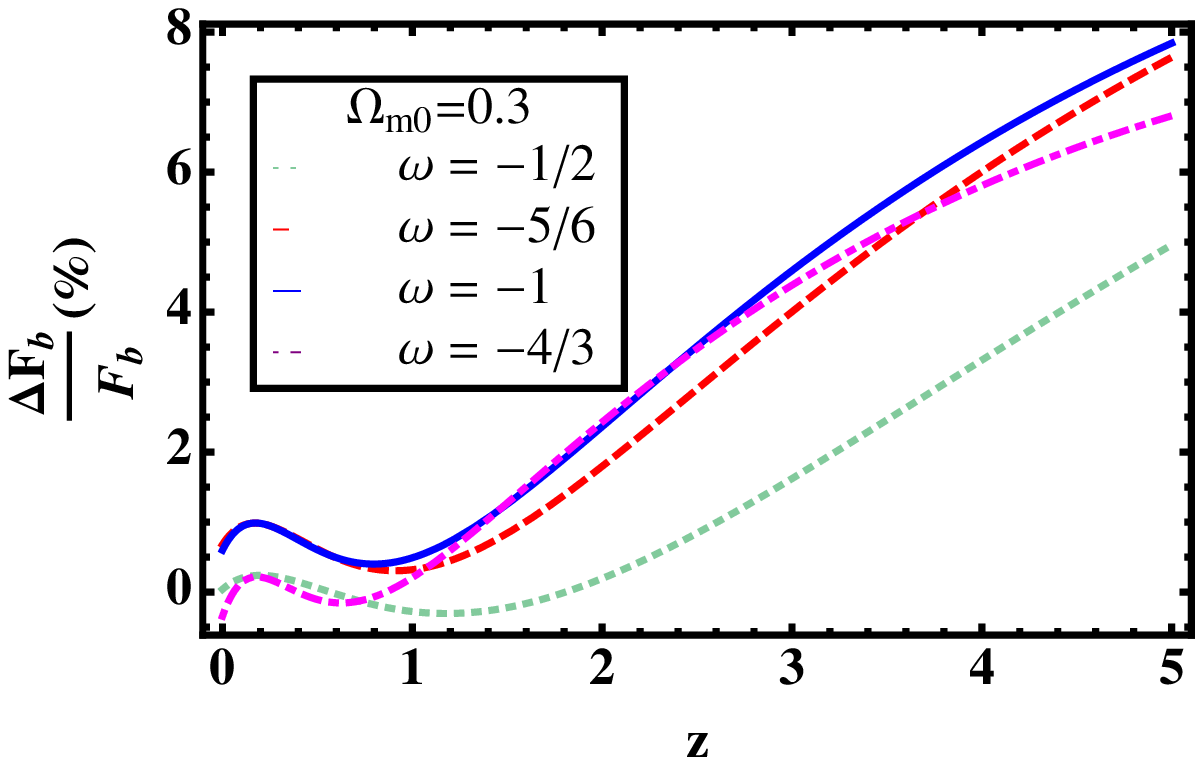,width=0.5\linewidth,clip=} &
\epsfig{file=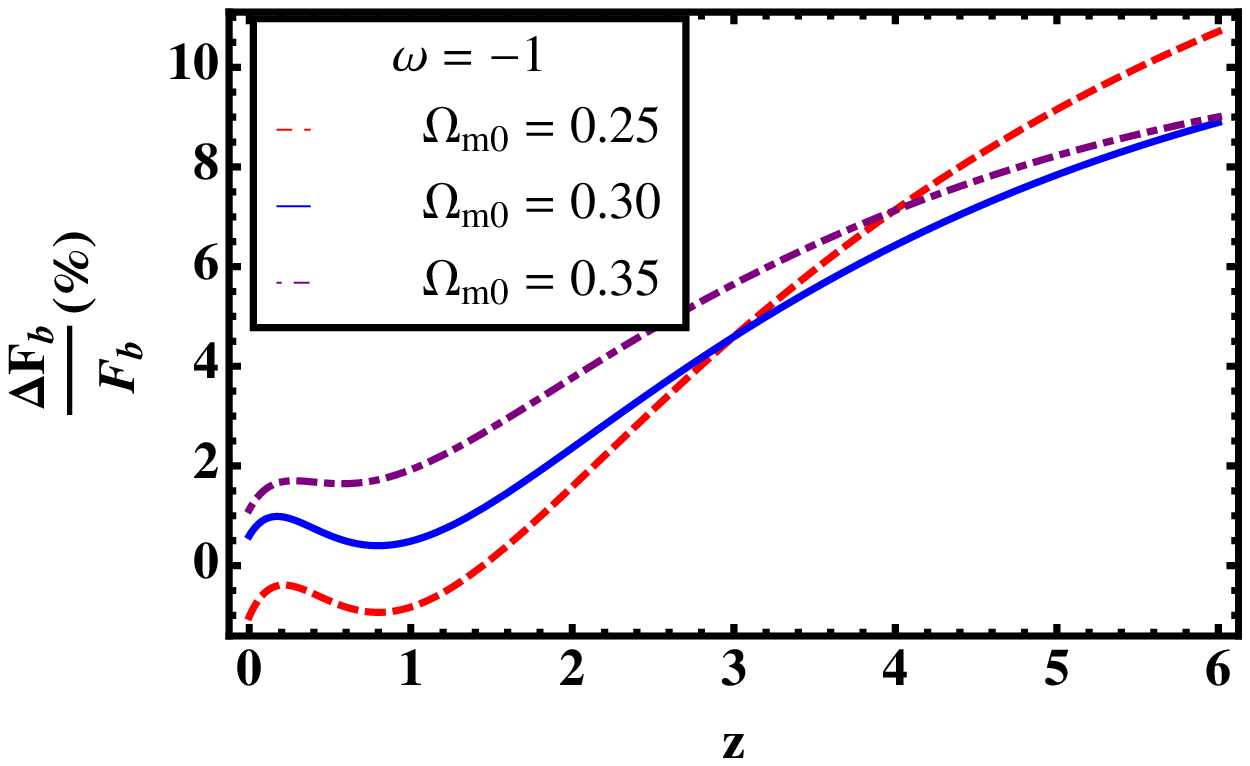,width=0.5\linewidth,clip=} \\
\end{tabular}
\vspace{-0.5cm}
\caption{{\bf First row} : The evolution of $F_{b}$ for the different $\omega$CDM models. a) $F_{b}(z)$ with the different values of $\omega$ when $\Omega_{m0} = 0.3$. The dotdashed, solid, dashed, and dotted lines correspond to $\omega = -\fr{4}{3}, -1, -\fr{5}{6}$, and $-\fr{1}{2}$, respectively. b) $F_{b}(z)$ with the different value of $\Omega_{m0}$ for $\omega = -1$. The dotdashed, solid, and dashed lines represent $\Omega_{m0} = 0.35, 0.30$, and $0.25$, respectively. {\bf Second row} : The errors of the $F_{b}$ fitting form as a function of $z$. c) $\fr{\Delta F_{b}}{F_{b}}$ for the different values of $\omega$ when $\Omega_{m0} = 0.3$. d) $\fr{\Delta F_{b}}{F_{b}}$ for the different values of $\Omega_{m0}$ when $\omega = -1$.} \label{fig9}
\end{figure}

We also investigate the dependence of $F_a$ on the cosmological parameters at the specific $z$. In the first row of Fig. \ref{fig8}, we show the values of $F_a$ as a function of $\omega$ and $\Omega_{m0}$ at the specific redshift $z$. In the left panel, we fix the redshift $z = 1$ and check the dependence of $F_a$ on $\omega$ for the different values of $\Omega_{m0}$. The dashed, solid, and the dotdashed lines correspond to $\Omega_{m0} = 0.25, 0.30$, and $0.35$, respectively. For $\Omega_{m0} = 0.35$, $F_a$ varies from -0.023 to -0.040 when $\omega$ changes from -0.5 to -1.2. We also obtain $-0.039 \leq F_a \leq -0.018$ for $-1.2 \leq \omega \leq -0.5$ when $\Omega_{m0} = 0.25$. Thus, we can conclude that $F_a$ dependence on $\omega$ becomes weaker as $\Omega_{m0}$ increases. In the right panel, we show the dependence of $F_a$ on $\Omega_{m0}$ for the different $\omega$ models at $z = 1$. Again, the dotted, dashed, solid, and dotdashed lines correspond to $\omega = -\fr{1}{2}, -\fr{5}{6}, -1$, and $-\fr{4}{3}$, respectively. For $\omega = -\fr{1}{2}$, $F_a$ varies from -0.018 to -0.023 when $\Omega_{m0}$ changes from 0.25 to 0.35. $F_a$ changes from -0.039 to -0.040 when $\Omega_{m0}$ changes from 0.25 to 0.35 for $\omega = -\fr{4}{3}$. This case $F_a$ is almost constant for the different values of $\Omega_{m0}$. Thus, we can conclude that $F_a$ dependence on $\Omega_{m0}$ becomes weaker as $\omega$ decreases. All of these properties are same as those of $E$. We also investigate the errors on the fitting form at $z = 1$ for the different cosmological models. In the second row of Fig. \ref{fig8}, we show the errors on the fitting form as a function of $\omega$ and $\Omega_{m0}$. In the first column, we show the dependence of errors on $\omega$ for the different values of $\Omega_{m0}$. The errors are about less than $2$ \% for $-1.2 \leq \omega \leq - 0.7$. In the second column, we show the errors as a function of $\Omega_{m0}$ for the different $\omega$ models at $z = 1$. The errors are less than 2 \% for all $\Omega_{m0}$ except $\omega = -\fr{1}{2}$.

\begin{figure}
\centering
\vspace{1.5cm}
\begin{tabular}{cc}
\epsfig{file=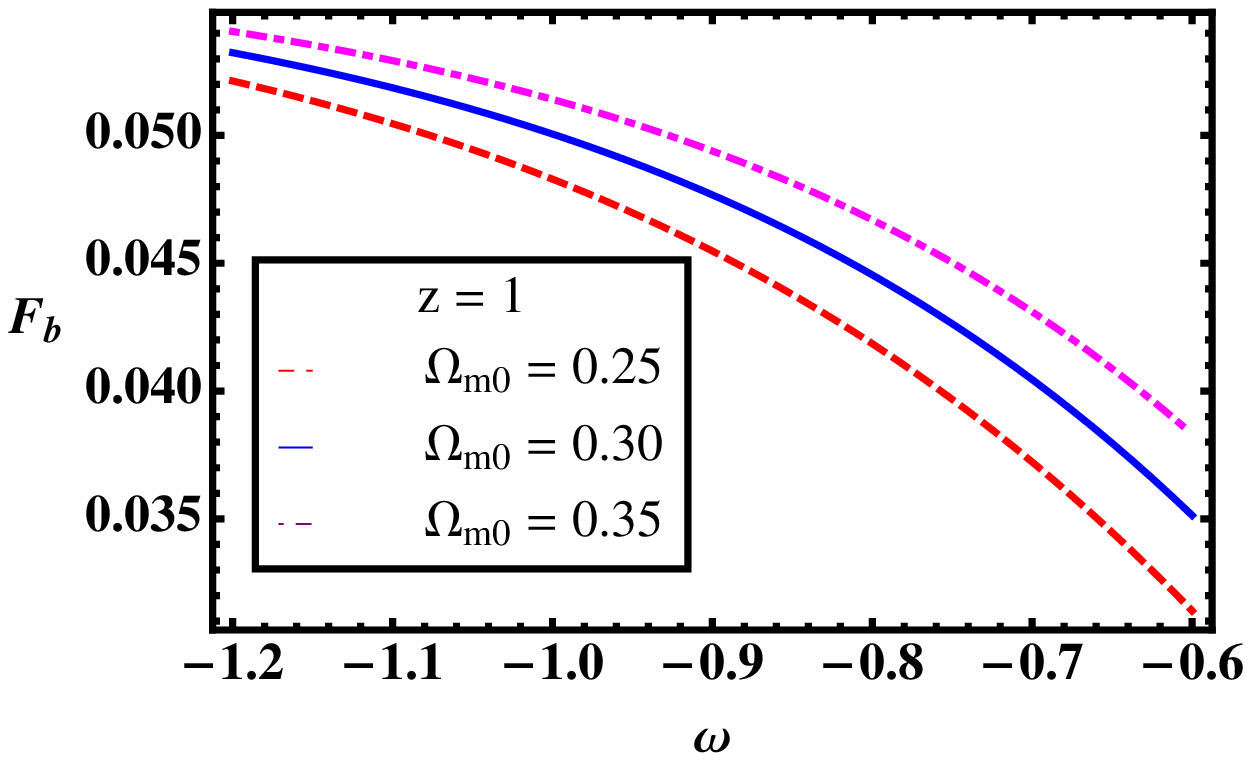,width=0.5\linewidth,clip=} &
\epsfig{file=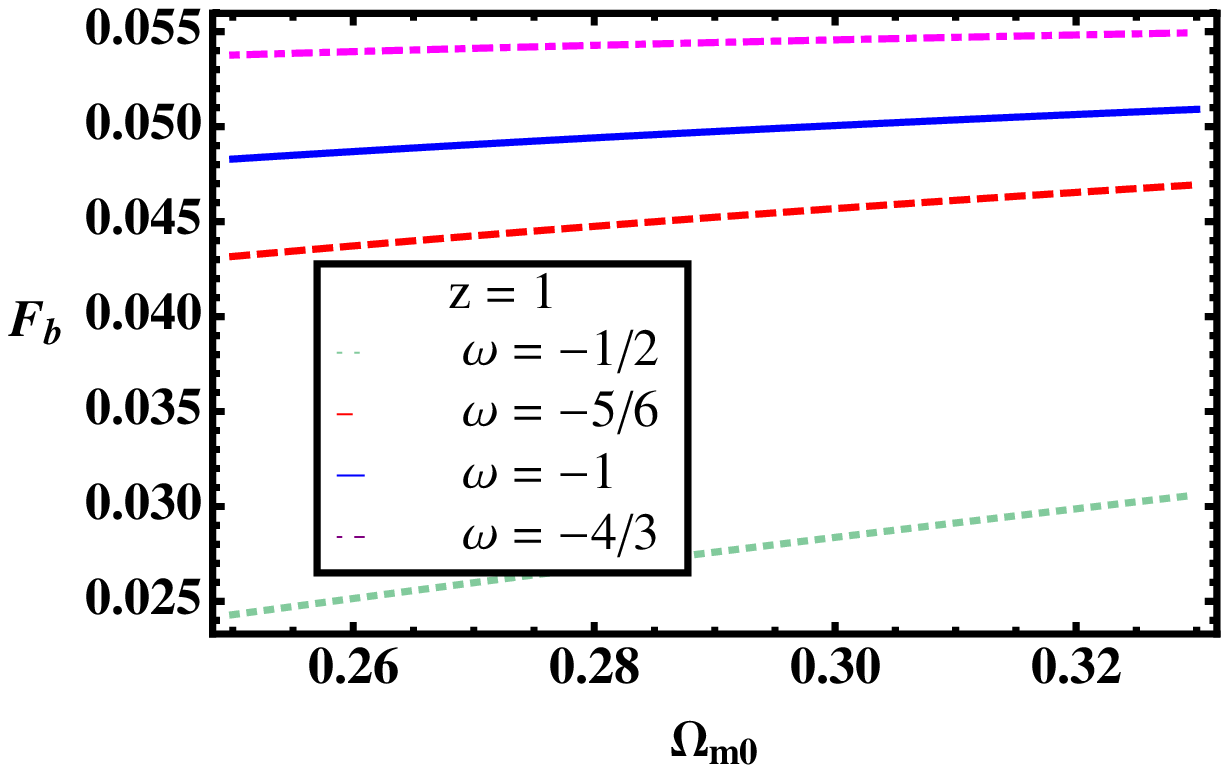,width=0.5\linewidth,clip=} \\
\epsfig{file=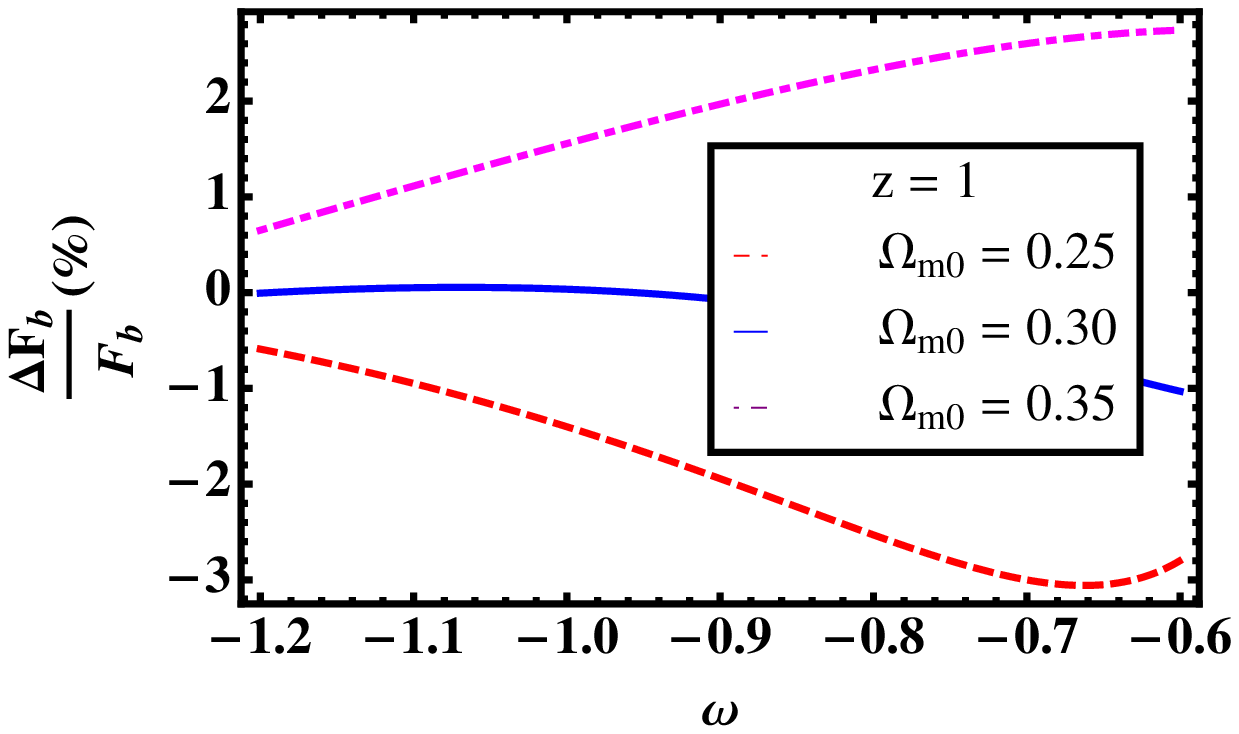,width=0.5\linewidth,clip=} &
\epsfig{file=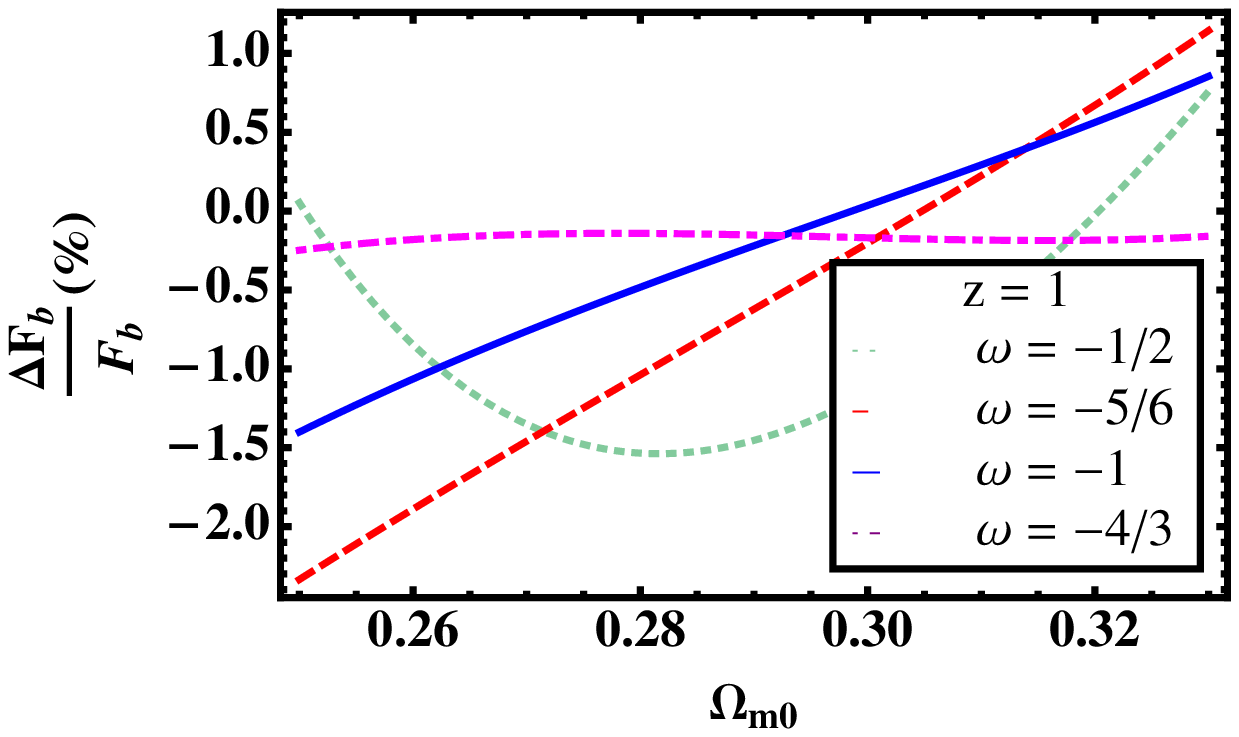,width=0.5\linewidth,clip=} \\
\end{tabular}
\vspace{-0.5cm}
\caption{{\bf First row} : The values of $F_b$ for the different cosmological parameters at $z=1$. a) $F_a$ dependence on $\omega$ for the different values of $\Omega_{m0}$. Dotdashed, solid, and dashed lines correspond to $\Omega_{m0} = 0.35, 0.30$, and $0.25$, respectively. b) $F_a$ dependence on $\Omega_{m0}$ for the different values of $\omega$. Dotted, dashed, solid, and dotdashed lines correspond to $\omega = -\fr{1}{2}, -\fr{5}{6}, -1$, and $-\fr{4}{3}$, respectively. {\bf Second row} : The errors on the fitting form at $z = 1$ for the different cosmological models. c) The dependence of errors on $\omega$ for the different values of $\Omega_{m0}$. d) The errors as a function of $\Omega_{m0}$ for the different values of $\omega$.} \label{fig10}
\end{figure}
One can rewrite the Eq. (\ref{GFbEq}) as
\be Y \fr{d^2 F_b}{d Y^2} + \Biggl[ 1 + \fr{1}{6 \omega} - \fr{1}{2(Y+1)} \Biggr] \fr{d F_b}{d Y} - \fr{1}{6 \omega^2} \fr{1}{Y+1} F_{b} = - \fr{1}{3 \omega^2} \fr{1}{Y+1} D \Bigl( E - D^2 \Bigr) \, . \label{DEFbyGDE} \ee
Now we can repeat to the same process as before to obtain the fitting form of $F_{b}$ as
\be F_b (a) = B_{Fb} Y^{P_{Fb}} (1+Y)^{Q_{Fb}} \, , \label{Fbwa} \ee where
\ba B_{Fb} &=& (-6 \omega)^{0.126} - 1.222 -0.439 A \label{BFb} \, , \\
P_{Fb} &=& -\Bigl( (- 5.79 + 3.98 A) \omega \Bigr)^{-0.72 - \fr{A}{6}} \label{PFb} \, , \\
Q_{Fb} &=& - \Bigl( ( -1.11 - \fr{A}{2}) \omega \Bigr)^{-1.24 + \fr{A}{6}} \label{QFb} \, . \ea

From the above fitting form Eq. (\ref{Fbwa}), we obtain several properties of the fastest growing solution $F_{b}$. First, the signature of $F_{b}$ is same as that of $D$. Thus, it increases as a function of time. Second, $F_{b}$ increases as $\omega$ decreases. Third, as $\Omega_{m0}$ increases so does $F_{b}$. These properties are similar to those of $D$ in the previous subsection and opposite to those of $E$ or $F_{a}$. It is easy to understand because $F_{b}$ has the opposite sign of the source term compared to those of $E$ or $F_{a}$. We investigate the behaviors of $F_{b}$ for the various cosmological parameters. We show the time evolution of $F_{b}$ for the different cosmological models in the first row Fig. \ref{fig9}. In the left panel, we show $F_{b}(z)$ for the different values of $\omega$ when $\Omega_{m0} = 0.3$. The dotted, dashed, solid, and dotdashed lines correspond to $\omega = -\fr{1}{2}, -\fr{5}{6}, -1$, and $-\fr{4}{3}$, respectively. The present values of $F_{b}$ vary from $0.104$ to $0.271$ for $\omega = -\fr{1}{2}$ and $-\fr{4}{3}$, respectively. In the right panel, we show the evolutions of $F_{b}$ for the different values of $\Omega_{m0}$ when we consider the $\Lambda$CDM model. The dotted, solid, and dotdashed lines correspond $\Omega_{m0} = 0.25, 0.30$, and $0.35$, respectively. We obtain $0.198 \leq F_b \leq 0.249$ for $0.25 \leq \Omega_{m0} \leq 0.35$ at present. The fitting form of $F_{b}$ is given by Eqs. (\ref{Fbwa}) - (\ref{QFb}). We investigate the accuracy of this fitting form for the different
cosmological models. In the second row of Fig.\ref{fig9}, we show the errors of the fitting form as a function of the redshift $z$ for the different models.
In the first column, we show the $\fr{\Delta F_{b}}{F_{b}}$ for the different values of $\omega$ when we fix $\Omega_{m0} = 0.3$.
The errors are
less than $4$ \% for all models when we consider $z \leq 3$. In the second column, we check the errors of the fitting form for the different
$\Omega_{m0}$ values when we consider the $\Lambda$CDM model. Again, the errors are less than $4$ \% for all the considered $\Omega_{m0}$ values up to $z \leq 3$.

We also investigate the dependence of $F_b$ on the cosmological parameters at the specific $z$. In the first row of Fig. \ref{fig10}, we show the values of $F_b$ as a function of both $\omega$ and $\Omega_{m0}$ at the specific redshift $z$. In the left panel, we fix the redshift $z = 1$ and check the dependence of $F_b$ on $\omega$ for the different values of $\Omega_{m0}$. The dashed, solid, and the dotdashed lines correspond to $\Omega_{m0} = 0.25, 0.30$, and $0.35$, respectively. For $\Omega_{m0} = 0.35$, $F_b$ varies from 0.032 to 0.055 when $\omega$ changes from -0.5 to $-\fr{4}{3}$. We also obtain $0.024 \leq F_b \leq 0.054$ for $-\fr{4}{3} \leq \omega \leq -\fr{1}{2}$ when $\Omega_{m0} = 0.25$. Thus, we can conclude that $F_b$ dependence on $\Omega_{m0}$ becomes weaker as $\omega$ decreases. In the right panel, we show the dependence of $F_b$ on $\Omega_{m0}$ for the different $\omega$ models at $z = 1$. Again, the dotted, dashed, solid, and dotdashed lines correspond to $\omega = -\fr{1}{2}, -\fr{5}{6}, -1$, and $-\fr{4}{3}$, respectively. For $\omega = -\fr{1}{2}$, $F_b$ varies from 0.024 to 0.032 when $\Omega_{m0}$ changes from 0.25 to 0.35. $F_b$ changes from 0.054 to 0.055 when $\Omega_{m0}$ changes from 0.25 to 0.35 for $\omega = -\fr{4}{3}$. $F_b$ is almost constant for the different values of $\Omega_{m0}$ when $\omega = -\fr{4}{3}$. Thus, we can conclude that $F_b$ dependence on $\Omega_{m0}$ becomes weaker as $\omega$ decreases. The errors on the fitting form at $z = 1$ for the different cosmological models are shown in the second row of Fig. \ref{fig10}. In the first column, we show the dependence of errors on $\omega$ for the different values of
$\Omega_{m0}$. The errors are about less than $3$ \% for $-1.2 \leq \omega \leq - 0.7$. In the second column, the errors of the fitting form as a function of $\Omega_{m0}$ are shown for the different $\omega$
models at $z = 1$. The errors are less than 2 \% for all $\Omega_{m0}$ including $\omega = -\fr{1}{2}$.

The Eq. (\ref{GFTEq}) for the transverse mode in this model becomes
\be Y \fr{d^2 F_{T}}{d Y^2} + \Biggl[ 1 + \fr{1}{6 \omega} - \fr{1}{2(Y+1)} \Biggr] \fr{d F_{T}}{d Y} = -\fr{1}{6 \omega^2} \fr{1}{Y + 1} D^3 \, . \label{DEFTyGDE} \ee The homogeneous solution for the above equation is given by
\be F_{T}^{(h)} = f_{a}^{(T)} + f_{b}^{(T)} \fr{1}{3 \omega -1} \Bigl(\fr{1}{A} Y \Bigr)^{\fr{3\omega-1}{6\omega}} \sqrt{\fr{1}{1-\Omega_{m0}}} F \Bigl[\fr{1}{2}, \fr{1}{2} - \fr{1}{6\omega}, \fr{3}{2} - \fr{1}{6\omega}, - Y \Bigr] \, . \label{FTh} \ee Even though, one can obtain the homogeneous solution of the transverse mode, we need to find the fitting form of the fastest growing solution $F_{T}$ which can be given by
\be F_{T}(a) = B_{F_{T}} Y^{P_{F_{T}}} (1+Y)^{Q_{F_{T}}} \label{FTwa} \, , \ee
where
\ba B_{F_{T}} &=& - (-0.169 \omega)^{0.053} - 0.147A + 0.909  \label{BFT} \, , \\
P_{F_{T}} &=& -\Bigl( -3.77 \omega \Bigr)^{-0.80} \label{PFT} \, , \\
Q_{F_{T}} &=& -\Bigl( -1.4
4 \omega \Bigr)^{-1.15} \label{QFT} \, . \ea
\begin{figure}
\centering
\vspace{1.5cm}
\begin{tabular}{cc}
\epsfig{file=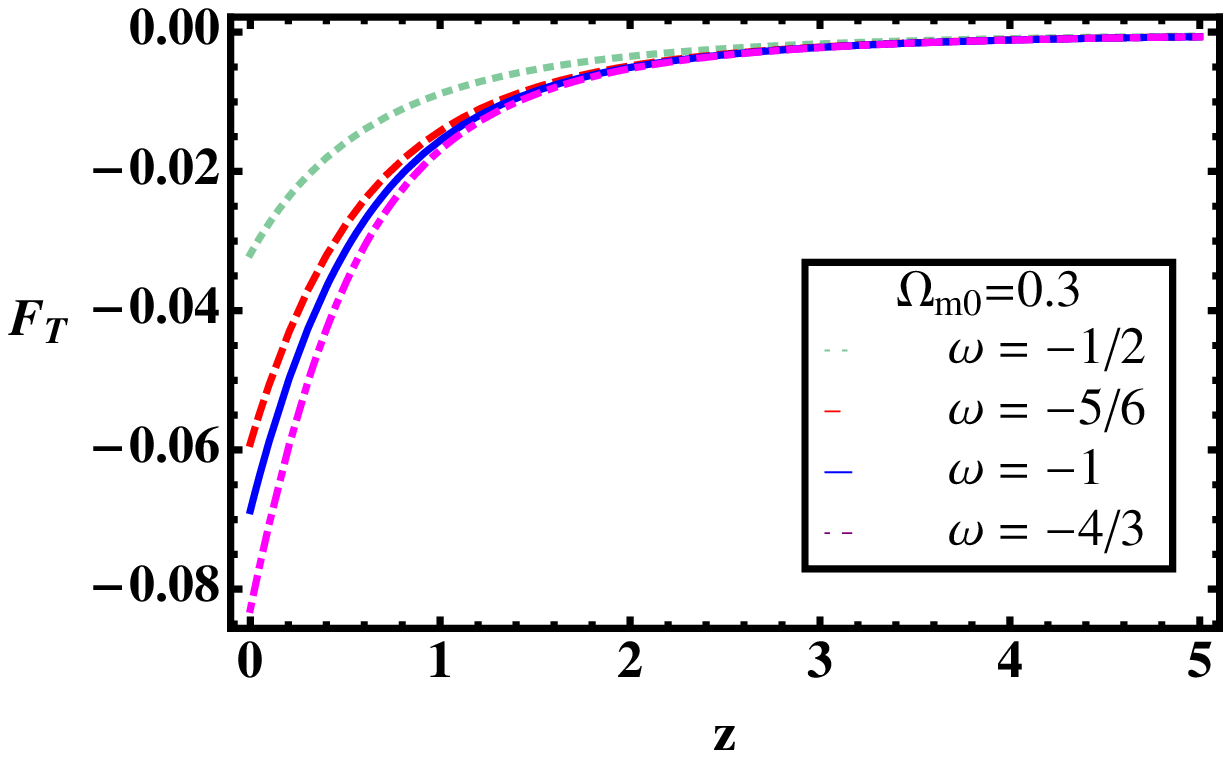,width=0.5\linewidth,clip=} &
\epsfig{file=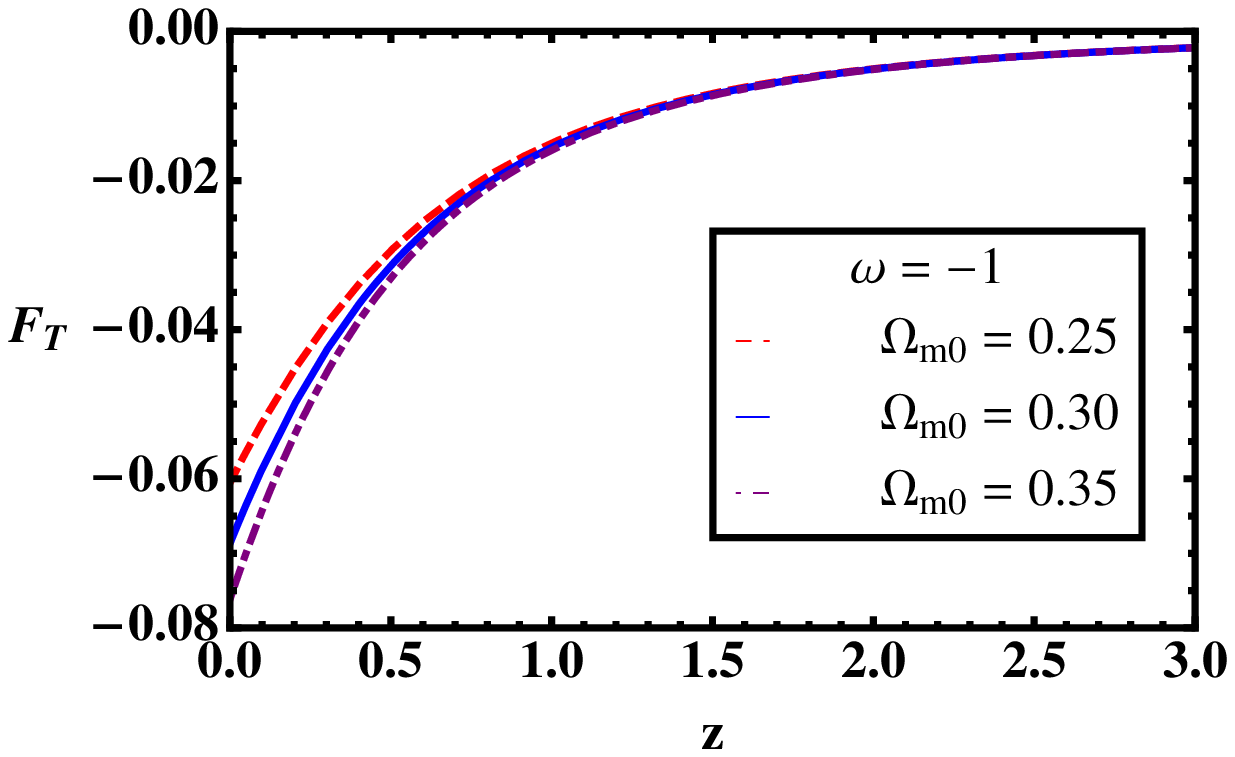,width=0.5\linewidth,clip=} \\
\epsfig{file=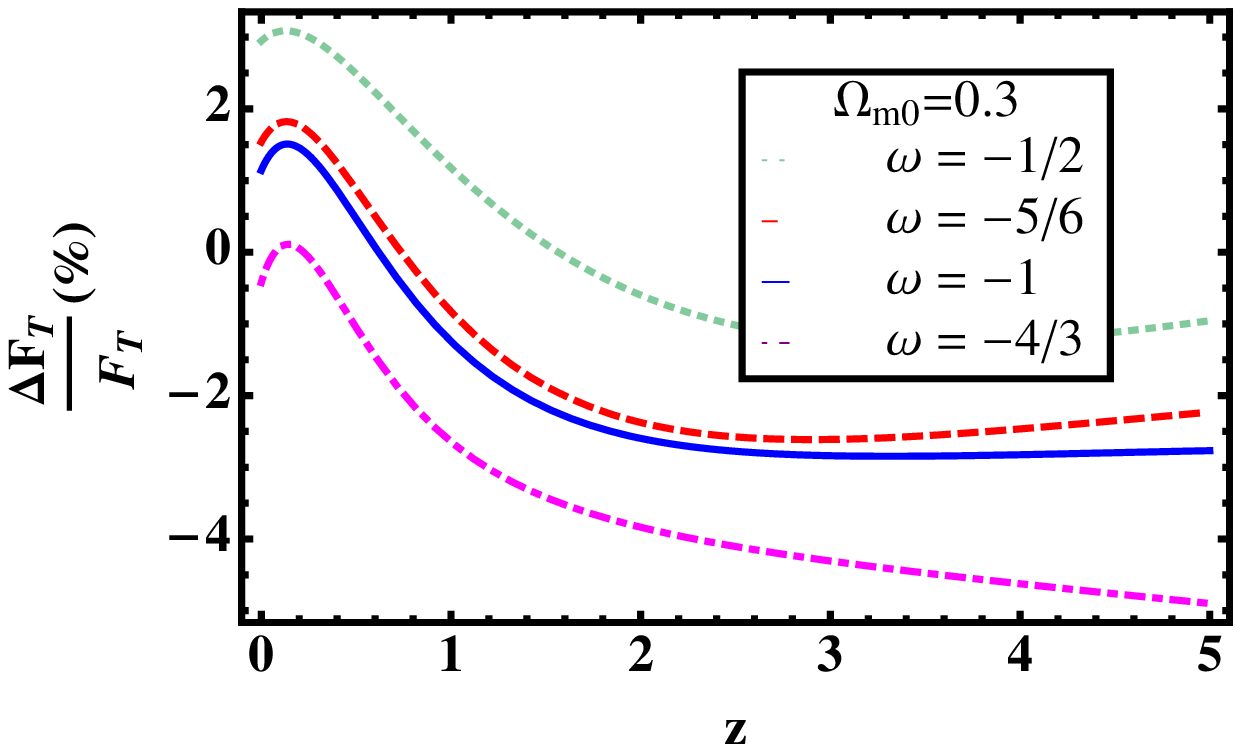,width=0.5\linewidth,clip=} &
\epsfig{file=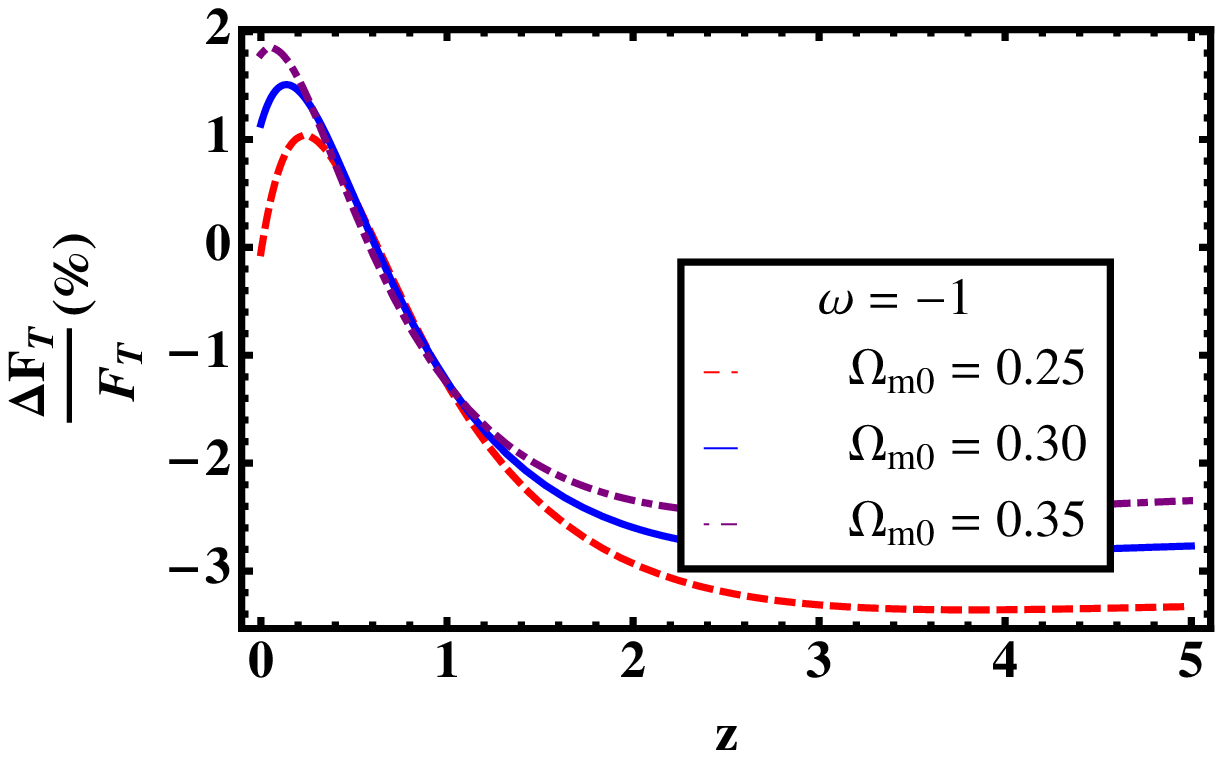,width=0.5\linewidth,clip=} \\
\end{tabular}
\vspace{-0.5cm}
\caption{{\bf First row} : The evolution of $F_{T}$ for the different $\omega$CDM models. a) $F_{T}(z)$ with the different values of $\omega$ when $\Omega_{m0} = 0.3$. The dotdashed, solid, dashed, and dotted lines correspond to $\omega = -\fr{4}{3}, -1, -\fr{5}{6}$, and $-\fr{1}{2}$, respectively. b) $F_{T}(z)$ with the different value of $\Omega_{m0}$ for $\omega = -1$. The dotdashed, solid, and dashed lines represent $\Omega_{m0} = 0.35, 0.30$, and $0.25$, respectively. {\bf Second row} : The errors of the $F_{T}$ fitting form as a function of $z$. c) $\fr{\Delta F_{T}}{F_{T}}$ for the different values of $\omega$ when $\Omega_{m0} = 0.3$. d) $\fr{\Delta F_{T}}{F_{T}}$ for the different values of $\Omega_{m0}$ when $\omega = -1$.} \label{fig11}
\end{figure}

From the above fitting form Eqs. (\ref{FTwa}) - (\ref{QFT}), one can obtain several properties of the fastest growing solution $F_{T}$ which is similar to those of $F_{a}$. It is natural because both $F_{T}$ and $F_{a}$ have the source terms which is proportional to $D^3$.
First, the signature of $F_{T}$ is opposite to that of $D$. Thus, it decreases as a function of time. Second, as $\omega$ decreases,
so does $F_{T}$. Third, $F_{T}$ decreases as $\Omega_{m0}$ increases.  We investigate the behaviors
of $F_{T}$ for the various cosmological parameters. We show the time evolution of $F_{T}$ for the different cosmological models in the first row of Fig. \ref{fig11}.
In the left panel,  we show $F_{T}(z)$ for the different values of $\omega$ when $\Omega_{m0} = 0.3$. The dotted, dashed, solid, and dotdashed lines
correspond to $\omega = -\fr{1}{2}, -\fr{5}{6}, -1$, and $-\fr{4}{3}$, respectively. The present values of $F_{T}$ vary from $-0.03$ to $-0.8$ for
$\omega = -\fr{1}{2}$ and $-\fr{4}{3}$, respectively. In the right panel, we show the evolutions of $F_{T}$ for the different values of $\Omega_{m0}$ when we consider the $\Lambda$CDM model.
The dotted, solid, and dotdashed lines correspond $\Omega_{m0} = 0.25, 0.30$, and $0.35$, respectively. We obtain $-0.18 \leq F_a \leq -0.14$ for
$0.25 \leq \Omega_{m0} \leq 0.35$. We investigate the accuracy of this fitting form for the different
cosmological models. In the second row of Fig.\ref{fig11}, we show the errors of the fitting form as a function of the redshift $z$ for the different models.
In the first column, we show the $\fr{\Delta F_{T}}{F_{T}}$ for the different values of $\omega$ when we fix $\Omega_{m0} = 0.3$.
The errors are less than about $4$ \% for all models when we consider $z \leq 5$. In the second column, we check the errors of the fitting form for the different $\Omega_{m0}$ values when we consider the $\Lambda$CDM model. Again, the errors are less than $3$ \% for all the considered $\Omega_{m0}$ values up to $z \leq 5$.

\begin{figure}
\centering
\vspace{1.5cm}
\begin{tabular}{cc}
\epsfig{file=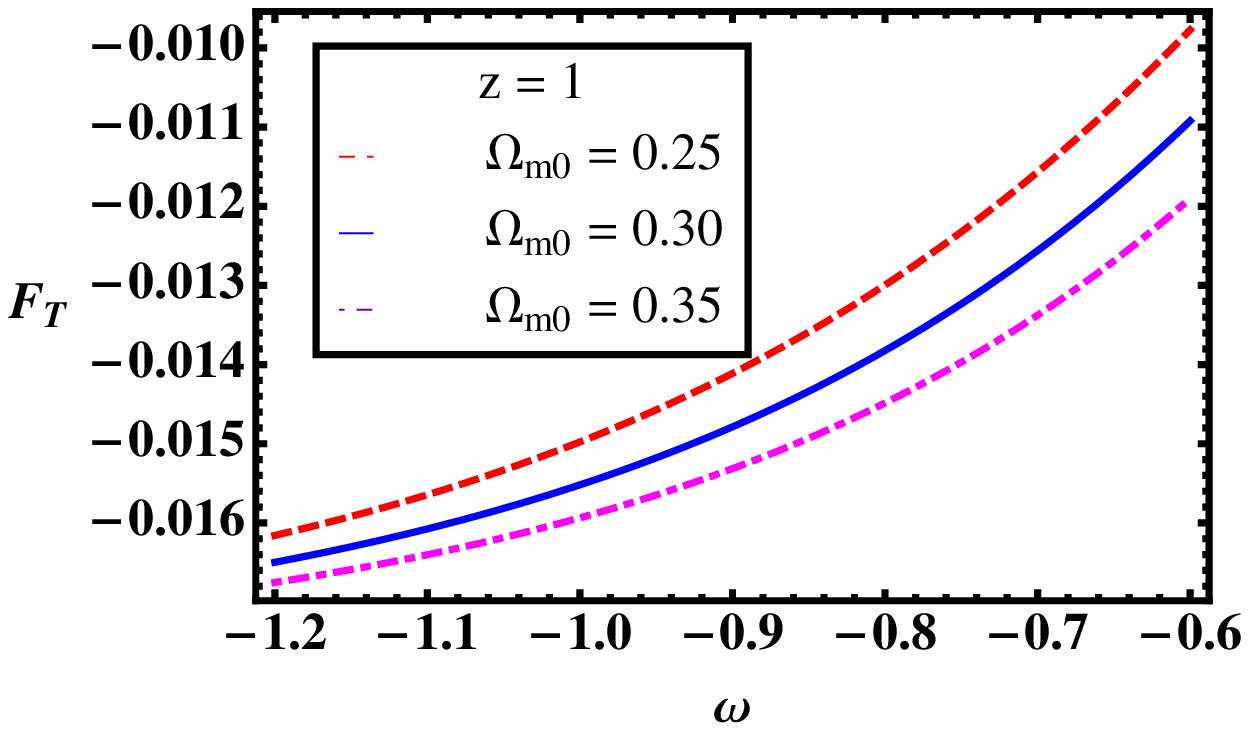,width=0.5\linewidth,clip=} &
\epsfig{file=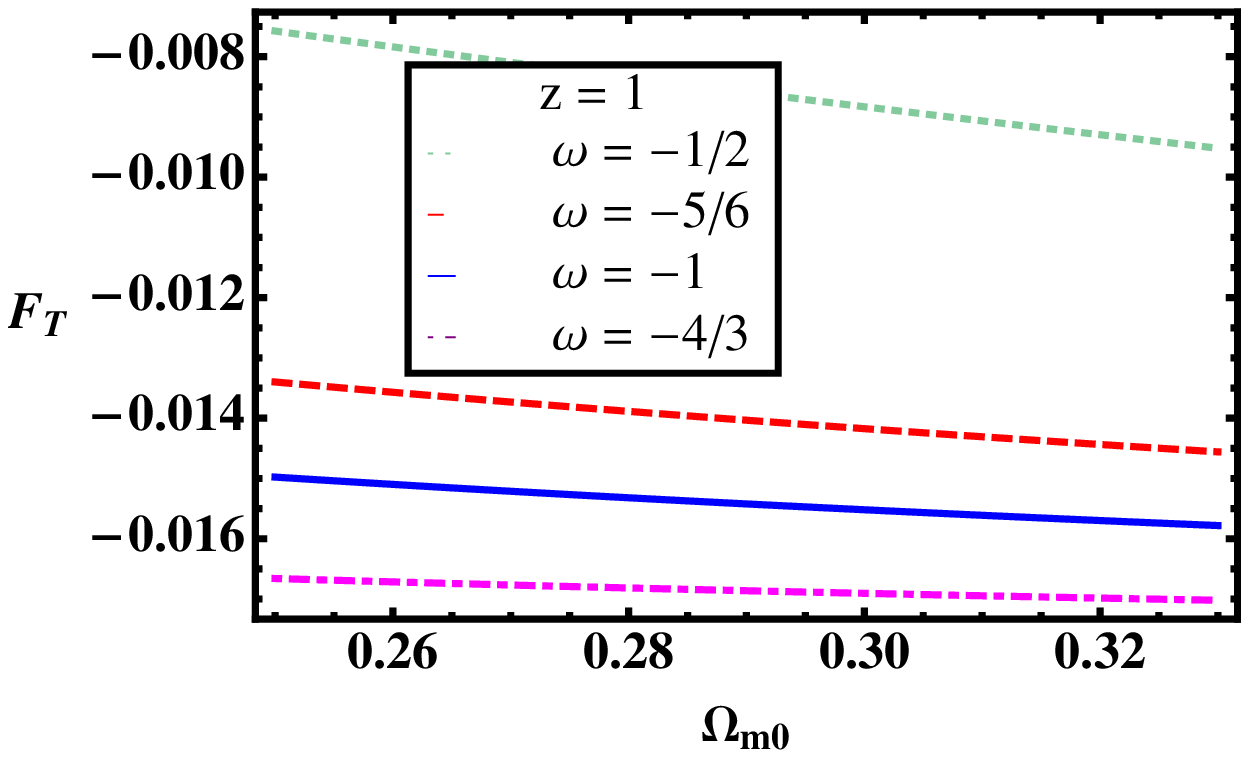,width=0.5\linewidth,clip=} \\
\epsfig{file=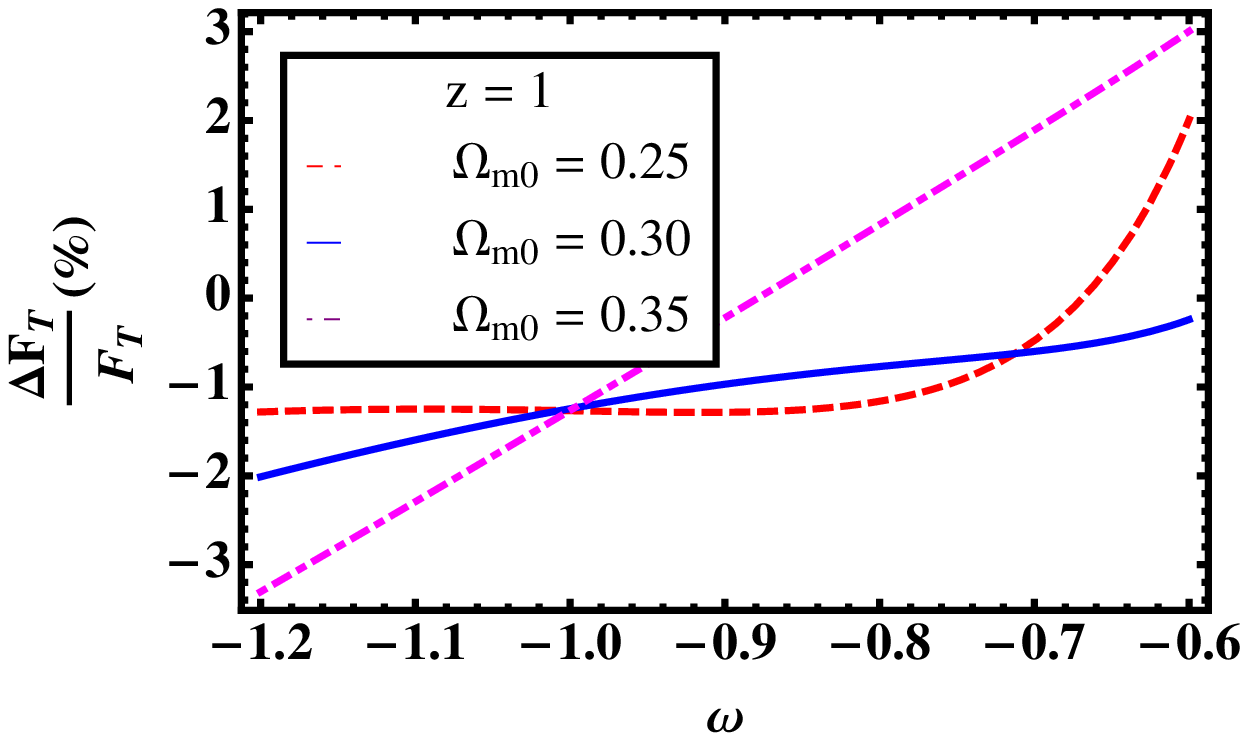,width=0.5\linewidth,clip=} &
\epsfig{file=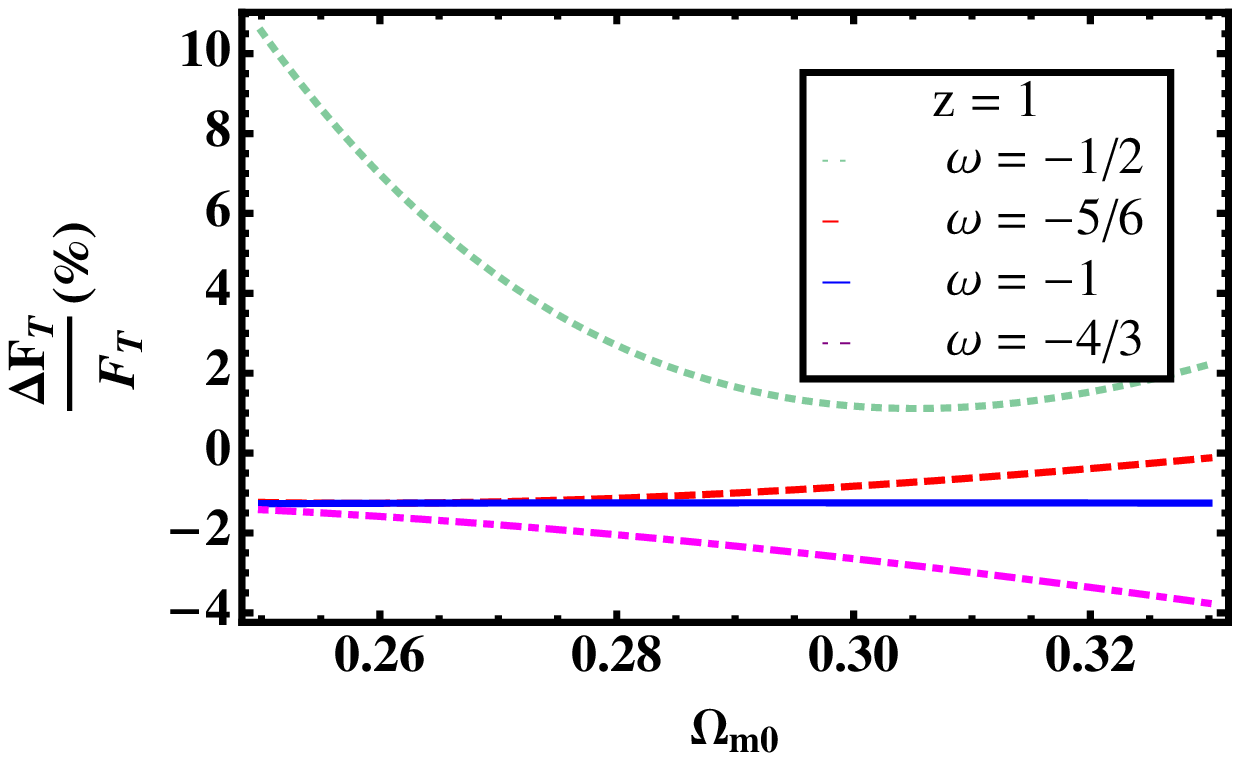,width=0.5\linewidth,clip=} \\
\end{tabular}
\vspace{-0.5cm}
\caption{{\bf Firs row} : The values of $F_T$ for the different cosmological parameters at $z=1$. a) $F_T$ dependence on $\omega$ for the different values of $\Omega_{m0}$. Dotdashed, solid, and dashed lines correspond to $\Omega_{m0} = 0.35, 0.30$, and $0.25$, respectively. b) $F_T$ dependence on $\Omega_{m0}$ for the different values of $\omega$. Dotted, dashed, solid, and dotdashed lines correspond to $\omega = -\fr{1}{2}, -\fr{5}{6}, -1$, and $-\fr{4}{3}$, respectively. {\bf Second row} : The errors on the fitting form at $z = 1$ for the different cosmological models. c) The dependence of errors on $\omega$ for the different values of $\Omega_{m0}$. d) The errors as a function of $\Omega_{m0}$ for the different values of $\omega$.} \label{fig12}
\end{figure}
We also investigate the dependence of $F_T$ on the cosmological parameters at the specific $z$. In the first row of Fig. \ref{fig12}, we show the values of $F_T$ as a function of $\omega$ and $\Omega_{m0}$ at the specific redshift $z$. In the left panel, we fix the redshift $z = 1$ and check the dependence of $F_T$ on $\omega$ for the different values of $\Omega_{m0}$. The dashed, solid, and the dotdashed lines correspond to $\Omega_{m0} = 0.25, 0.30$, and $0.35$, respectively. For $\Omega_{m0} = 0.35$, $F_T$ varies from -0.01 to -0.017 when $\omega$ changes from -0.5 to -1.2. We also obtain $-0.016 \leq F_T \leq -0.008$ for $-1.2 \leq \omega \leq -0.5$ when $\Omega_{m0} = 0.25$. In the right panel, we show the dependence of $F_T$ on $\Omega_{m0}$ for the different $\omega$ models at $z = 1$. Again, the dotted, dashed, solid, and dotdashed lines correspond to $\omega = -\fr{1}{2}, -\fr{5}{6}, -1$, and $-\fr{4}{3}$, respectively. For $\omega = -\fr{1}{2}$, $FTa$ varies from -0.008 to -0.01 when $\Omega_{m0}$ changes from 0.25 to 0.35. $F_T$ changes from -0.017 to -0.016 when $\Omega_{m0}$ changes from 0.25 to 0.35 for $\omega = -\fr{4}{3}$. This case $F_T$ is almost constant for the different values of $\Omega_{m0}$. Thus, we can conclude that $F_a$ dependence on $\Omega_{m0}$ becomes weaker as $\omega$ decreases. All of these properties are same as those of $F_{a}$. We also investigate the errors on the fitting form at $z = 1$ for the different cosmological models. In the second row of Fig. \ref{fig12}, we show the errors on the
fitting form as a function of $\omega$ and $\Omega_{m0}$. In the first column, we show the dependence of errors on $\omega$ for the different values of $\Omega_{m0}$. The errors are about less than $3$ \% for $-1.2 \leq \omega \leq - 0.7$. In the second column, the errors are depicted as a function of $\Omega_{m0}$ for the different $\omega$ models at $z = 1$. The errors are less than 4 \% for all $\Omega_{m0}$ except $\omega = -\fr{1}{2}$.

\section{Discussion and Conclusions}
We reinvestigate solutions for the Lagrangian perturbation theory of an irrotational fluid up to the third order for the Einstein-de Sitter and open universe. With the correct initial conditions and the proper consideration for the fastest growing mode solutions, we correct the known solutions for these models. For the first time, we obtain the analytic approximate solutions for the general dark energy with the constant equation of state. These fitting forms have less than 5 \% errors compared to the numerical solutions for all orders up to $z \leq 3$.

So far, one have used the EdS solutions for the time component to incorporate the power spectrum or higher order moments even when one adopt the cosmology dominated by the dark energy at present epoch. This is an inaccurate approximation. Thus, with our analytic approximate solutions one can consider the correct dark energy dependent on observable \cite{SL}. Even though these solutions are obtained for the constant equation of state dark energy models, one can apply these solution to the time varying dark energy models by interpolating between models with the constant equation of states \cite{09072108}.

These analytic solutions provide us the tools to the systematic study for the dependence of the solutions for each order on both the matter energy density and the dark energy equation of state. We are also able to investigate the time dependence of models to scan for large parameter spaces with solutions.

\section{Acknowledgements}
We would like to thank Cornelius Rampf for fruitful discussion and comments on the manuscript. We specially thanks to the anonymous referee for the useful comments. We also thank KIAS Center for Advanced Computation for providing computing resources.

\section{Appendix}
\setcounter{equation}{0}

\subsection{Einstein de Sitter Universe}

Then, we can rewrite the Friedmann equation as
\be H^2 = \Bigl(\fr{\alpha}{a^3} \fr{da}{d \tau} \Bigr)^2 = \fr{8 \pi G \rho_{m0}}{3} a^{-3} \,\, \rightarrow \,\, \fr{da}{d \tau} = \pm \sqrt{\fr{8 \pi G \rho_{m0}}{3 \alpha^2}} a^{\fr{3}{2}} \, . \label{FrddS} \ee (Using $\int a^{-\fr{3}{2}} da = -\fr{2}{\sqrt{a}}$) We obtain
\be -2 \Bigl(\fr{1}{\sqrt{a}} - \fr{1}{\sqrt{a_i}} \Bigr) = \pm \sqrt{\fr{8 \pi G \rho_{m0}}{3 \alpha^2}} (\tau - \tau_i) \, . \label{tauadS} \ee If we use the fact that $a_{i} \rightarrow \infty$ as $\tau_{i} \rightarrow 0$, then we obtain
\be -\fr{2}{\sqrt{a}} =  \sqrt{\fr{8 \pi G \rho_{m0}}{3 \alpha^2}} \tau \,\, \rightarrow \,\, \fr{1}{\sqrt{a}} =  \fr{\tau}{\tau_0} \, , \label{tauadS2} \ee where we adopt $a_{0} =1$, then $\tau_0 = -\fr{2}{\sqrt{\fr{8 \pi G \rho_{m0}}{3 \alpha^2}}}$.

\subsection{Particular solution}

Let us consider the nonhomogeneous second order differential equation
\be f''[y] + P(y) f'[y] + Q[y] f[y] = G[y] f_{h}^2 \, , \label{inh2DE} \ee where $f_{h}$ means the homogeneous solution of the above equation. We can put $f_h(y) = c_{1} f_1 + c_2 f_2$. Then, we obtain the equation as
\be c_{1} (f_1'' + P f_1' + Q f_1) = c_{2} (f_2'' + P f_2' + Q f_2 ) = 0 \, . \label{h2DE} \ee Now we consider the nonhomogeneous (particular) solution as $f_{p} = ( c_{3} f_1 + c_{4} f_2 )^2$ and plug this trial solution into Eq. (\ref{inh2DE}) to get
\be 2 (c_3 f_1 + c_4 f_2)(c_3 f_1'' + c_4 f_2'') + 2(c_3 f_1' + c_4 f_2')^2 + 2 P (c_3 f_1 + c_4 f_2) (c_3 f_1' + c_4 f_2') + Q (c_3 f_1 + c_4 f_2)^2 = G (c_1 f_1 + c_2 f_2)^2 \label{inh2DE2} \ee After we rearrange the equation by using the homogeneous solution, then we obtain
\be 2 (c_3 f_1' + c_4 f_2')^2 = Q (c_3 f_1 + c_4 f_2)^2  + G (c_1 f_1 + c_2 f_2)^2 \, . \label{inh2DE3} \ee We can find the relation between $c_1, c_2$ and $c_3, c_4$ in EdS universe where $Q = G = -\fr{6}{y^2}$ with $f_{1} = y^{-2}$ and $f_{2} = y^{3}$. Then, the above Eq. (\ref{inh2DE3}) becomes
\be 2 (-2 c_3 y^{-3} + 3 c_4 y^{2})^2 = -\fr{6}{y^2} \Biggl[ ( c_3 y^{-2} + c_4 y^{3})^2 + (c_1 y^{-2} + c_2 y^{3})^2 \Biggr) \label{inh2DE4} \ee  The above equation produces
\be c_3^2 = -\fr{3}{7} c_{1}^2 \,\,\,\, , \,\,\,\, c_4^2 = -\fr{1}{4} c_{2}^2 \,\,\,\, , \,\,\,\, c_3 c_4 = c_1 c_2 \label{c3c1c4c2} \ee Thus, the particular solution becomes
\be f_{p}(y) = -\fr{3}{7} c_1^2 f_{1}^2 + 2 c_1 c_2 f_{1} f_2 -\fr{1}{4} c_2^2 f_2^2 \label{fpy} \ee

\end{document}